\documentclass[twocolumn,showpacs,preprintnumbers,amsmath,amssymb]{revtex4}

\usepackage{graphicx}

\usepackage{dcolumn}

\usepackage{bm}

\begin{document}

\def\eps{$\epsilon$-expansion}
\def\xp{\xi_\perp}
\def\xz{\xi_z}
\def\xpnl{\xi^\perp_{NL}}
\def\xznl{\xi^z_{NL}}
\def\nhat{${\hat n}(\vec r)$ }


\preprint{APS/123-QED}

\title{Smectics $A$ and $C$, and the Transition between them, in Uniaxial
Disordered Environments}
\author{Leiming Chen and John Toner}
\affiliation{Department of Physics and Institute of Theoretical
Science, University of Oregon, Eugene, OR 97403}
\date{\today}

\begin{abstract}
We present a theory of the elasticity and fluctuations of the
Smectic A and C phases in  uniaxial, anisotropic disordered
environments, e.g., stretched aerogel. We find that, bizarrely,
the {\it low-temperature}, {\it lower-symmetry} Smectic $C$ phase
is {\it less} translationally ordered than the {\it
high-temperature}, {\it higher-symmetry} Smectic $A$ phase, with
short-ranged ``$m=1$ Bragg glass'' and algebraic ``$XY$ Bragg
glass'' order, respectively.  The $AC$ phase transition belongs to
a new universality class, whose fixed points and exponents we find
in a $d=5-\epsilon$ expansion. We give very detailed predictions
for the very rich light scattering behavior of both phases, and
the critical point.

\end{abstract}
\pacs{61.30.Dk, 64.60.Fr,
64.70.Md, 82.70.-y}
\maketitle


\section{Introduction}

Randomly pinned elastic media occur in many contexts, including
disordered superconductors \cite{SC}, charge density waves
\cite{CDW}, Josephson junction arrays \cite{J} and Helium in
aerogel \cite{Helium}. By far the most exotic phenomenon (at least
in the - admittedly biased -  opinion of the current authors) in
such systems occurs in liquid crystals in aerogel:  anomalous
elasticity.  That is, many liquid crystals in aerogel exhibit
scalings of their elastic energies that differ radically
(specifically, by non-trivial power laws) from those found in the
absence of pinning.

However, despite the considerable amount of prior work done on
these problems, there
had, until recently \cite{us},
been no previous theoretical work on phase
transitions in pinned liquid crystal systems and little on phase
transitions in {\it any} pinned elastic system.  In this paper we
remedy this  by treating the smectic $A$ to smectic $C$
(hereafter $AC$ ) transition in an {\it anisotropic}, {\it
uniaxial} disordered environment. A brief summary of a few of our
results
has already appeared in \cite{us}. Such an environment could be
realized, e.g.,
by absorbing the liquid crystal in uniaxially stretched aerogel. For
brevity,
we will hereafter refer to the special uniaxial direction as the
$z$-axis of
our co-ordinate system.

The $AC$ phase transition
in such a  system
separates the two novel, anomalously elastic glassy phases
discovered (in a totally different context) and treated in
reference\cite{Karl}. The high temperature phase ($T>T_{AC}$) is the
glassy analog of the smectic $A$ phase of the pure problem, in that
both the layer normal $\hat{N}$ and nematic director $\hat{n}$ lie,
on average, along the $z$-axis. This phase is in the universality
class of the random field $XY$ model\cite{XY}; hence, following
\cite{Karl}, we call it the ``random field $XY$ smectic Bragg
glass'', or $XYBG$ for short.

The low temperature phase is the
glassy analog of the smectic $C$ phase, in that both the layer
normal $\hat{N}$ and nematic director $\hat{n}$ tilt from
$z$-axis. This tilting obeys the ``geometrical constraint''   that
$\hat{N}$, $\hat{n}$ and $\hat{z}$ are in the same plane, with
$\hat{z}$ between $\hat{N}$ and $\hat{n}$ and the angles between
these three vectors non-fluctuating, albeit temperature dependent.
Hence, as in the smectic C phase
in an {\it isotropic} environment, the only new ``Goldstone mode''
degree of
freedom associated with the tilting is the overall azimuthal angle of
rotation
of the vectors
$\hat{N}$ and $\hat{n}$ about $\hat{z}$.

The temperature dependence of the angles $\theta_{L} (T)$ between
$\hat{N}$ and $\hat{z}$ and $\theta_{n} (T)$ between $\hat{n}$ and
$\hat{z}$ are the same near $T_c$; that is, the ratio $\theta_L /
\theta_n$ is a (negative) constant near $T_c$. Hence, we are free to
choose either the layer normal tilting angle $\theta_L(T)$,  or the
molecular tilting angle $\theta_n(T)$, both of which can be measured
experimentally, as the magnitude of the order parameter for the AC
transition.

The smectic C phase is in the universality class of the ``$m=1$
smectic Bragg glass'' phase studied in \cite{Karl}.

We call both the A and C phases
   ``glassy'' because both lack
long-ranged translational order due to the disordering effect of
the random environment (i.e. the aerogel). The extent of this
destruction, however, differs greatly between the two phases.
Strikingly, it is the {\it low}-temperature, higher-symmetry,
``Smectic $C$ glass'' that has {\it less} translational order. In
the ``glassy smectic $A$'' or ``$XYBG$ '' phase, translational
correlations are ``quasi-long-ranged'', by which we mean they
decay as power laws with distance. In the ``glassy smectic $C$''
or ``$m=1$ Bragg glass'' phase, these correlations are
short-ranged. These differences in the translational correlations
lead to radically different X-ray scattering signatures in the two
phases which we will now describe.

In the ``glassy $A$'' or ``$XYBG$'' phase, the X-ray scattering
intensity $I(\vec{q})$ diverges near the smectic Bragg peaks,
which occur at $\vec{q}=nq_0\hat{z}$ for all $n$ is integer, where
$q_0={{2\pi}/ a}$, with $a$ the smectic layer spacing. Because of
the lack of true, long-ranged translational order, this divergence
is {\bf not} in the form of a delta-function; rather, it is an
integrable power-law divergence:
\begin{eqnarray}
I\left(\vec{q}\right)\propto [\left(q_z-nq_0\right)^2+\alpha
q_{\perp}^2] ^{{-3+.55n^2}\over 2}, \label{Intensity}
\end{eqnarray}
where $\alpha$ is a non-universal constant of order 1 and
$q_{\perp}$ is the magnitude of the projection of $\vec{q}$
perpendicular to $\hat{z}$. Note that the power law $-3+.55n^2$
characterizing the divergence of the $n$th peak depends on which
peak we are considering. Indeed, only the first 2 peaks ($n=1$ and
$n=2$) actually diverge.

In contrast, in the ``glassy $C$'' or ``$m=1$ $BG$'' phase, the
peaks in the X-ray scattering intensity are broad, with
$I(\vec{q})$ finite for all $\vec{q}$.

As $T\to T_{AC}$ from above (i.e., on the $A$ side), the sharpness
of the peak disappears in an unusual way. The peaks look broad and
qualitatively Lorentzian for $\vec{q}$ 's sufficiently far from the
Bragg peak position $nq_0\hat{z}$, while for $\vec{q}$ 's
sufficiently close to the peak, it follows the power law divergence
(\ref{Intensity}). ``Sufficiently close'', in this context, means
that {\it both} $|\vec{q}_{\perp}|\ll \delta q_{\perp}^c(n,T)$, {\it
and} $|q_z-nq_0|\ll {\delta q_z^c(T)}$, where the crossover
wavevectors $\delta q_{\perp}^c(n,T)$, $\delta q_z^c(n,T)$ are given
by
\begin{eqnarray}
\delta q^c_{\perp}(n,
T)&\propto&\left(\xi_{\perp}^c\right)^{-{n^2\over
{3-0.55n^2}}}\, ,\\
\delta q^c_z(n, T)&\propto&\left(\xi_z^c\right)^{-{n^2\over
{3-0.55n^2}}}\label{},
\end{eqnarray}
where $n$ again is the index of the Bragg peak, and
$\xi_{\perp}^c$ and $\xi_z^c$ are correlation lengths along and
perpendicular to the smectic layers respectively that both diverge
extremely strongly as $T\to T_{AC}$. Specifically:
\begin{eqnarray}
\xi_{\perp, z}^c\propto \exp(A\left|T-T_{AC}\right|^{-\Omega})
\label{xic},
\end{eqnarray}
where $\Omega$ is a universal exponent which we have calculated  in the
$\epsilon$-expansion discussed below, and $A$ is a non-universal
constant.

\begin{figure}
   \includegraphics[width=0.35\textwidth,angle=90]{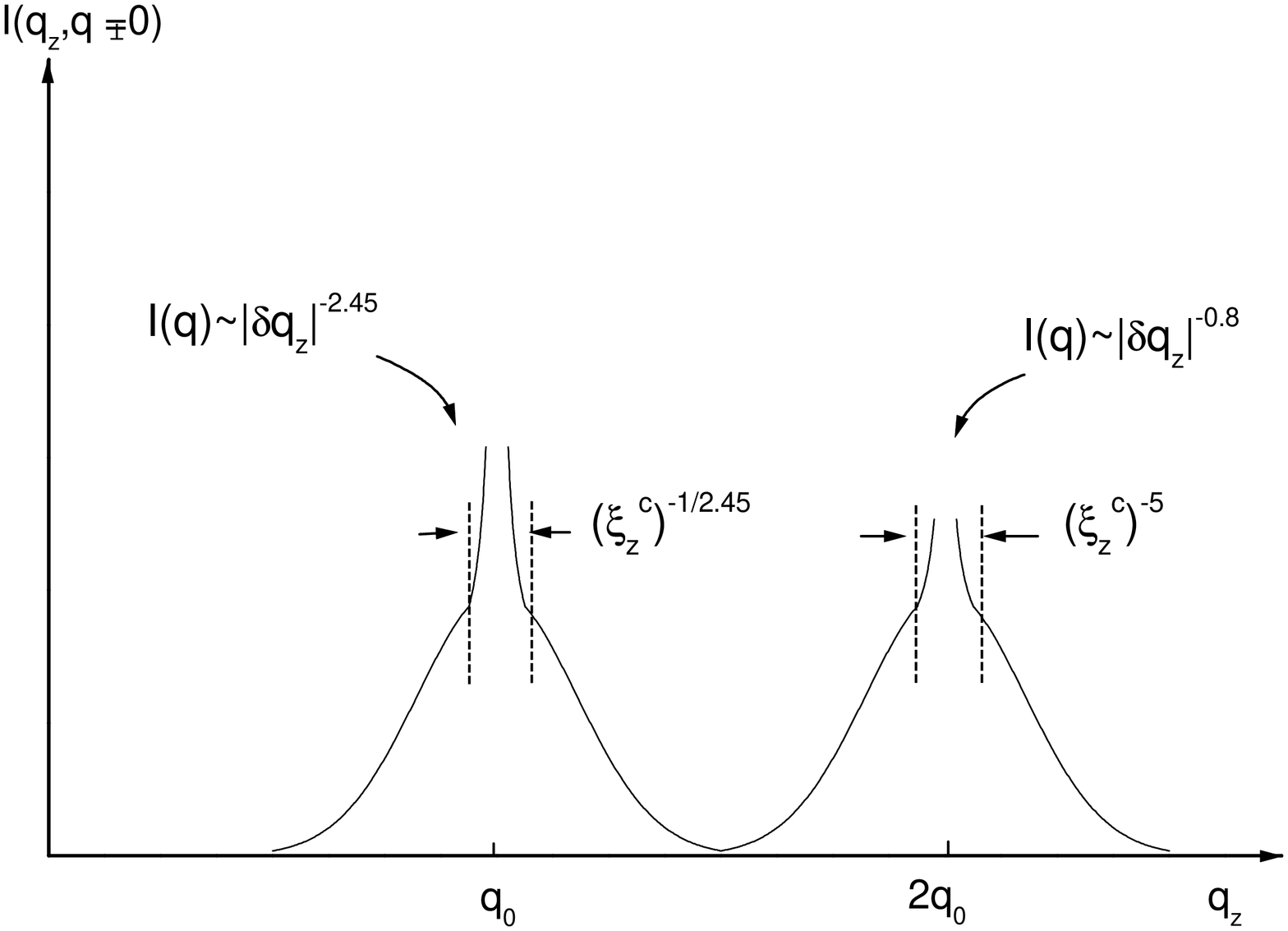}
   \caption{\label{fig: Xrayz}The $q_z$-dependence of the X-ray scattering intensity
   for $q_{\perp}=0$ in the smectic $A$ phase. In the $C$ phase, the
   sharp, power law peaks disappear, leaving only the broad peak.}
   \end{figure}

\begin{figure}
\includegraphics[width=0.35\textwidth,angle=-90]{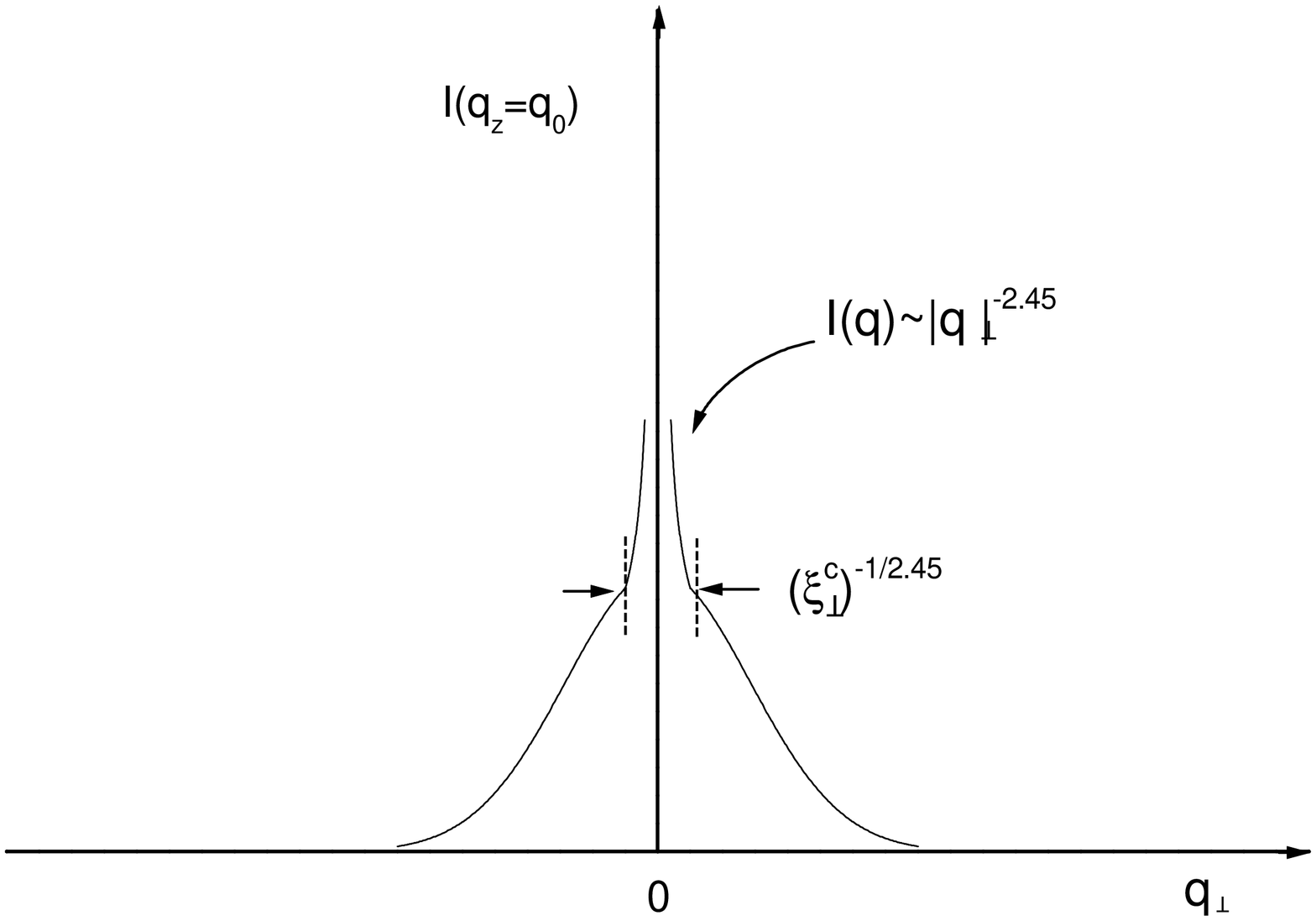}
\caption{\label{fig: Xrayperp}The $q_{\perp}$-dependence of the
X-ray scattering intensity for $q_z=q_0$ in the smectic $A$ phase.
Again, the sharp peak vanishes in the $C$ phase.}
\end{figure}

These X-ray scattering predictions are illustrated in Fig.
\ref{fig: Xrayz} and \ref{fig: Xrayperp}. The divergence of
$\xi_{{\perp}, z}^c$ implies that, as $T\to T_{AC}^+$, the
algebraic ``spikes'' on the top of the broad quasi-long-ranged
peaks get narrower and less intense, vanishing completely at
$T_{AC}$. Lowering temperature further leads only to the broad
peaks of the Smectic $C$ phase. This entire scenario of sharp
peaks at high temperature and broad peaks at low temperature is
completely counterintuitive, and contrary to the behavior seen in
almost every other translationally ordered system \cite{foot}.

We turn now to the smectic $C$ phase, and in particular to the
question of why this phase, though of lower symmetry than the
smectic $A$ phase, is {\it less} translationally ordered. In fact,
it is precisely the new broken symmetry of the smectic $C$
phase-that is, the tilt of the layer normal and nematic
director-that causes this. This is because, while the
energetically preferred layer normal $\hat{N}$ and nematic
director $\hat{n}$ in the smectic $A$ phase are {\it unique}-they
must point {\it along} $\hat{z}$-there are {\it infinitely} {\it
many} energetically preferred orientations of $\hat{N}$ and
$\hat{n}$ in the $C$ phase: $\hat{N}$ can lie anywhere on a cone
making an angle $\theta_{L}(T)$ with $\hat{z}$, which, combined
with the ``geometrical constraint'' (described in the third
paragraph) on the directions of $\hat{N}$ and $\hat{n}$, also
determines the similar cone of directions $\hat{n}$ can point.
This {\it exact} symmetry of the elastic energy of the smectic $C$
phase means that the direction perpendicular to the
$\hat{z}-<\hat{N}>$ plane (where $<\hat{N}>$ is the mean of
$\hat{N}$; i.e., the direction of spontaneous tilt) becomes
``soft'': that is, an easy direction for layer displacements to
vary in. Precisely such softness occurs (for different reasons) in
the ``$m=1$ smectic'' studied in \cite{Karl} and, indeed,
expanding our elastic Hamiltonian around the new broken symmetry,
tilted smectic $C$ ground state, we obtain an elastic Hamiltonian
describing small positional fluctuations about the tilted smectic
$C$ ground state, which, after some manipulation, proves to be
identical to that studied for ``$m=1$ smectic'' in \cite{Karl}.
Thus, we can simply transcribe the results of \cite{Karl} to this
problem.

The most striking result of reference \cite{Karl} is that this phase
exhibits a bizarre phenomenon known as ``anomalous elasticity''.
That is, the elastic moduli that characterize the elastic response
of this phase can no longer be called ``elastic constants'', because
they are {\it not} constants. Rather, they become {\it singular}
functions of the length-scale, or, equivalently, the wavevector, at
which they are measured.
%
%
%

The elastic constants that characterize the smectic C phase in
uniaxial disordered media are defined in (\ref{H_C}), which gives
the ``replicated'' elastic Hamiltonian describing the disordered,
glassy smectic C phase. Of these,
%
%
%
%
%
%
%
%
%
%
%
%
%
the bend modulus $\tilde{K}$, and the disorder variances or
effective disorder strengths $\Delta_{s'}$ and $\Delta_{x'}$
diverge, while the in-plane tilt modulus $\gamma$ (which gives the
energy cost for changing the angle between $\hat{N}$ and $\hat{z}$
from $\theta_L(T)$) vanishes, as $|\vec{q}|\to 0$, according to
%
%
%
%
%
%
%
%
\begin{widetext}
\begin{eqnarray}
\tilde{K}\cong\left\{
\begin{array}{ll}
\tilde{K}_c(T) (\xi_\perp q_{s'})^{-{\tilde{\eta}_K}}, &(\xi_\perp
q_{s'})^{\tilde{\zeta}_{x'}}\gg \xp q_{x'}, (\xp
q_{s'})^{\tilde{\zeta}_{z'}}\gg \xi_z q_{z'}
\\
\tilde{K}_c(T) (\xi_\perp
q_{x'})^{-{\tilde{\eta}_K}/\tilde{\zeta}_{x'}}, &\xp q_{x'}\gg (\xp
q_{s'})^{\tilde{\zeta}_{x'}}, \xp q_{x'}\gg (\xz
q_{z'})^{\tilde{\zeta}_{x'}\over\tilde{\zeta}_{z'}}
\\
\tilde{K}_c(T) (\xz q_{z'})^{-{\tilde{\eta}_K}/\tilde{\zeta}_{z'}},
&\xz q_{z'}\gg (\xp q_{s'})^{\tilde{\zeta}_{z'}}, \xz q_{z'}\gg (\xp
q_{x'}) ^{\tilde{\zeta}_{z'}\over\tilde{\zeta}_{x'}}
\end{array}\right. ,
\label{KanomC}
\end{eqnarray}
\end{widetext}

\begin{widetext}
\begin{eqnarray}
\Delta_{s',x'}\cong\left\{
\begin{array}{ll}
\Delta_{s',x'}^c(T) (\xi_\perp q_{s'})^{-{\tilde{\eta}_{s',x'}}},
&(\xi_\perp q_{s'})^{\tilde{\zeta}_{x'}}\gg \xp q_{x'},
(\xp q_{s'})^{\tilde{\zeta}_{z'}}\gg \xi_z q_{z'}
\\
\Delta_{s',x'}^c(T) (\xi_\perp
q_{x'})^{-{\tilde{\eta}_{s',x'}}/\tilde{\zeta}_{x'}}, &\xp q_{x'}\gg
(\xp q_{s'})^{\tilde{\zeta}_{x'}}, \xp q_{x'}\gg
(\xz q_{z'})^{\tilde{\zeta}_{x'}\over\tilde{\zeta}_{z'}}
\\
\Delta_{s',x'}^c(T) (\xz
q_{z'})^{-{\tilde{\eta}_{s',x'}}/\tilde{\zeta}_{z'}},
&\xz q_{z'}\gg (\xp q_{s'})^{\tilde{\zeta}_{z'}},
\xz q_{z'}\gg (\xp q_{x'})
^{\tilde{\zeta}_{z'}\over\tilde{\zeta}_{x'}}
\end{array}\right. ,
\label{DeltaanomC}
\end{eqnarray}
\end{widetext}
\begin{widetext}
\begin{eqnarray}
\gamma\cong\left\{
\begin{array}{ll}
\gamma_c(T) (\xi_\perp q_{s'})^{{\tilde{\eta}_\gamma}},
&(\xi_\perp q_{s'})^{\tilde{\zeta}_{x'}}\gg \xp q_{x'},
(\xp q_{s'})^{\tilde{\zeta}_{z'}}\gg \xi_z q_{z'}
\\
\gamma_c(T) (\xi_\perp q_{x'})^{{\tilde{\eta}_\gamma}/\tilde{\zeta}_{x'}},
&\xp q_{x'}\gg
(\xp q_{s'})^{\tilde{\zeta}_{x'}}, \xp q_{x'}\gg
(\xz q_{z'})^{\tilde{\zeta}_{x'}\over\tilde{\zeta}_{z'}}
\\
\gamma_c(T) (\xz
q_{z'})^{{\tilde{\eta}_\gamma}/\tilde{\zeta}_{z'}}, &\xz q_{z'}\gg
(\xp q_{s'})^{\tilde{\zeta}_{z'}}, \xz q_{z'}\gg (\xp q_{x'})
^{\tilde{\zeta}_{z'}\over\tilde{\zeta}_{x'}}
\end{array}\right. ,
\label{gammaanomC}
\end{eqnarray}
\end{widetext}
where the wavevector-independent quantities $\tilde{K}_c(T)$,
$\Delta_{s',x'}^c(T)$, and $\gamma_c(T)$ are the ``half-dressed''
(i.e., renormalized by the critical fluctuations we will discuss in
a moment, but unrenormalized by smectic C fluctuations) values of
the corresponding elastic moduli and disorder variances;  we've
defined a new, {\it non-orthogonal} set of wavevector coordinates:
\begin{eqnarray}
q_{x'}&\equiv& q_x-\Gamma q_z  ,\label{qx}\\
q_{s'}&\equiv& q_s,\label{qs}\\
q_{z'}&\equiv& q_z\label{qz}
\end{eqnarray}
where   $q_s$ is the component of $\vec{q}$ perpendicular to the
$\hat{z}$-$<\hat{N}>$ plane, and $q_x$ is the component of $\vec{q}$
within the $\hat{z}$-$<\hat{N}>$ plane but orthogonal to $\hat{z}$,
$\xp (T)$ and $\xz (T)$ are temperature-dependent lengths, and
$\Gamma (T)$ is a dimensionless temperature-dependent constant. The
temperature dependences of $\xp (T)$, $\xz (T)$, $\tilde{K}_c(T)$,
$\Delta_{s',x'}^c(T)$, $\gamma_c(T)$ and $\Gamma (T)$ are all
singular at the transition temperature $T_{AC}$, with $\xp (T)$,
$\xz (T)$, $K_c(T)$, $\Delta_{s',x'}^c(T)$, and $\Gamma(T)$
diverging, and $\gamma_c(T)$ vanishing, as $T \rightarrow T_{AC}$.
These scaling laws governing their respective divergences and
vanishings will be given later in this introduction,  in section
\ref{sec: Cfunctions}, and in Appendix \ref{sec: Table}.

Here and throughout the rest of this paper, a tilde ( $\tilde{}$ )
over an exponent denotes an exponent describing anomalous
elasticity of the C phase.


The {\it universal} exponents ${\tilde{\eta}_K},
\tilde{\eta}_{\gamma}, \tilde{\eta}_{s', x'}, \tilde{\zeta}_{x'}$
and $\tilde{\zeta}_{z'}$ appearing in the above expressions  were
calculated in \cite{Karl}, using an approach that yielded
numerical estimates and error bars for them. We will review the
logic of their approach in Appendix \ref{sec: Average}; here, we
merely quote their results:
\begin{eqnarray}
\tilde{\eta}_K&=&0.50\pm0.03\, ,\label{Cnumexp1}\\
\tilde{\eta}_{\gamma}&=&0.26\pm0.12\, ,\label{Cnumexp2}\\
\tilde{\eta}_{s'}&=&0.132\pm0.002\, ,\label{Cnumexp3} \\
\tilde{\zeta}_{x'}&=&1.62\pm0.08 \, ,\label{Cnumexp4}\\
\tilde{\zeta}_{z'}&=&1.75\pm0.02 \, . \label{Cnumexp5}
\end{eqnarray}

These exponents also obey the {\it exact} scaling relations (in
$d=3$):
\begin{eqnarray}
\tilde{\zeta}_{x'}&=&2-\big({{\tilde{\eta}_{\gamma}+\tilde{\eta}_K}\over
2}\big)\label{zetax'}\, ,\\
\tilde{\zeta}_{z'}&=&2-{\tilde{\eta}_K\over 2}\label{zetaz'}\, ,\\
\tilde{\eta}_{s'}&=&{\tilde{\eta}_{\gamma}\over2}+ 2\tilde{\eta}_K-1
\label{scaling1}\, .
\end{eqnarray}

In addition to the above results derived in reference\cite{Karl},
we derive in section \ref{sec: ACTransition} an additional {\it
exact} scaling relation for $\tilde{\eta}_{x'}$:
\begin{eqnarray}
 \tilde{\eta}_{x'}=2+\tilde{\eta}_{s'}-\tilde{\eta}_K-\tilde{\eta}_{\gamma}\ ,
 \label{scaling3}
\end{eqnarray}
from which it follows, upon inserting the numerical results Eqns.
(\ref{Cnumexp1}, \ref{Cnumexp2}, \ref{Cnumexp3}) that
\begin{eqnarray}
\tilde{\eta}_{x'}&=&1.37\pm0.15\
\label{etaxnum}\ .
\end{eqnarray}
%
%
%
%

Having described the $A$ and $C$ phases
%
%
%
%
in this uniaxially disordered system,
%
%
%
%
we now turn to the transition between them.
%
%
%
%
We find that the upper critical dimension $d_{uc}$ for this
transition, below which it is no longer accurately described by a
purely Gaussian theory, is $d_{uc} = 5$.
%
%
%
%
We have studied this phase transition in an $\epsilon = 5 - d$
expansion, where $d$ is the dimension of the space filled by the
liquid crystal (i.e., $d = 3$ is the case of physical interest),
and find that there {\it is} a stable renormalization-group fixed
point, the existence of which implies a second-order phase
transition with {\it universal} critical behavior.
%
%
%
%
%
%
Of course, since $d_{uc}= 5$, $\epsilon = 2$ in the physically interesting
case $d = 3$, which is far too large an $\epsilon$ for the results for
exponents that we quote to be quantitatively reliable in $d=3$. However, we
expect the {\it qualitative} features of the transition that we find to be
robust down to
$d=3$ (although we will discuss some rather strong caveats to this statement
later).

The second-order nature of the transition is, as usual, manifested
in {\it universal} power-law dependence of many physical
observables on reduced temperature $T-T_{AC}$.
%
%
%
%
In particular, the layer normal and director tilt angle
$\theta_{L,n}(T)$ obeys
\begin{eqnarray}
\theta_{L,n}(T) = A_{L,n}\left(T_{AC}-T\right)^{\beta} \label{OP},
\end{eqnarray}
where the order parameter exponent $\beta$ {\it is} {\it
universal}, and given, to leading order in $\epsilon = 5 - d$, by
\begin{eqnarray}
\beta = {1 \over 2} - {\epsilon \over 10} +
O\left(\epsilon^2\right) \label{beta},
\end{eqnarray}
the amplitude $A_{L,n}$ are non-universal and depend on the
system.

We have also calculated the specific heat exponent
\begin{eqnarray}
\alpha = - {\epsilon \over 10} + O\left(\epsilon^2\right) \, .
\label{alpha}
\end{eqnarray}

The order parameter $\vec{N}_{\perp}$ for this transition is the
projection of the smectic layer normals $\hat{N}$ onto the $xy$
plane (i.e., perpendicular to the direction along which the
aerogel is stretched ). Above $T_{AC}$, the two point real space
correlations of $\vec{N}_{\perp}$ decay rapidly with distance,
with correlation lengths $\xi_z$ and $\xi_{\perp}$ parallel and
perpendicular to $z$-axis respectively which behave quite
differently as $T \to T_{AC}^+$. Specifically, both diverge as
power laws in $( T - T_{AC} )$:
%
%
%
%
\begin{equation}
\xi_{\perp, z} \propto |T -
T_{AC}|^{-\nu_{\perp, z}}
\label{xis}
\end{equation}
%
%
%
%
%
but with exponents $\nu_\perp$ and $\nu_z$ that differ from each
other due to the strong anisotropy of the problem.
%
%
%
 From our
$\epsilon$-expansion, we find
\begin{eqnarray}
\nu_{\perp} &=& {1 \over 2} + {3\epsilon \over 20} +
O\left(\epsilon^2\right) \label{no perp},\\
\nu_z &=& 1 + {3\epsilon \over 10} + O\left(\epsilon^2\right)
\label{no z}.
\end{eqnarray}

%
%
%
%
For $T < T_{AC}$, i.e., in the C phase, the correlation lengths $\xi_z$ and
$\xi_{\perp}$ can still be defined, now by looking at the {\it connected}
correlations of
${\hat N}$. They continue to be given by the scaling law (\ref{xis}), with the
same exponents (\ref{no perp} ) and (\ref{no z}).

In addition to their role as correlation lengths for ${\hat N}$, the
lengths $\xi_z$ and $\xi_{\perp}$ are also the ones that appear in
the scaling laws Eqns. (\ref{KanomC}), (\ref{DeltaanomC}) and
(\ref{gammaanomC}) for the anomalous elasticity in the C phase.

We also find that, as in the $C$ phase, the system exhibits
anomalous elasticity right at $T_{AC}$ as well. Specifically, right
at $T_{AC}$, the smectic layer bend modulus $K$ becomes strongly
wavevector-dependent, vanishing as $\vec{q} \rightarrow 0$ according
to the scaling laws
\begin{widetext}
\begin{eqnarray}
K\left(\vec{q}, T=T_{AC}\right) = K_0(\xpnl q_{\perp})^{-\eta_K} f_K
\left({(\xznl q_z ) \over (\xpnl q_{\perp})^{\zeta} }\right)
\approx\left\{
\begin{array} {ll}
K_0 (\xpnl q_{\perp})^{-\eta_K}, &\xznl q_z \ll (\xpnl q_{\perp})^{\zeta} \\
K_0 (\xznl q_{z})^{-{\eta_K\over \zeta}}, &\xznl q_z \gg (\xpnl
q_{\perp})^{\zeta}\\
\end{array}\right.
\label{K},
\end{eqnarray}
\end{widetext}
where $\eta_K<0$, and the anisotropy exponent

\begin{equation}
\zeta = 2 - {\eta _K
\over 2}
\label{critanis}
\end{equation}
This anisotropy exponent also obeys
\begin{equation}
\zeta=
{\nu_z/\nu_{\perp}}
\label{nuanis}.
\end{equation}

In Eq. (\ref{K}), the constant $K_0$ is the ``completely bare''
(i.e., unrenormalized by {\it any} fluctuations) value of the bend
modulus (i.e., the value it would have in the absence of the
aerogel). While weakly temperature dependent, like all elastic
moduli, $K_0$ has {\it no} critical singularity near $T_{AC}$.

Also in Eq. (\ref{K}) $\xi_{NL}^{\perp, z}$ are the length scales
in the $\perp$ and $z$ directions beyond which the elasticity
becomes anomalous. These lengths depend on the disorder strength,
and hence the aerogel density, and remain finite (and non-zero) as
$T\to T_{AC}$, although both diverge as the aerogel density
$\rho_A\to 0$. In three dimensions these length scales are given
as
\begin{eqnarray}
 \xi_{NL}^{\perp} &=& \left(64\pi\over 3\right)^{1/2} {K_0^{5/4}\over
 B_0^{1/4} {\Delta_t^0}^{1/2}}\, ,\label{NonlinearPerp}\\
 \xi_{NL}^z &=& {64\pi\over 3}{K_0^2\over \Delta_t^0}\label{NonlinearZ}\, ,
\end{eqnarray}
where $\Delta_t^0$ is the ''completely bare'' value of disorder
variance $\Delta_t$, which is discussed below.

The disordering effect of the random aerogel matrix can be
quantified by disorder variances $\Delta_t$ and $\Delta_c$
describing tilt and compressive stresses respectively. The former
arise due to random torques exerted by the aerogel on the smectic
layers, causing them to tilt, while the latter are caused by
random compression (or stretching) of the smectic layer spacing
induced by the aerogel, causing layers to move closer together (or
further apart). These variances also become anomalous at $T=T_{AC}$, obeying
\begin{widetext}
\begin{eqnarray}
\Delta_{t,c}\left(\vec{q}, T=T_{AC}\right) = \Delta_{t, c}^0(\xpnl
q_{\perp})^{-\eta_{t,c}} f_{t, c} \left({(\xznl q_z ) \over (\xpnl
q_{\perp})^{\zeta} }\right) \approx\left\{
\begin{array} {ll}
\Delta_{t,c}^0 (\xpnl q_{\perp})^{-\eta_{t,c}}, &\xznl q_z \ll
(\xpnl q_{\perp})^{\zeta} \\ \Delta_{t,c}^0 (\xznl
q_{z})^{-{\eta_{t,c}\over \zeta}}, &\xznl q_z
\gg (\xpnl q_{\perp})^{\zeta}\, .\\
\end{array}\right.
\label{Deltacrit}
\end{eqnarray}
\end{widetext}
where $\Delta_c^0$ is the ``complete bare'' value of $\Delta_c$.

Because our model at $T=T_{AC}$ proves remarkably similar
(although not quite identical) to that for a smectic $A$ in
isotropic disordered media, we expected, when we began our study,
to find anomalous behavior for the smectic layer compression
modulus $B$, with it {\it vanishing} as $\vec{q}\rightarrow 0$
according to
\begin{widetext}
\begin{eqnarray}
B\left(\vec{q}, T=T_{AC}\right) = B_0(\xpnl q_{\perp})^{\eta_B} f_B
\left({(\xznl q_z ) \over (\xpnl q_{\perp})^{\zeta} }\right)
\approx\left\{
\begin{array} {ll}
B_0 (\xpnl q_{\perp})^{\eta_B}, &\xznl q_z \ll (\xpnl q_{\perp})^{\zeta}\\
B_0 (\xznl q_{z})^{{\eta_B\over \zeta}}, &\xznl q_z \gg (\xpnl
q_{\perp})^{\zeta}\, ,\\
\end{array}\right.
\label{B}
\end{eqnarray}
\end{widetext}
where $B_0$ is the ``complete bare'' value of $B$, as found
\cite{LC} in the isotropic disordered smectic $A$ problem. To our
surprise, we found that, in this problem, $B$ exhibits no such
anomaly, remaining constant as $\vec{q}\rightarrow 0$.  This result
is {\it exact}, not an artifact of the $\epsilon=5-d$ expansion, and
we expect it to hold in $d = 3$. $f_{K, t, c, B}(x)$ are universal
scaling functions, which have the property:
\begin{eqnarray}
 f_{B, K, t, c}(x)=\left\{
 \begin{array}{ll}
 1, &x\ll 1\\
 x^{\eta_B, -\eta_{K, t, c}/\zeta}, &x\gg 1
 \end{array}\right.
 \label{ScalingFunction1}.
\end{eqnarray}

The exponents $\eta_K$, $\eta_t$ and $\eta_c$ {\it are} non-zero,
however. Both $\eta_K$ and $\eta_t$ {\it are} zero to
$O(\epsilon)$, but non-zero to $O(\epsilon^2)$, and given by
\begin{eqnarray}
\eta_K &=& C_K \epsilon^2+O(\epsilon^3) \label{eta_K}\, ,\\
\eta_t &=& C_{\Delta} \epsilon^2+O(\epsilon^3) \label{eta_Delta}
\end{eqnarray}
with $C_K = {\left(32\ln\left(4/3\right) - 10\right)/ 225}\cong
-.00353$ and $C_{\Delta} = {\left(12\ln\left(4/3\right) - {1\over
3}\right)/ 225}\cong .01386$. Note that $C_K < 0$, which implies
that $K$ {\it vanishes} as $\vec{q}\to 0$. This is another
unexpected result: $K$ {\it diverges} as $\vec{q}\to 0$ in every
other problem of this type \cite{LC,Karl,John}previously studied.
Of course, whether $\eta_K$ remains negative all the way down from
$d=5$ to $d=3$ remains an open question.

The remaining exponent
\begin{eqnarray}
\eta_c=2-{\epsilon\over5}+O(\epsilon^2). \label{eta_c}
\end{eqnarray}

\begin{figure}
\includegraphics[width=0.35\textwidth,angle=-90]{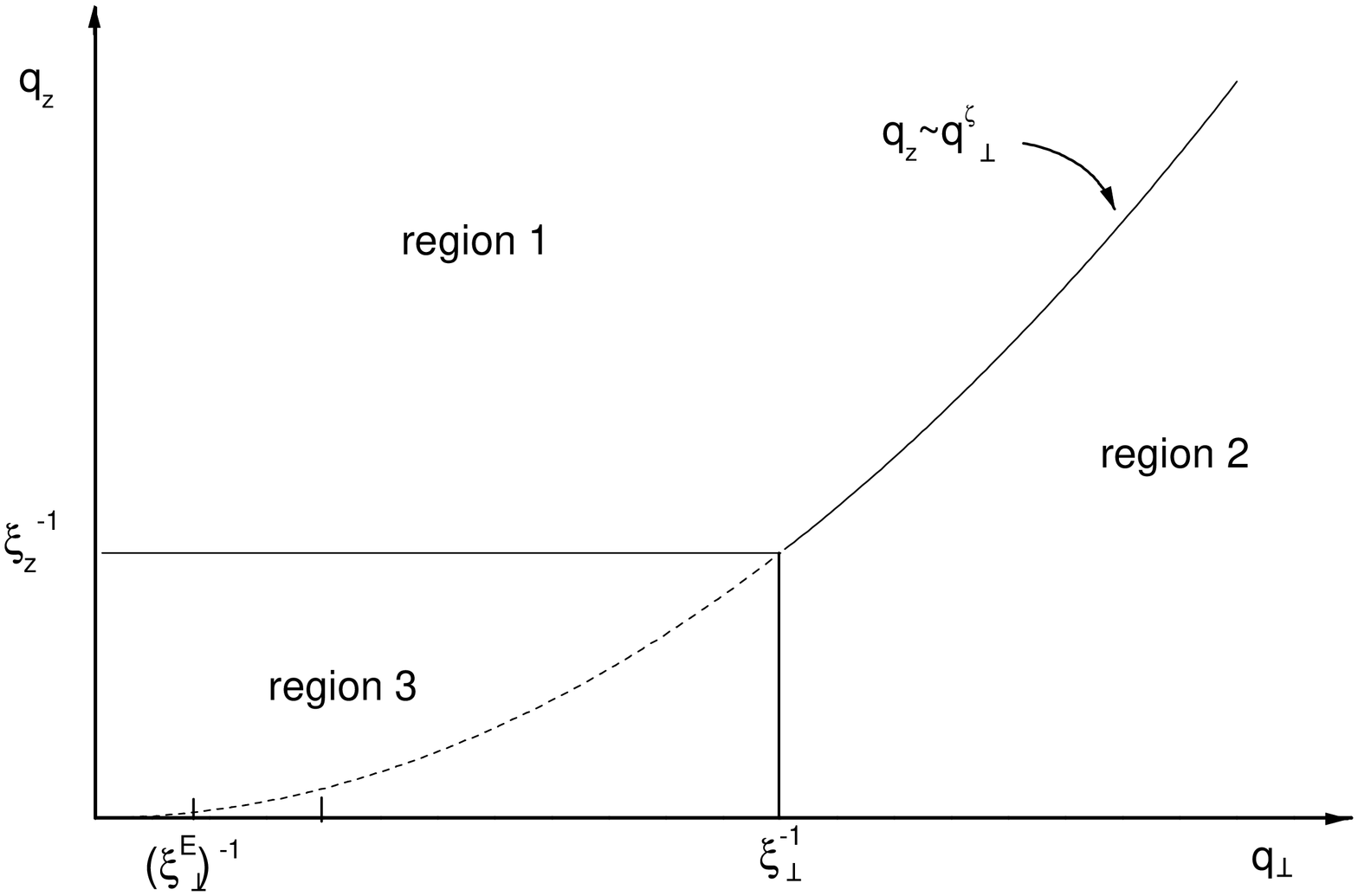}
\caption{\label{fig: 3regions}Illustration of the three distinct
regions in wavevector space with different wavevector-dependences of
$\Delta_{t, c}$ and $K$, which is given in Eq. (\ref{DeltaT}). }
\end{figure}

For $T$ bigger than $T_{AC}$ the (wavevector $\vec{q}$ and
temperature $T$-dependent) disorder variance $\Delta_{t, c}
\left(\vec{q}, T\right)$ and layer bend modulus $K \left(\vec{q},
T\right)$ are given by their $T = T_{AC}$ forms equations (\ref{K})
and (\ref{Deltacrit}) if {\it either} $q_{\perp}\xi_{\perp} \gg 1$
or $q_z\xi _z\gg 1$.  Otherwise (i.e., if {\it both}
$q_{\perp}\xi_{\perp} \ll 1$ and $q_z\xi _z\ll 1$), $\Delta_{t, c}
\propto \xi^{\eta_{t, c}}_{\perp}\propto\left(T - T_{AC}
\right)^{-\nu_{\perp}\eta_{t, c}}$ and $K
\propto\xi^{\eta_K}_{\perp} \propto \left(T
-T_c\right)^{-\nu_{\perp}\eta_K}$, results which can be (and, in
fact, were) straightforwardly obtained by matching the behavior for
{\it either} $q_{\perp}\xi_{\perp} \gg 1$ or $q_z\xi _z\gg 1$ to the
wavevector-independent behavior expected in the A phase when  {\it
both} $q_{\perp}\xi_{\perp} \ll 1$ and $q_z\xi _z\ll 1$. This can be
summarized as follows:
\begin{eqnarray}
\Delta_{t, c}\left(\vec{q}, T \right),  \ \ K \left(\vec{q}, T
\right)\propto \left\{
\begin{array}{ll}
q_z^{-\eta_{t, c, K}/\zeta} &\mbox{region 1}\\
q_{\perp}^{-\eta_{t, c, K}} &\mbox{region 2}\\
\xi_{\perp}^{\eta_{t, c, K}} &\mbox{region 3}
\end{array}\right.
\label{DeltaT},
\end{eqnarray}
where $\xi_{\perp}\propto\left(T - T_{AC} \right)^{-\nu_{\perp}}$,
and the three regions in $\vec{q}$ space are illustrated in Fig.
\ref{fig: 3regions}.

Having defined all of the critical exponents associated with this
problem, we can now give several {\it exact} scaling relations
between these exponents:
\begin{eqnarray}
\alpha &=& 2 - \nu _{\perp} \left(d - 1 + {\eta _K \over 2} - \eta
_t \right) \label{alpha scale},\\
\beta &=& \nu _{\perp} \left(2d - 6 + 3 \eta _K - 2 \eta _t
\right)/4 \label{beta scale},\\
\nu _z &=& \zeta\nu _{\perp} \label{nu z scale}.
\end{eqnarray}
Note that $\alpha$ does {\it not} obey the usual anisotropic
hyperscaling relation $\alpha = 2 - (d - 1) \nu _{\perp} - \nu_z$;
this is due to the strongly relevant disorder.

All of these exponents can be deduced experimentally from
measurements either for $T>T_{AC}$ or $T<T_{AC}$. The specific
heat can, of course, be measured by the usual thermodynamic
measurements on either side of the transition.

The spontaneous tilt angle $\theta_L(T)$ is another matter.  In
pure systems, this is simply the angle between the sharp Bragg
spot of the smectic $C$ and $\hat{z}$-axis. However, as discussed
earlier, in the Smectic $C$ glass phase the X-ray scattering peak
is broad, with a width that remains {\it finite} as $T \rightarrow
T^-_{AC}$. Hence, close to $T_{AC}$, the broad peak strongly
overlaps $\hat{z}$-axis, rendering measurement of $\theta_L(T)$ by
this approach impossible. Unfortunately this is {\it precisely}
the temperature range one needs to study to measure $\beta$.

Fortunately, an alternative measure of $\theta_L(T)$ is available
in the disorder averaged mean value of dielectric or diamagnetic
susceptibility tensors $\chi_{ij}$ and $\epsilon_{ij}$. In the $A$
phase, one of the principal axes of both tensors is along
$\hat{z}$-axis.  In the $C$ phase, this axis spontaneously rotates
away from $\hat{z}$-axis due to the spontaneous tipping of the
smectic layers and nematic directors. The rotation angle is
proportional to $\theta_L(T)$.

The best experimental probe of the critical phenomena (in
particular, the correlation lengths $\xp$ and $\xz$), as well as
of the anomalous elasticity  and disorder-induced fluctuations in
both the A and the C phases, is light scattering, which probes
fluctuations in both the dielectric($\epsilon_{ij}$) and
diamagnetic($\chi_{ij}$) susceptibility tensors.
%
%
%
%
%
%
%
%
%
%
%
%
%
%
%
%

Specifically,the large biaxial
fluctuations on {\it both} sides of the transition
%
%
%
%
%
%
%
lead to fluctuations in both $\epsilon_{ij}$ and $\chi_{ij}$ that
are proportional to \cite{Lubensky}
$C_{ij}(\vec{q})\equiv\overline{\left<N^{\perp}_i
\left(\vec{q}\right)N^{\perp}_j\left(-\vec{q}\right)\right>}$,
where $\vec{N}_{\perp}$ is the order parameter, which we remind
the reader is the projection of the local layer normal
perpendicular to the $\hat{z}$-axis.
%
%
%
%

The behavior of $C_{ij}$ is radically different in the two phases, and quite
involved (as the intrepid reader is about to discover) in both. Here we'll
begin by summarizing the behavior in the ({\it relatively}) simpler A phase,
and then proceed to a description of the {\it even} more complicated
behavior in the C phase.
%
%
%
%
%
%

In the A phase ($T>T_{AC}$), there are three important contributions to
$C_{ij}(\vec{q})$: two arise from the previously discussed random
tilt ($\Delta_t(\vec{q})$) and compression ($\Delta_c(\vec{q})$)
disorders. The third is caused by random {\it positional} pinning
%
%
%
$\Delta_p$,
%
%
%
%
%
%
which reflects the tendency of the aerogel to
pin the smectic layers in particular {\it positions}. This type of
disorder is {\it irrelevant} for $T\leq T_{AC}$, but becomes
important at sufficiently long distance (or small $\vec{q}$) in the A
phase (i.e., above
$T_{AC}$).
%
%
%
Indeed, as we will see in section \ref{sec: Cfunctions}, this
random positional pinning $\Delta_p$ actually {\it dominates} the
fluctuations at small ${\vec q}$ in the A phase; furthermore, it
is precisely this pinning that leads to the destruction of
long-ranged translational order, and the replacement of sharp,
delta-function Bragg peaks with power-law peaks, in the A phase.

Combining the contributions from
these three distinct types of disorder gives
\begin{widetext}
\begin{eqnarray}
C_{ij}(\vec{q})&=& L^{\perp}_{ij}(\hat{q})\left({\Delta_t
\left(\vec{q}, T \right)q_{\perp}^4 \over \left(Bq^2_z+D(T)q^2_{\perp}
+K\left(\vec{q}, T
\right)q^4_{\perp}\right)^2}+{\Delta_c\left(\vec{q}, T
\right)q_z^2 q_{\perp}^2 \over\left(Bq^2_z+D(T)q^2_{\perp}
+K\left
(\vec{q}, T\right)q^4_{\perp}\right)^2}\right.\nonumber\\
&+&\left.{C{B^{1\over 2}Dq_{\perp}^2}\over
\left(Bq^2_z+D(T)q^2_{\perp}+K\left(\vec{q},
T\right)q_{\perp}^4\right) ^{3\over2}q_0^2}\right) \label{LSI},
\end{eqnarray}
\end{widetext}
where $L^{\perp}_{ij}(\hat{q}) \equiv {q^{\perp}_i q^{\perp}_j /
q^2_{\perp} }$ is the projection operator along $\vec{q}_{\perp}$,
the compression modulus $B$ remains constant as $\vec{q}\to 0$,
$C$ is a universal, $O(1)$ constant ($C\approx 1.10\pi^2$),
\begin{eqnarray}
D(T)  \propto \left\{
\begin{array}{ll}
(T-T_{AC})q_z^{-\eta_D/\zeta} &\mbox{region 1}\\
(T-T_{AC})q_{\perp}^{-\eta_D} &\mbox{region 2}\\
\xi^{\eta_K-2}_{\perp} &\mbox{region 3}
\end{array}\right.
\label{D1},
\end{eqnarray}
where $\eta_D=2-\eta_K-1/\nu_{\perp}$, and the (wavevector $\vec{q}$
and temperature $T$-dependent) disorder variances $\Delta_t
\left(\vec{q}, T\right)$, $\Delta_c \left(\vec{q}, T\right)$ and
layer bend modulus $K \left(\vec{q}, T\right)$ are given by Eq.
(\ref{DeltaT}). The $C$-term comes from the periodic random pinning
potential. The three regions in (\ref{D1}) is again illustrated in
Fig. \ref{fig: 3regions}.

Clearly, the expression (\ref{LSI}) for $C_{ij}$ is somewhat
complicated, due to the fact that {\it three} distinct types of
disorder contribute to layer normal fluctuations. The situation is
further exacerbated by the strong wavevector and temperature
dependence of the elastic constants $D$ and $K$. Fortunately, there
are well-defined regions of wavevector ${\vec q}$, in each of which
one of the three terms (i.e., the  $\Delta_t$, the $\Delta_c$, or
$\Delta_p$-embodied in the C-term) dominate $C_{ij}$, thereby
simplifying the expression for it. Each of these regions can be
further subdivided according to which of the terms $Bq^2_z$,
$D(T)q^2_{\perp}$, and $K\left(\vec{q}, T \right)q^4_{\perp}$
dominates the common denominator $Bq^2_z+D(T)q^2_{\perp}
+K\left(\vec{q}, T \right)q^4_{\perp}$ of all three terms. In
addition, $K({\vec q}, T)$, $\Delta_{c,t}\left(\vec{q}, T \right)$
and $D(T)$ themselves have the crossovers embodied in Eqs.
(\ref{DeltaT}, \ref{D1}), which further subdivide some of the
regions of ${\vec q}$-space into subregions of distinct behavior.

\begin{figure}[h]
\includegraphics[width=0.35\textwidth, angle=-90]{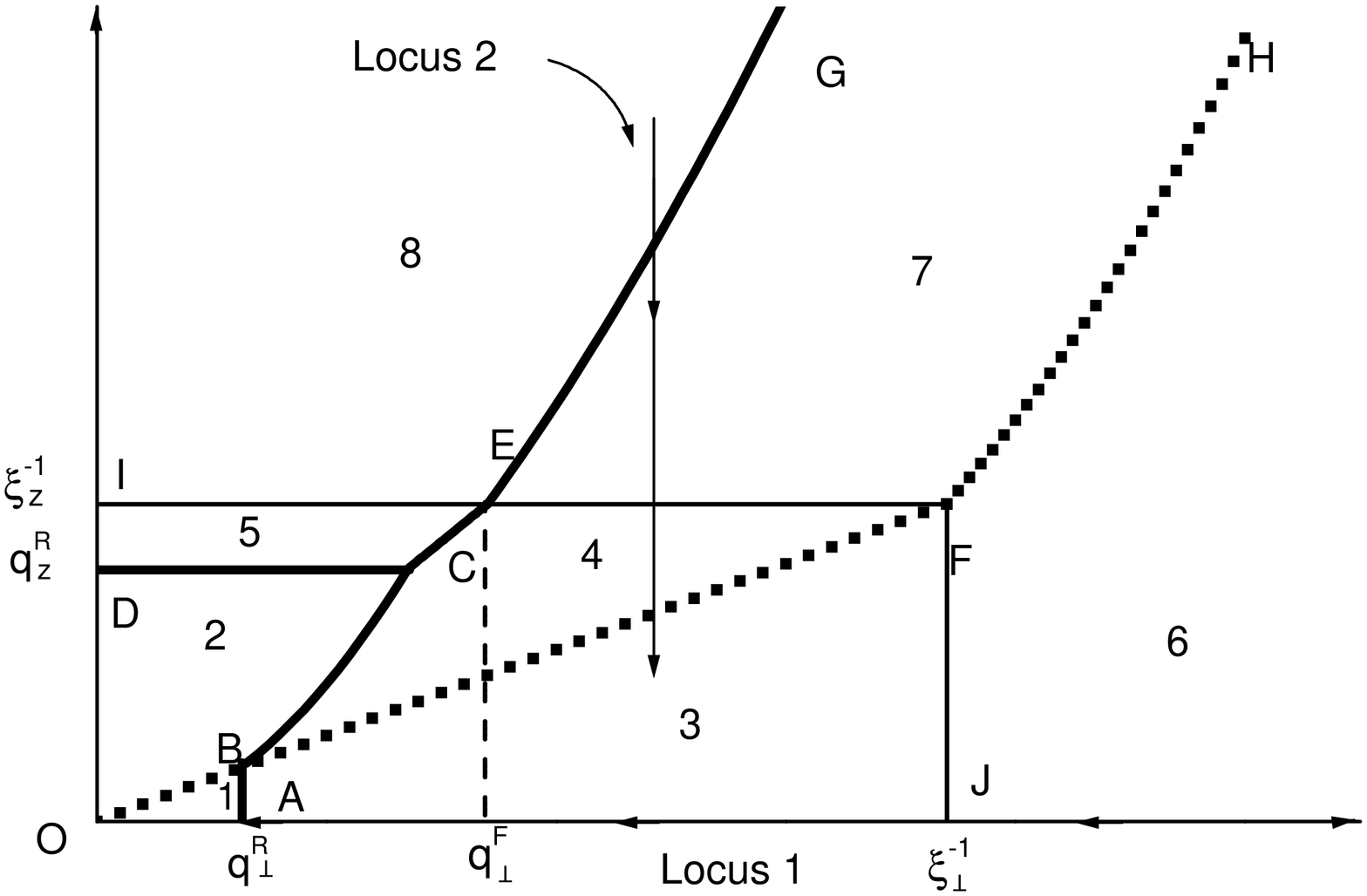}
\caption{\label{fig: 8regions}Illustration of the eight regions
listed in (\ref{light}), each of which exhibits different scaling of
the light scattering intensity in the $A$ phase  as $T\to T^+_{AC}$.
The light scattering intensity is dominated by random field fluctuations in
regions 1 and 2, by random tilt fluctuations in regions 3, 4,  6 and 7,
and by random compression fluctuations in regions 5 and 8. Above locus $OFH$ the common denominator in
formula (\ref{LSI}) is dominated by $Bq_z^2$, while below $OFH$ it is
dominated by $Kq_{\perp}^4$ in region 6 and by $Dq_{\perp}^2$ in regions 1 and
3, respectively.}
\end{figure}

Painstakingly, but essentially straightforwardly, sorting out these
different regions leads to the identification of {\it eight} (count
'em!) distinct regions of qualitatively different
wavevector-dependence for $C_{ij}$, which are illustrated in Fig.
\ref{fig: 8regions}. The leading wavevector and temperature
dependence of $C_{ij}$ in each of these regions is:
%
%
%
%
%
%
%
\begin{eqnarray}
C_{ij}(\vec{q})\sim
\left\{\begin{array}{ll}{\xi_{NL}^{\perp}\over{\lambda
q_0^2}}\left(\xi_{NL}^{\perp}\over\xi_{\perp}\right)^{{\eta_K\over2}-1}{1
\over
q_{\perp}},&\mbox{region 1}\\
{\lambda^2\over
{q_0^2\left(\xi_{NL}^{\perp}\right)^2}}\left(\xi_{NL}^{\perp}\over\xi_{\perp
}\right)^{2-\eta_K}{q_{\perp}^2\over
q_z^3},&\mbox{region 2}\\
\lambda\left(\xi_{NL}^{\perp}\right)^2\left(\xi_{NL}^{\perp}\over\xi_{\perp}
\right)^{2\eta_K-\eta_t-4},&\mbox{region
3}\\
{\lambda^5\over\left(\xi_{NL}^{\perp}\right)^2}\left(\xi_{NL}^{\perp}\over
\xi_{\perp}\right)^{-\eta_t}\left(q_{\perp}\over
q_z\right)^4, &\mbox{region 4}\\
{\lambda^5\Delta_c^0\over\left(\xi_{NL}^{\perp}\right)^2\Delta_t^0}
\left(\xi_{NL}^{\perp}\over\xi_{\perp}\right)^{-\eta_c}\left(q_{\perp}\over
q_z\right)^2, &\mbox{region 5}\\
{\lambda\over\left(\xi_{NL}^{\perp}\right)^2}\left(\xi_{NL}^{\perp}q_{\perp}
\right)^{2\eta_K-\eta_t}{1\over
q_{\perp}^4},&\mbox{region 6}\\
{\lambda^5\over\left(\xi_{NL}^{\perp}\right)^2}\left(q_z\xi_{NL}^z\right)^
{-{\eta_t\over\zeta}}\left(q_{\perp}\over
q_z\right)^4,&\mbox{region 7}\\
{\lambda^5\Delta_c^0\over\left(\xi_{NL}^{\perp}\right)^2\Delta_t^0}
(q_z\xi_{NL}^z)^{-{\eta_c\over\zeta}}\left(q_{\perp}\over
q_z\right)^2,&\mbox{region 8}
\end{array}\right.
\label{light}.
\end{eqnarray}
In (\ref{light}) we have used the standard ``smectic
penetration length'' $\lambda \equiv \sqrt{K_0/B_0}$, where $K_0$
and $B_0$ are the ``bare'' smectic layer bending modulus and
compression modulus respectively; i.e., the bending modulus and
compression modulus for the smectic in the absence of disorder.
The loci separating the regions in Fig. \ref{fig: 8regions} are:
\begin{eqnarray}
FH:&&q_z\xi_{NL}^z =
\left(q_{\perp}\xi_{NL}^{\perp}\right)^{2-{\eta_K\over2}}\label{cross1}\\
EG:&&q_z\xi_{NL}^z =
\left(q_{\perp}\xi_{NL}^z\sqrt{\Delta_t^0/\Delta_c^0}\right)^A\label{cross2}\\
CE:&&q_z = \sqrt{\Delta_t^0/\Delta_c^0}\left(\xi_{NL}^{\perp}\over
\xi_{\perp}\right)^{1-\left({\eta_K+\eta_3}\over
2\right)}q_{\perp}\label{cross3}\\
AB:&&q_{\perp} = q_{\perp}^R =
{1\over{q_0^2\lambda^2\xi_{NL}^{\perp}}}\left(\xi_{\perp}\over
\xi_{NL}^{\perp}\right)^{{3\eta_K\over 2}-\eta_t-3}\label{cross5}\\
DC:&&q_z = q_z^R =
{\Delta_t^0\over{q_0^2\lambda^3}\Delta_c^0}\left(\xi_{NL}^{\perp}\over
\xi_{\perp}\right)^{4+\eta_t-2\eta_K-\eta_3}\label{cross4}\\
OF:&&q_z =
{\lambda\over\xi_{NL}^{\perp}}\left(\xi_{\perp}\over\xi_{NL}^{\perp}\right)
^{{\eta_K\over2}-1}q_{\perp}\\
BC:&&q_z =
q_0^2\lambda^3\left(\xi_{\perp}\over\xi_{NL}^{\perp}\right)^{2+\eta_t-\eta_K}q_{\perp}^2,
\label{locii}
\end{eqnarray}
where $q_0 \equiv {2\pi/ a}$, $\xi_{NL}^z \equiv
{\left(\xi_{NL}^{\perp}\right)^2/\lambda}$, $A \equiv
{\zeta/(1+{\eta_3/2})}$,  and $a$ is the smectic layer spacing.
Here the positive definite universal exponent $\eta_3$ is defined
in Eq. (\ref{eta_3}). We have calculated it in the
$\epsilon=5-d$-expansion and found:
\begin{eqnarray}
\eta_3= {\epsilon \over 5}  + O(\epsilon^2) .
\label{eta3eps}
\end{eqnarray}

%
%
%
%
%
%
%
%
%
%
%
%
%
%
%
%

Of course, none of these crossover lines is sharp, since none of
the crossovers is abrupt. Rather, as is the nature of all
crossovers, these are loci near which two terms start to become
comparable. Hence, these crossover boundaries are only defined up
to $O(1)$ factors.

The fairly complicated scaling behavior embodied in (\ref{LSI})
simplifies considerably when we take $q_z = 0$ and vary $q_{\perp}$
(i.e., move along locus 1 in Fig. \ref{fig: 8regions}). The second
term in (\ref{LSI}) then vanishes. For $q_{\perp}\gg
\xi_{\perp}^{-1}$ , $Kq_{\perp}^4$ dominates in the denominator of
the first term in (\ref{LSI}), which term itself dominates the third
term, and using (\ref{light}) we obtain $\overline{\left<N^{\perp}_i
\left(\vec{q}\right)N^{\perp}_j\left(-\vec{q}\right)\right>}\propto
   q_{\perp}^{-4+2\eta_K-\eta_t}$. For $q_{\perp}\ll \xi_{\perp}^{-1}$,
the first term dominates at short wavelength, and the third term
dominates at long wavelengths. The crossover between these two is at
\begin{eqnarray}
q_{\perp}=\left(\xi^R_{\perp}\right)^{-1}\propto\xi_{\perp}^{-3+{3\over2}\eta_K-\eta_t}\,
. \label{}
\end{eqnarray}
$\xi_{\perp}^R$ is much larger than $\xi_{\perp}$ as $T\to
T_{AC}^+$, provided $3+\eta_t-{3\eta_K\over 2}>1$, a
condition we can show to be satisfied.
Clearly, for
$q\gg(\xi^R_{\perp})^{-1}$, if there were {\it no} anomalous
elasticity of $K$ (i.e., {\it if } $\eta_K = 0$), the light
scattering lineshape would be a Lorentzian squared, with
correlation length $\xi_{\perp}$.  In reality, the lineshape is
{\it quantitatively} different (e.g., its tails scale like
$q_{\perp}^{-4+2\eta_K-\eta_t}$), but still qualitatively like a
Lorentzian squared, with linewidth $\xi^{-1}_{\perp}$. For
$q\ll(\xi^R_{\perp})^{-1}$the light scattering diverges as a power
law. More precisely,
\begin{widetext}
\begin{eqnarray}
C_{ij}(\vec{q_\perp , q_z = 0})&=&
L^{\perp}_{ij}\left(\hat{q}\right) \left({\Delta_t
\left(\vec{q}_{\perp}, q_z = 0, T \right) \over\left[D(T) +
K\left( \vec{q}_{\perp}, q_z = 0, T\right)q_{\perp}^2\right]^2}
+{{CB^{1\over2}}D\over\left[D(T) + K\left(\vec{q}_{\perp},
q_z = 0,T\right)q_{\perp}^2\right]q_{\perp}q_0^2}\right)\nonumber\\
&\sim &\left\{
\begin{array}{ll}
{\lambda\over\left(\xi_{NL}^{\perp}\right)^2}\left(\xi_{NL}^{\perp}q_{\perp}
\right)^{2\eta_K-\eta_t}{1\over q_{\perp}^4}, &q_{\perp}
\gg\xi_{\perp}^{-1}\, ,\\
\lambda\left(\xi_{NL}^{\perp}\right)^2\left(\xi_{NL}^{\perp}\over\xi_{\perp}
\right)^{2\eta_K-\eta_t-4},&\left(\xi^R_{\perp}\right)^{-1}\ll
q_{\perp}\ll\xi_{\perp}^{-1}\, ,\\
{\xi_{NL}^{\perp}\over{\lambda
q_0^2}}\left(\xi_{NL}^{\perp}\over\xi_{\perp}\right)^{{\eta_K\over2}-1}
{1\over q_{\perp}},&q_{\perp}\ll\left(\xi^R_{\perp}\right)^{-1}\, .
\end{array}\right.
\label{no number}
\end{eqnarray}
\end{widetext}
Comparison of this expression with light scattering data should
allow easy determination of $\xi_{\perp}(T)$ and the combination
of exponents $2\eta_K - \eta_t$.  Fitting the $T$-dependence of
$\xi_{\perp}(T)$ to $\left(T - T_{AC} \right)^{-\nu_{\perp}}$ then
determines $\nu_{\perp}$.

Another simple locus is
%
%
%
%
obtained by
%
%
%
%
%
%
taking
%
%
%
%
%
$q_{\perp}$ fixed in the range
%
%
%
%
%
%
\cite{wise-ass} $q_{\perp}^F\ll q_{\perp} \ll\xi_{\perp}^{-1}$ and
varying $q_z$ (i.e., moving along locus 2 in Fig. \ref{fig:
8regions}), which gives
\begin{widetext}
\begin{eqnarray}
C_{ij}(\vec{q})\sim\left\{
\begin{array}{ll}
{\lambda^5\Delta_c^0\over\Delta_t^0\left(\xi_{NL}^{\perp}\right)^2}
\left(q_z\xi_{NL}^z\right)^{-{\eta_c\over\zeta}}\left(q_{\perp}\over
q_z\right)^2,&q_z\gg\left(q_{\perp}\xi_{NL}^z\right)^A\left(\xi_{NL}^z\right
)^{-1}\\
{\lambda^5\over\left(\xi_{NL}^{\perp}\right)^2}\left(q_z\xi_{NL}^z\right)^{-
{\eta_t\over\zeta}}\left(q_{\perp}\over
q_z\right)^4,&\left(\xi_z\right)^{-1}\ll
q_z\ll\left(q_{\perp}\xi_{NL}^z\right)^A\left(\xi_{NL}^z\right)^{-1}\\
{\lambda^5\over\left(\xi_{NL}^{\perp}\right)^2}\left(\xi_{NL}^{\perp}\over
\xi_{\perp}\right)^{-\eta_t}\left(q_{\perp}\over q_z\right)^4,
&{\lambda\over\xi_{NL}^{\perp}}\left(\xi_{\perp}\over\xi_{NL}^{\perp}\right)
^{{\eta_K\over2}-1}q_{\perp}\ll q_z
\ll\left(\xi_z\right)^{-1}\\
\lambda\left(\xi_{NL}^{\perp}\right)^2\left(\xi_{NL}^{\perp}\over\xi_{\perp}
\right)^{2\eta_K-\eta_t-4},&0\ll
q_z\ll
{\lambda\over\xi_{NL}^{\perp}}\left(\xi_{\perp}\over\xi_{NL}^{\perp}\right)^
{{\eta_K\over2}-1}q_{\perp}
\end{array}\right.
\label{},
\end{eqnarray}
\end{widetext}
where
$q_{\perp}^F\sim\xi_z^{-1}\left(\xi_{NL}^{\perp}/\xi_{\perp}\right)^{{{\eta_K
+\eta_3}\over 2}-1}$. From this locus $\xi_z$ can be read from the
point where a log-log plot of $I\left(\vec{q} \right)$ vs $q_z$
for fixed $\vec{q}_{\perp}$ changes slope from $-{\eta_t \over
\zeta} - 4$ to $-4$. Again, once $\xi_z(T)$ is known, $\nu_z$ can
be determined.  Note that measurements of the exponents ${\eta_t
\over \zeta} + 4$ and ${\eta_c\over \zeta} + 2$,  taken in
conjunction with knowledge of $\eta_t - 2\eta_K$ from locus 1,
determine $\eta_K$, $\eta_t$ and $\eta_c$ (recall that the
anisotropy exponent $\zeta = 2 - {\eta_K \over 2}$).

To summarize our light scattering predictions for the %
%
%
%
$A$-side of the
%
%
%
%
%
%
%
%
critical regime: measurements along locus 1 and 2 can be used to
determine the exponents $\nu_{\perp}$, $\nu_z$, $\eta_K$, and
$\eta_t$, as well as testing that $\eta_B = 0$ (which was assumed
in the above discussion). These exponents can then be
quantitatively compared with our predictions, and, furthermore,
test the {\it exact} scaling relation (\ref{nu z scale}), and
combined with specific heat measurements to test the {\it exact}
scaling relation (\ref{alpha scale}) for $\alpha$, and with the
aforedescribed dielectric measurement determination of $\beta$ to
test the equally {\it exact} scaling relation (\ref{beta scale}).

%
%
%
%
%
We now turn to the smectic C ($T<T_{AC}$).
%
%
%
%
%
%
%
%
%
%
%
%
For $q_ {\perp}\xi_{\perp} \gg 1$ or $q_z\xi _z\gg 1$, the light
scattering is the same as in the high temperature case, since this
range of wavevectors corresponds to the critical regime, in which
one cannot tell whether one is on the high- or low-temperature
side of the transition.
%
%
%
%
%
In the other limit -
namely, when both
$q_{\perp}\xi_{\perp}
\ll 1$ and $q_z\xi _z\ll 1$,
%
%
%
%
%
%
%
%
%
%
%
we are in the smectic C-phase range of wavevectors, and $C_{ij}$ is given by:

%
%
%

%
%
%
%
\begin{widetext}
\begin{eqnarray}
C_{ij}(\vec{q}) &=& L^{\perp}_{ij}(\hat{q})\left({{\Delta_{s'}
\left(\vec{q}\,', T \right)q_{s'}^2q_ {\perp}^2}\over
\left(\gamma(\vec{q}\,', T)q_{x'}^2 +
\tilde{B}q^2_{z'}+\tilde{K}\left (\vec{q}\,',
T\right)q^4_{s'}\right)^2}+{\Delta_{z'}\left(
T\right)q_{z'}^2q_{\perp}^2 \over \left(\gamma(\vec{q}\,',
T)q_{x'}^2 + \tilde{B}q^2_{z'}+\tilde{K}\left(\vec{q}\,', T
\right)q^4_{s'}\right)^2}\right)\nonumber\\&+&L^{\perp}_{ij}(\hat{q})
{\Delta_{x'}\left(\vec{q}\,', T \right)q_{x'}^2q_{\perp}^2 \over
\left(\gamma(\vec{q}\,', T)q_{x'}^2 +
\tilde{B}q^2_{z'}+\tilde{K}\left(\vec{q}\,', T
\right)q^4_{s'}\right)^2} \label{LST},
\end{eqnarray}
\end{widetext}
where $q_{x'}$, $q_{s'}$, and $q_{z'}$ are defined in Eqs.
(\ref{qx}, \ref{qs}, \ref{qz}) respectively, $\tilde{K}$, $\gamma$
and $\Delta_{s', x'}$ are anomalous and given in Eqs. (\ref{KanomC},
\ref{gammaanomC}, \ref{DeltaanomC}), and $\Delta_{z'}$ is
%
%
%
%
%
wavevector $\vec{q}$-independent. In addition, the temperature
dependent quantity $\Gamma(T)$ defined in the coordinate
transformations Eqs. (\ref{qx}, \ref{qs}, \ref{qz}) has the critical
scaling:
\begin{eqnarray}
\Gamma(T)\propto\left(\xi_{\perp}\over\xi_{NL}^{\perp}\right)^{1-({{\eta_K
+\eta_3}\over 2})}
\label{Gammacrit}.
\end{eqnarray}
Since the azimuthal symmetry about
the $\hat{z}$-axis has been broken due to the tilting of the layers, the light
scattering
%
%
%
%
%
%
also loses azimuthal symmetry, and becomes fully three-dimensional.
%
%
%
%
%
%
%
%
%
%
It is most conveniently described in the  transformed,
non-orthogonal wavevector coordinates (i.e., $\vec{q}\,'$-space),
whose relation to the lab wavevector coordinates (i.e.,
$\vec{q}$-space) are given in Eqs. (\ref{qx}, \ref{qs}, \ref{qz}).
%
%
%
%
%
%
%
%
%
%
%

\begin{figure}[h]
\includegraphics[width=0.35\textwidth, angle=-90]{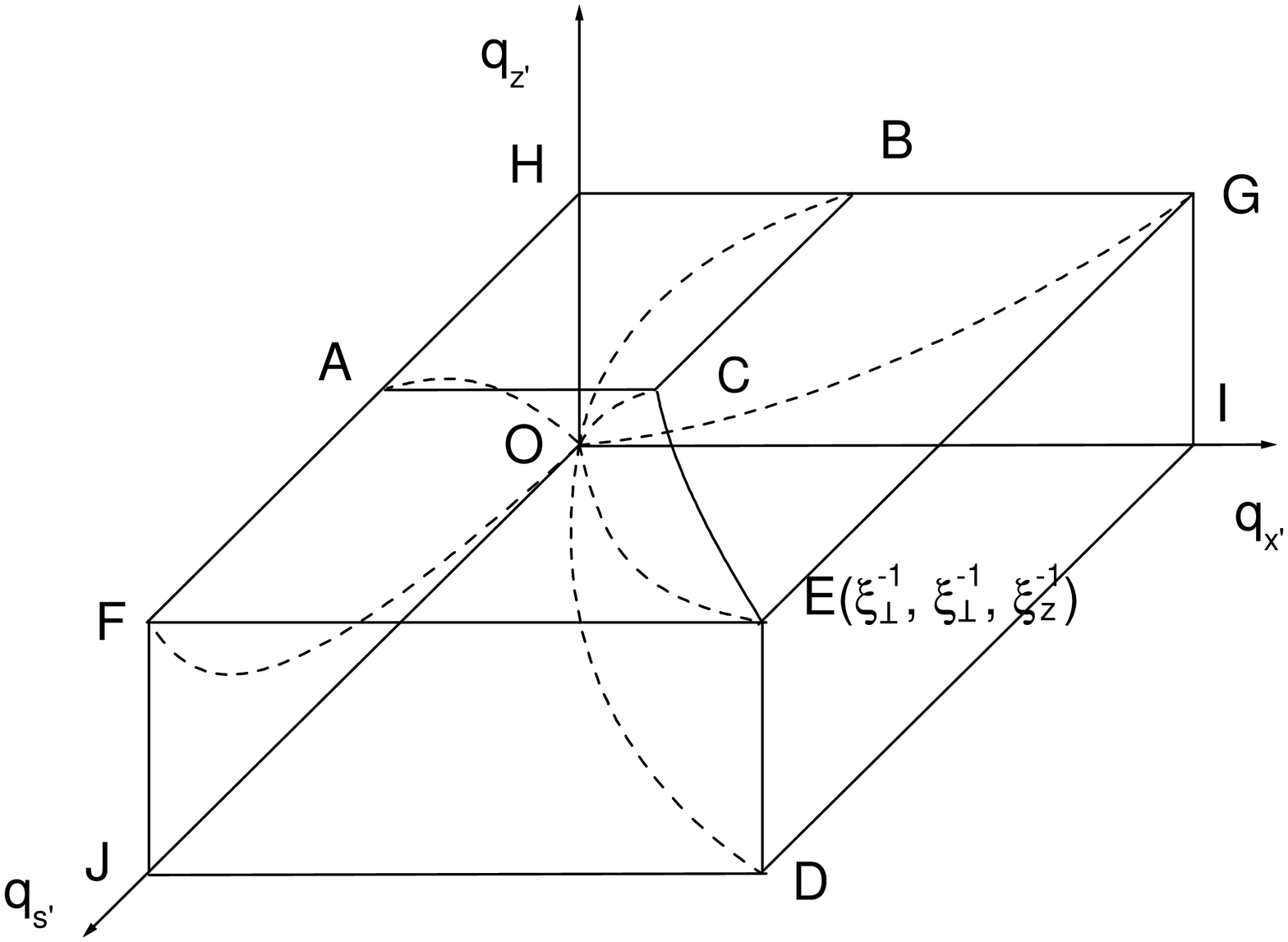}
\caption{\label{fig: 5regions}3D picture in the transformed
reciprocal space ($\vec{q}\, '$-space) illustrating the five
distinct regions with different wavevector-dependences of
$C_{ij}(\vec{q})$ in the $C$ phase, which is given in
(\ref{Cij5Regions}).}
\end{figure}

As in the A-phase, in the C-phase the light scattering displays
many different regimes of very different scaling with wavevector
and temperature, due, as there, to both the crossovers between
domination by different types of disorder, and by the crossovers
between dependences on different components of ${\vec q}$. Here,
the three-dimensional $\vec {q}\, '$ space is divided into the
five regions illustrated in Fig. \ref{fig: 5regions},
%
%
%
%
%
%
in which the light scattering intensity is given by
\begin{widetext}
  \begin{eqnarray}
  C_{ij}(\vec{q})\approx
 L_{ij}^{\perp}q_{\perp}^2\left\{\begin{array}{ll}
 {\Delta_n^0\lambda^5\over\Delta_t^0(\xi_{NL}^{\perp})^2}
 \left(\xi_{\perp}\over\xi_{NL}^{\perp}\right)^{\eta_c}
 \left(1\over q_{z'}\right)^2, &\mbox{region O-ABCH}\\
 {\lambda^5\over(\xi_{NL}^{\perp})^2}
 \left(\xi_{\perp}\over\xi_{NL}^{\perp}\right)^{\eta_t}
 \left(\xi_zq_{z'}\right)^{-\tilde{\eta}_{s'}/\tilde{\zeta}_{z'}}
 \left(q_{s'}\over q_{z'}^2\right)^2, &\mbox{region O-AFEC}\\
 {\lambda^5\over(\xi_{NL}^{\perp})^2}
 \left(\xi_{\perp}\over\xi_{NL}^{\perp}\right)^{\eta_t}
 \left(\xi_zq_{z'}\right)^{-\tilde{\eta}_{x'}/\tilde{\zeta}_{z'}}
 \left(q_{x'}\over q_{z'}^2\right)^2, &\mbox{region O-CEGB}\\
 {\lambda\over(\xi_{NL}^{\perp})^2}
 \left(\xi_{\perp}\over\xi_{NL}^{\perp}\right)^{\eta_t-2\eta_K}
 \left(\xi_{\perp}q_{s'}\right)^{2\tilde{\eta}_K-\tilde{\eta}_{s'}}
 \left(1\over q_{s'}\right)^6, &\mbox{region O-FJDE}\\
 {\lambda(\xi_{NL}^{\perp})^2}
 \left(\xi_{\perp}\over\xi_{NL}^{\perp}\right)^{\eta_t-2\eta_K+4}
 \left(\xi_{\perp}q_{x'}\right)^{\left(-2\tilde{\eta}_{\gamma}-\tilde{\eta}_{x'}\right) /\tilde{\zeta}_{x'}}
 \left(1\over q_{x'}\right)^2, &\mbox{region O-EDIG}\,
  \end{array}\right.\, .
  \label{Cij5Regions}
  \end{eqnarray}
\end{widetext}

%
%
%
%
%
%
%
%
%
%
%
%
The crossovers between these five regions obey
\begin{eqnarray}
 OAC:&&q_{z'}\xi_{z}=\left[\sqrt{\Delta_t^0\over\Delta_n^0}\left(\xi
 _{\perp}\over
 \xi_{NL}^{\perp}\right)^{\eta_3/2}\left(\xi_{NL}^{\perp}\over\lambda\right)
 \right.
 \nonumber\\
 &&~~~~~~~~~\left.\times
 \left(q_{s'}\xi_{\perp}\right)\right]^{\tilde\phi_{s'z'}}\, ,
 \end{eqnarray}
 \begin{eqnarray}
 OFE:&&q_{z'}\xi_z=\left(q_{s'}\xi_{\perp}\right)^{\tilde{\zeta}_{z'}}\, ,
 \end{eqnarray}
 \begin{eqnarray}
 OBC:&&q_{z'}\xi_{z}=\left[\sqrt{\Delta_t^0\over\Delta_n^0}\left(\xi
 _{\perp}\over
 \xi_{NL}^{\perp}\right)^{\eta_3/2}\left(\xi_{NL}^{\perp}\over\lambda\right)
 \right.
 \nonumber\\
 &&~~~~~~~~~\left.\times
 \left(q_{x'}\xi_{\perp}\right)\right]^{\tilde\phi_{x'z'}}\ , ,\\
 OGE:&&q_{z'}\xi_z=\left(q_{x'}\xi_{\perp}\right)^{\tilde{\zeta}_{z'}/{\tilde{\zeta}_{x'}}}\, ,\\
 ODE:&&q_{x'}\xi_{\perp}=\left(q_{s'}\xi_{\perp}\right)^{\tilde{\zeta}_{x'}}\, ,\\
 OEC:&&q_{x'}=\left(q_{z'}\xi_z\right)^{\tilde\phi_{s'x'}}q_{s'}\, ,
\end{eqnarray}
where we've defined the C-phase crossover exponents:
\begin{eqnarray}
\tilde\phi_{s'z'}=2\tilde{\zeta}_{z'}/\left(2\tilde{\zeta}_{z'}+\tilde{\eta}_{s
'}\right)\, , \label{phi_sz}
\\
\tilde\phi_{x'z'}=2\tilde{\zeta}_{z'}/\left(2\tilde{\zeta}_{z'}+\tilde{\eta}_{x
'}\right)\, , \label{phi_xz}
\\
\tilde\phi_{s'x'}=\left(2-\tilde{\eta}_K-\tilde{\eta}_{\gamma}\right)/2\tilde{\zeta}_{z'}\,
, \label{phi_sx}
\end{eqnarray}
and the quantity
$\Delta_n^0=\Delta_c^0+\Delta_t^0g_0^2\left(\xi_{NL}^{\perp}\right)^2/
w_0K_0$, where $g_0$ and $w_0$  are the ``complete bare'' values of
of the coefficients of the anharmonic terms in the Hamiltonian
(\ref{H_AC}).

Now let us narrow our discussion to wavevectors in the $q_z$-$q_x$
plane and measure $C_{ij}(\vec{q})$ given by
%
%
%
%
%
%
\begin{eqnarray}
C_{xx}(\vec{q}) &=&{{(\Delta_{z'}q_{z'}^2 +{\Delta_{x'}
\left(\vec{q}, T\right)q_{x'}^2)q_{x}^2}}\over \left(\gamma
(\vec{q}, T)q_{x'}^2 + \tilde{B}q^2_{z'}\right)^2} \label{}\, ,
\end{eqnarray}
%
%
%
%
where we alert the reader to the fact that this expression contains {\it both}
the {\it rotated} wavevector component $q_{x'}$ and the {\it un}-rotated
wavevector component $q_{x}$. This is {\it not} a typo!

%
%
%


This two-dimensional $q_z$-$q_x$ plane can also be divided into
different regions with different wavevector-dependences of
$C_{xx}(\vec{q})$, including (for illustration) the critical regime
(i.e., $q_z \gg \xz^{-1}$ or $q_\perp \gg \xp^{-1}$). (Note that we
are now working with the {\it un}-transformed $\vec{q}$\,'s). This
leads to the 10 regions which are illustrated in Fig. \ref{fig:
10regions}. In these regions, the light scattering results are:
%
%
%
%
%
%
\begin{widetext}
\begin{eqnarray}
C_{xx}(\vec{q})\sim
\left\{\begin{array}{ll}{\lambda^5\Delta_c^0\over\left(\xi_{NL}^{\perp
}\right)^2
\Delta_t^0}\left(q_z\xi_{NL}^z\right)^{-\eta_c/\zeta}\left(q_x\over
q_z
\right)^2,&\mbox{region 1}\\
{\lambda^5\over\left(\xi_{NL}^{\perp}\right)^2
}\left(q_z\xi_{NL}^z\right)^{-\eta_t/\zeta}\left(q_x\over q_z
\right)^4,&\mbox{region 2, 3}\\
{\lambda^5\Delta_c^0\over\left(\xi_{NL}^{\perp}\right)^2\Delta_t^0
}\left(\xi_{\perp}\over\xi_{NL}^{\perp}\right)^{\eta_c}\left
(q_x\over q_z\right)^2,&\mbox{region 4}\\
{\lambda^5\over\left(\xi_{NL}^{\perp}\right)^2}\left(q_z\xi_z\right)
^{-\tilde{\eta}_{x'}/\tilde{\zeta}_{z'}}\left(\xi_{\perp}\over\xi_
N^{\perp}\right)
^{\eta_t}\left(q_x\over q_z\right)^4, &\mbox{region5, 6}\\
{\lambda\xi_{\perp}^4\over\left(\xi_{NL}^{\perp}\right)^2}\left(q_x\xi_{\perp}\right)
^{-(\tilde{\eta}_{x'}+2\tilde{\eta}_{\gamma})/\tilde{\zeta}_{x'}}
\left(\xi_{\perp}\over\xi_{NL}^ {\perp}\right)^{-2\eta_K+\eta_t},
&\mbox{region 7,
8}\\
{\lambda\over\left(\xi_{NL}^{\perp}\right)^2}\left(q_x\xi_{NL}^{\perp}
\right)^{2\eta_K-\eta_t}q_x^{-4}, &\mbox{region 9, 10}\\
\end{array}\right..
\label{Cxx}
\end{eqnarray}
\end{widetext}
At short wavelengths (i.e., $q_x\gg\xi_{\perp}^{-1}$ or
$q_z\gg\xi_z^{-1}$),
%
%
%
%
%
that is, in the critical regime,  the crossovers between
%
%
%
%
%
%
the different regions remain the same as those of high temperature
phase, which were given in equations (\ref{cross1}, \ref{cross2}).
At long wavelengths (i.e., both $q_z\ll\xi_z^{-1}$ and
$q_x\ll\xi_{\perp}^{-1}$) the crossovers are given by:
\begin{eqnarray}
&&OE,OE':q_z\xi_z=\left[\sqrt{\Delta_t^0\over\Delta_n^0}\left(\xi_z\over
\Gamma\right)\left( q_x-\Gamma q_z\right)\right]^{\tilde{\phi}_{x'z'}}\, ,\nonumber\\
&&OF,OF':\left(q_{z}\xi_z\right)^{\tilde{\zeta}_{x'}}=\left(q_x\xi_{\perp}
\right)^{\tilde{\zeta}_{z'}} \nonumber.
\end{eqnarray}
%
%
%
%
%
%
%
%
%
%



\begin{figure}[h]
\includegraphics[width=0.35\textwidth, angle=-90]{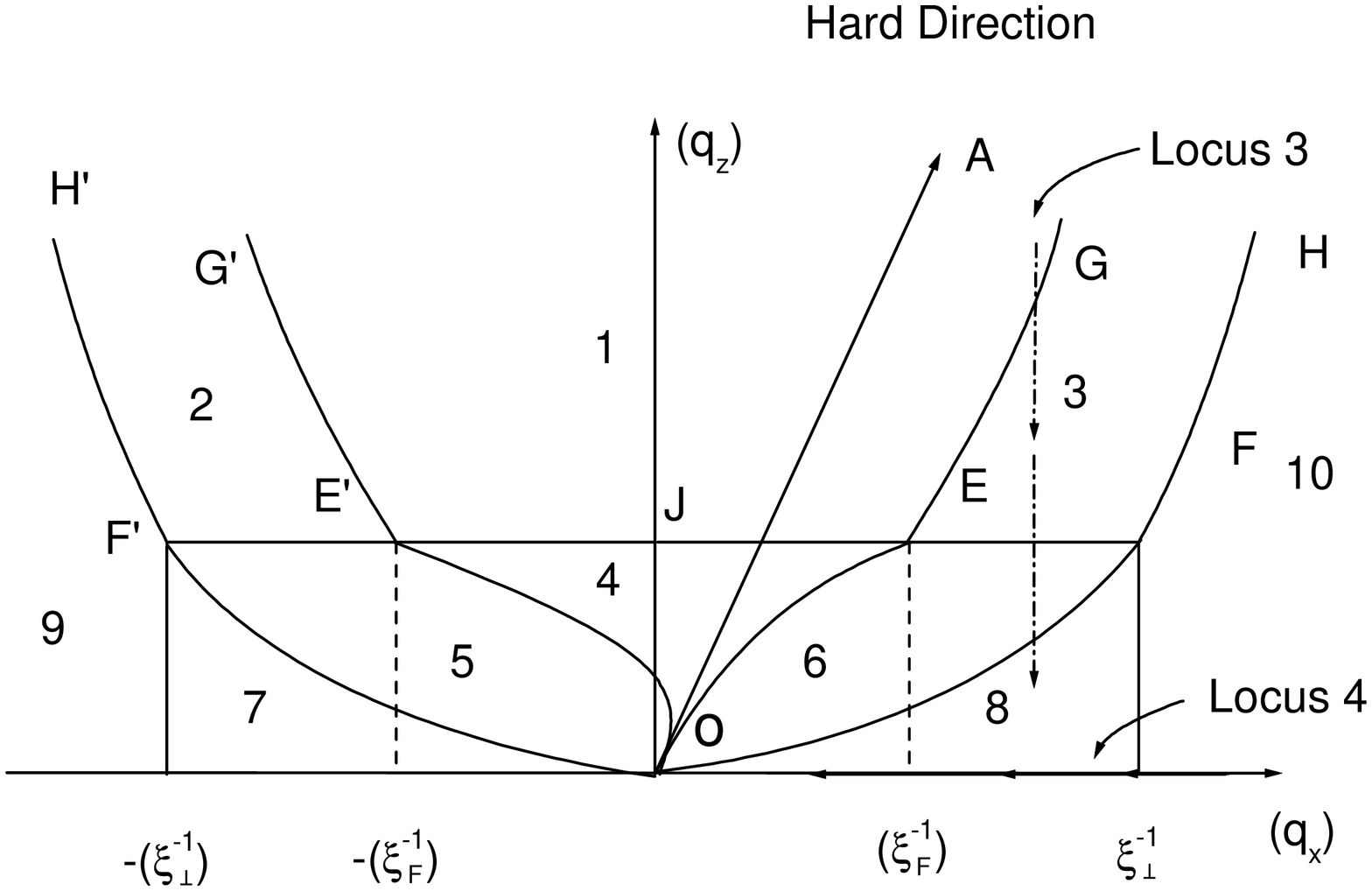}
\caption{\label{fig: 10regions}Illustration of the 10 distinct
regions in $q_z$-$q_x$ plane with different wavevector-dependences
of $C_{xx}(\vec{q})$, which is given in (\ref{Cxx}) for the $C$
phase. Line $OA$ is define by $q_{x'}=0$.}
\end{figure}

%
%
%
%
%
%
Two possible experimental loci through the $q_x$-$q_z$ plane are
shown in Fig. \ref{fig: 10regions}. By holding 3,  $q_x$ fixed and
varying $q_z$ (locus 3), we can determine
$-{\tilde{\eta}_{x'}/\tilde{\zeta}_{z'}}-4$ from a log-log plot of
$I(\vec{q})$ vs $q_z$, and also deduce  $\xi_z$ from the location
of the point at which  the log-log plot changes slope from
$-{\tilde{\eta}_{x'}/\tilde{\zeta}_{z'}}-4$ to $-{\eta_t/\zeta}-4$
. Likewise
$-\left(\tilde{\eta}_{x'}+2\tilde{\eta}_{\gamma}\right)/\tilde{\zeta}_
{x'}$ and $\xi_{\perp}$ can be determined by moving along  locus 4
(the $q_x$-axis).
%
%
%
%
%

To completely determine the anomalous exponents for the low
temperature phase, one more experiment is necessary. One possibility
is to measure the light scattering at both $q_z=0$ and $q_x=0$,
varying $q_s$, for which (\ref{LST}) leads to
\begin{widetext}
\begin{eqnarray}
C_{ss}\left(\vec{q}\right)\sim \left\{\begin{array}{ll}
{\lambda\over\left(\xi_{NL}^{\perp}\right)^2}\left(\xi_{NL}^{\perp}q_
s\right)^{2\eta_K-\eta_t}\left(1\over q_s\right)^4,
&q_s\gg\xi_{\perp}\\
{\lambda\over\left(\xi_{NL}^{\perp}\right)^2}\left(\xi_{\perp}
\over\xi_{NL}^{\perp}\right)^{\eta_t-2\eta_K}\left(\xi_{\perp}q_
s\right)^{2\tilde{\eta}_K-\tilde{\eta}_{s'}}\left(1\over
q_s\right)^4. &q_s\ll\xi_{\perp}
\end{array}\right.
\label{}.
\end{eqnarray}
\end{widetext}
%
%
%
%
%
%
Such measurements along this experimental locus determine
$2\tilde{\eta}_K-\tilde{\eta}_{s'}-4$ and also measure
$\xi_{\perp}$. Combining this result with the data for
$-{\tilde{\eta}_{x'}/\tilde{\zeta}_{z'}}-4$,
$-\left(\tilde{\eta}_{x'}+2\tilde{\eta}_{\gamma}\right)/\tilde{\zeta}_
{x'}$ and the scaling relations (\ref{scaling1}), we can deduce the
exponents $\tilde{\eta}_{\gamma}$, $\tilde{\eta}_K$,
$\tilde{\eta}_{s'}$, $\tilde{\eta}_{x'}$ and, furthermore, check the
scaling relation (\ref{scaling3}).
%
%
%
%

The remainder of this paper is organized as follows. In section
\ref{sec: Model}, we derive our model for the smectic in stretched
aerogel, and
%
%
%
perform the ``replica trick'' on it.
%
%
%
%
In section \ref{sec: ACTransition}, we derive the renormalization
group recursion relations for the model near the $AC$ transition,
identify their stable fixed point, and calculate the thermodynamic
critical exponents in a $d=5-\epsilon$ expansion. In section \ref{sec: CriticalExponents},
we calculate the critical exponents. In sections
\ref{sec: AnomalousElasticity} and \ref{sec:
AnomalousElasticityIrrelevant}, we calculate the
wavevector-dependences of the elastic constants and disorder
variances. In section \ref{sec: ACPhases}, we treat the smectic
$A$ and $C$ phases of the model and show that the former is in the
``$XY$ Bragg glass'' universality class, while the latter is in the
``$m=1$ smectic Bragg glass'' \cite{Karl} universality class. In
section \ref{sec: Cfunctions}, we calculate correlation functions
near critical region. In section \ref{sec: Stability}, we discuss
the validity of our predictions for the critical behavior
%
%
%
%
%
against the unbinding of topological defects and the loss of
orientational order.

%
%

\section{\label{sec: Model}Model}

%
%
%

As in earlier treatments of smectics $A$ and $C$\cite{deGen} in {\it
clean} environments, we expect, even in the presence of disorder,
that the important fluctuating variables in our system the ``layer
displacement field'' $u(\vec {r})$, and the molecular axis ${\hat
n}(\vec{r})$. The  layer displacement field $u(\vec{r})$ is defined,
as usual, as the displacement along the mean layer normal of the
layers relative to some ideal reference configuration of flat,
uniformly spaced layers. We will choose the orientation of these
reference layers to be perpendicular to the stretch axis.

We will derive our starting Hamiltonian by beginning with that for
a smectic $A$ in a clean environment, and then modifying it to
reflect the effects of the aerogel, and to allow the molecular
axes \nhat to tilt away from the layer normal.

The Hamiltonian for the smectic $A$ phase in a clean environment is
\cite{deGen}:
\begin{eqnarray}
H_{cl} &\equiv& \int d^dr  \left[{ K \over 2}(\nabla^2_{\perp}u)^2
+ {B \over  2}(\partial_zu-{1\over2}|\vec{\nabla}_{\perp}u|^2)^2
\right]\nonumber\\&=&\int d^dr  \left[{ K \over
2}(\nabla^2_{\perp}u)^2 + {B \over  2}(\partial_zu)^2 - {g \over
2}(\partial_zu) |\vec{\nabla}_{\perp}u|^2
\right.\nonumber\\&+&\left.{w \over 8} \left|\vec{\nabla}_{\perp}u
\right|^4\right], \label{Hcl}
\end{eqnarray}
where, in the second equality, $g=w=B$. This equality of $g$, $w$,
and $B$, which is, of course, a consequence of simply expanding out
the square of the $B$-term in Eq. (\ref{Hcl}), is also a consequence
of the global rotation invariance of a smectic in the clean system,
since it is that invariance which forces the $B$-term to be
precisely the square of the rotation invariant quantity
$E\equiv\partial_zu-{1\over2}|\vec{\nabla}_{\perp}u|^2$ in the first
place.

Once the smectic is placed in a non-rotation invariant environment
like stretched aerogel, we no longer have the constraint of
rotation invariance. However, obviously, any term that was allowed
by symmetry in the rotation invariant case will certainly be
allowed when the symmetry is lowered.  Hence, the $B$, $g$, and
$w$ terms in the second equality of Eq. (\ref{Hcl}) are still {\it
allowed} when the smectic is put in anisotropic aerogel. But since
that system is no longer rotation invariant, there is no longer
any symmetry requiring that $B=g=w$. Thus, for a smectic in
anisotropic aerogel, part of the Hamiltonian will be just the
expression after the second equality in (\ref{Hcl}) with {\it no}
constraints relating $B$, $g$, and $w$; that is, all three will
now be {\it independent} parameters of our model.

There are additional terms not present in (\ref{Hcl}) that are
allowed in our Hamiltonian due to the absence of rotation
invariance in our problem. In particular, there can, and hence, in
general, {\it will} be terms picking out a preferred orientation
for both the layer normal $\hat N$ and the molecular director
$\hat n$. We will assume that both of these preferred directions
are along the stretch axis $\hat z$.

The simplest coupling that will accomplish this is:
\begin{equation}
H_{stretch} \equiv - \int d^dr \left[M({\hat N}\cdot {\hat z})^2
      +Q({\hat n}\cdot {\hat z})^2    \right] ,
\label{Hstretch}
\end{equation}
with $M$ and $N$  both $>0$.

In addition, the molecular axis $\hat n$ and the layer normal
$\hat N$ should couple. As noted by deGennes\cite{deGen}, the
simplest such coupling takes the form:
\begin{equation}
H_{Nn} \equiv P \int d^dr |\hat{N}-\hat{n}|^2 \, .\label{HNn}
\end{equation}

Using the well-known\cite{deGen} geometrical relation
\begin{eqnarray}
{\hat N} = {{{\hat z}-\vec{\nabla}u}\over |{\hat z} -
\vec{\nabla}u|} \label{Nhat}
\end{eqnarray}
between $\hat N$ and the layer displacement field $u$, defining
${\vec{\delta n}\equiv {\hat{n}-\hat{z}}}$, and dropping terms of
$O(|\vec{\nabla}_{\perp}u|^4 , |{\vec{\delta n}_{\perp}}|^4)$
(which can be shown to be absorbable into a shift of the quartic
coupling $w$ in Eq. (\ref{Hcl})), and adding terms reflecting the
randomness of the aerogel, we obtain our starting Hamiltonian:
$H=H_1+H_2$, with:
\begin{eqnarray}
H_1 &\equiv& \int d^dr  \left[{ K \over 2}(\nabla^2_{\perp}u)^2 +
{B \over  2}(\partial_zu)^2 - {g \over 2}(\partial_zu)
|\vec{\nabla}_{\perp}u|^2 \right.\nonumber\\&+&\left. {w \over 8}
\left|\vec{\nabla}_{\perp}u \right|^4
+\vec{h}\left(\vec{r}\right)\cdot\vec{\nabla}u +
V_p(u-\phi(\vec{r}))\right] \label{H1}
\end{eqnarray}
and
\begin{equation}
H_2 \equiv \int d^dr \left[M|\vec{\nabla}_{\perp}u|^2
+P|\vec{\nabla}_{\perp}u+{\vec{\delta n}_{\perp}}|^2
      +Q|{\vec{\delta n}_{\perp}}|^2\right]\, .
\label{H2}
\end{equation}
The $\vec{h}(\vec{r})$ in (\ref{H1}) is a quenched random field
which we take to be Gaussian, zero mean, and characterized by
short-ranged anisotropic correlations
\begin{eqnarray}
\overline{h_i\left(\vec{r}\right)h_j\left(\vec{r}\,^{\prime}\right)}
= \left[\Delta_t\delta
_{ij}^{\perp}+\Delta_c\delta_{ij}^z\right]\delta^d
\left(r-r^{\prime}\right) \label{5}.
\end{eqnarray}
One might reasonably question this assumption of short-ranged
correlations of the disorder, given the known fractal structure of
aerogel. Indeed, while reference \cite{John} argued that for
liquid crystals in aerogel the components of ${\vec h}$ orthogonal
to $z$ do not have such long-ranged correlations, $h_z(\vec{r})$
{\it does}. Hence, our results are not directly applicable to
these systems, although they could be applied to systems of higher
aerogel density, which would have shorter ranged correlations.

The field $\phi(\vec{r})$ in (\ref{H1}) is also a quenched random
field with only short-ranged correlations, and is uniformly
distributed between 0 and $a$, the smectic layer spacing. The
function $V_p(u-\phi)$ is periodic with period $a$.

The physical interpretation of the quenched random fields
$\vec{h}(\vec{r})$ and $V_p(u-\phi)$ is very simple. The random
field $\vec{h}$ incorporates random torques and random
compressions, coming from the $\perp$ and $z$ components of
$\vec{h}$, respectively. The function $V_p(u-\phi(\vec{r}))$
represents the tendency of the aerogel to pin the smectic layers
in a set of random positions $\phi(\vec{r})$, modulo the smectic
layer spacing $a$, which is why $V_p$ is periodic in its argument.
The Hamiltonian (\ref{H1}) is identical to the elastic theory of a
smectic $A$ in isotropic aerogel developed in \cite{John}, with
one crucial exception: in a smectic in isotropic aerogel, rotation
invariance requires that $g=w=B$, while for our problem,  even at
$T = T_{AC}$,  where softness is recovered, $g$ and $w$ are still
free \cite{GP}, because rotational invariance is still broken. The
remaining effects of the anisotropy are incorporated in the M and
Q terms in Eq. (\ref{H2}).
%
%
%
%
%
%
%
It is the $P$ term that actually drives the AC transition in this model.
%
%
%
%
%
%
When $P$ is positive, it is energetically
favorable for ${\hat n}$ to be normal to the layers (which implies
${\delta \vec{n}}=-\vec{\nabla}_{\perp}u$), while, as we shall see
in a moment, when $P$ is sufficiently negative, it is
energetically favorable for the molecules to tip relative to the
layers, thereby putting the system in the C phase.

%
%
%
To see this, note that
%
%
%
%
%
after a linear change of variable:
\begin{eqnarray}
\vec{\delta n'}_{\perp}=\vec{\delta n}_{\perp}
+R\vec{\nabla}_{\perp}u\label{2}
\end{eqnarray}
with $R\equiv{P/(Q+P)}$, (\ref{H2}) becomes
\begin{eqnarray}
H_2= \int d^dr \left[{D(T)\over
2}|\vec{\nabla}_{\perp}u|^2+(P+Q)|\vec{\delta n'}_{\perp}
|^2\right] , \label{H2'}
\end{eqnarray}
with $D(T) \equiv 2M+{2QP/(Q+P)}$. Assuming when temperature drops
both $M$ and $Q$ remain positive and $P$ decreases continuously
from positive to negative, which drives the system from $A$ phase
into $C$ phase, $D = M+{QP/(Q+P)}$ becomes negative first.
Therefore just below $T_{AC}$, at the new ground state,
$|\vec{\nabla}_{\perp}u|$ is non-zero, $\vec{\delta
n'}_{\perp}=0$, which, combined with (\ref{2}), gives
$\delta\vec{n}_{\perp}=-R\vec{\nabla}_{\perp}u$. This implies that
both the layers and the director tilt, but  in opposite
directions. Clearly,  if $Q\gg M$, the layers tilt more than the
molecules, while in the opposite limit ($M\gg Q$), the director
tilts more, as observed in \cite{LI}.

Since   the fluctuations of $\delta\vec{n'}_{\perp}$ are massive,
we can set $\delta\vec{n'}_{\perp} = 0$ in (\ref{H2'}) and use the
sum of (\ref{H1}) and (\ref{H2'}), which now only involves $u$, as
our complete Hamiltonian:
\begin{eqnarray}
H &=& \int d^dr  \left[{ K \over 2}(\nabla^2_{\perp}u)^2 + {B
\over  2}(\partial_zu)^2+{D\over 2} \left|\vec{\nabla}_{\perp}u
\right|^2 \right.\nonumber\\&-&\left. {g \over 2}(\partial_zu)
|\vec{\nabla}_{\perp}u|^2+{w \over 8}
\left|\vec{\nabla}_{\perp}u \right|^4\right.\nonumber\\
&+&\left.\vec{h}\left(\vec{r}\right)\cdot\vec{\nabla}u +
V_p(u-\phi(\vec{r}))\right] \label{H}.
\end{eqnarray}

Since $|\vec{\nabla}_{\perp} u|$ is proportional to the tilt angle
of the smectic layers, the coefficient $D(T)$ is positive in the
$A$ phase (favoring alignment of the layer normal with the stretch
axis), and negative in the $C$ phase (favoring tilt of the
layers). Hence, by continuity, at $T=T_ {AC}$ $D(T)$ vanishes. In
what follows, we will assume that $D(T) \propto T-T_{AC}$ near
$T_{AC}$.

We can now treat
%
%
%
the quenched disorder in this Hamiltonian using the replica trick
\cite{LC, John}. This trick starts by assuming that the actual
free energy and correlation functions measured in an experiment on
a single sample will be close to the average values of these
quantities when averaged over many realizations of the disorder,
with a weight for this average given by the distributions
described earlier. To compute, e.g., the average
%
%
%
Free energy $\overline{F}=-T \overline{\log Z}$, where $Z$ is the
partition function,
%
%
%
%
%
we use on the identity $\overline{\log Z} = \lim\limits_{n
\rightarrow 0}{\overline{Z^n} - 1 \over n}$.
$\overline{Z^n}$ can now be computed by doing a repeated functional
integral over $n$ ``replicas'' of the field $u$:

\begin{eqnarray}
\overline{Z^n}&=&\overline{\left(\int{Du\ \ e^{-\beta H(u,
\vec{h},
\phi)}}\right)^n}\nonumber\\
&=&\overline{\left(\int{Du_1\ \ e^{-\beta H(u_1, \vec{h},
\phi)}}\right)\left(\int{Du_2\ \ e^{-\beta H(u_2, \vec{h},
\phi)}}\right)}\nonumber\\
& &...........\overline{\left(\int{Du_n\ \ e^{-\beta H(u_n,
\vec{h}, \phi)}}\right)}\nonumber\\
&=&\overline{\int{\coprod_{\alpha=1}^{n}Du_{\alpha}\ \ e^{-\beta
\sum\limits_{\alpha}H(u_{\alpha}, \vec{h}, \phi)}}}
\end{eqnarray}
The advantage gained via the replica trick is that the average over the random
fields  $\vec h$ and $\phi$ (which, we note, are the same for all $n $
replicas), can now be done first, leading to an effective
``replicated'' Hamiltonian for the set of $n$ $u_\alpha$'s alone.
The average over the random fields  $\vec h$ and $\phi$ couples the previously
uncoupled replicas.

Details of how this averaging process is performed can be obtained
in reference\cite{John}. The calculations here are virtually
identical, the only differences being the slightly different form of
the starting Hamiltonian (\ref{H1}, \ref{H2}).
%
%
%
%
%
%
After replicating and integrating over the disorder
$\vec{h}(\vec{r})$ utilizing Eq. (\ref{5}) we obtain
%
%
\begin{widetext}
\begin{eqnarray}
H[u_{\alpha}] &=& {1 \over 2} \int d^dr \left(\sum^n_{\alpha = 1}
\left[K\left(\nabla_{\perp}^2u_{\alpha}\right)^2 + B\left(\partial
_zu_{\alpha}\right)^2 - g(\partial
_zu_{\alpha})\left(\nabla_{\perp}u_{\alpha}\right)^2 + {w \over 4}
\left|\vec{\nabla}_{\perp}u_{\alpha}\right|^4 +
D(T)\left|\nabla_{\perp}u_{\alpha}\right|^2
\right]\right.\nonumber\\&-&\left. \sum^n_{\alpha, \beta =
1}\left[{\Delta_t\over 2T}  \nabla_{\perp}u_{\alpha} \cdot
\nabla_{\perp}u_{\beta} +{\Delta_c\over 2T}
\partial_zu_{\alpha} \cdot \partial_zu_{\beta}+{1\over
T}\Delta_p(u_{\alpha}-u_{\beta})\right]\right),
\label{CompleteH_AC}
\end{eqnarray}
\end{widetext}
where $\Delta_p(u_{\alpha}-u_{\beta})$ is a periodic function
with period $a$, the smectic layer spacing.

\section{\label{sec: ACTransition}Renormalization Group}
%
%
%
%
%
In this section we derive the RG flow equations for studying the
phase transition. The basic idea of the RG is to eliminate the
short-length degrees of freedom of $u$ and see how this affects the
long-length physics. At length scales shorter than the correlation
length, since the tilt term in Hamiltonian (\ref{CompleteH_AC}) is
negligible compared to the bend term, our model is effectively
similar to that for smectic $A$ in {\it isotropic} aerogel, in which
the tilt term is absent due to the symmetry. In the latter,
reference \cite{LC} showed that the random field disorder
$\Delta_p(u_{\alpha}-u_{\beta})$ is irrelevant compared to the
random tilt disorder in $d<5$. Repeating the (virtually identical)
calculation for this problem, we also find that
$\Delta_p(u_{\alpha}-u_{\beta})$ is irrelevant in our problem at
length scales shorter than the correlation length. However, it
becomes relevant at longer length scales in the $A$ phase, which will be discussed
in section \ref{sec: ACPhases}.
%
%
%
%
%
%
In addition, the random compression term $\Delta_c$ can also be
shown to be irrelevant in the renormalization group sense.
However, while it does {\it not} affect the renormalization group
recursion relations to lowest order in $\epsilon$,  it {\it does}
prove to be important for the correlation functions. We will
therefore treat it in detail later in section \ref{sec: AnomalousElasticityIrrelevant}.

With these simplifications ,
  %
%
%
%
%
%
we can study  the $AC$ phase transition using the following
truncated Hamiltonian:
\begin{widetext}
\begin{eqnarray}
H[u_{\alpha}] &=& {1 \over 2} \int d^dr \sum^n_{\alpha = 1}
\left[K\left(\nabla_{\perp}^2u_{\alpha}\right)^2 + B\left(\partial
_zu_{\alpha}\right)^2 - g(\partial
_zu_{\alpha})\left(\nabla_{\perp}u_{\alpha}\right)^2
 +{w
\over 4} \left|\vec{\nabla}_{\perp}u_{\alpha}\right|^4 +
D(T)\left|\nabla_{\perp}u_{\alpha}\right|^2 \right]\nonumber\\&-&
{\Delta_t\over 2T}\int d^dr \sum^n_{\alpha, \beta = 1}
\nabla_{\perp}u_{\alpha} \cdot \nabla_{\perp}u_{\beta}
\label{H_AC}.
\end{eqnarray}
\end{widetext}
This Hamiltonian's noninteracting propagator $G_{\alpha
\beta}(\vec{q})\equiv <u_{\alpha}(\vec{q})u_{\beta}(-\vec{q})>_0$
is found to be
\begin{eqnarray}
G_{\alpha \beta}(\vec{q})=TG(\vec{q})\delta_{\alpha\beta}+\Delta_t
q_{\perp}^2G(\vec{q})^2 \label{propagator}
\end{eqnarray}
with
\begin{eqnarray}
G(\vec{q})={1\over (Bq_z^2+Kq_{\perp}^4)}\, .
\end{eqnarray}

To qualitatively estimate the effect of the anharmonic terms in
Hamiltonian (\ref{H_AC}), we perform the ordinary perturbation theory. 
\begin{figure}
   \includegraphics[width=0.30\textwidth, angle=-90]{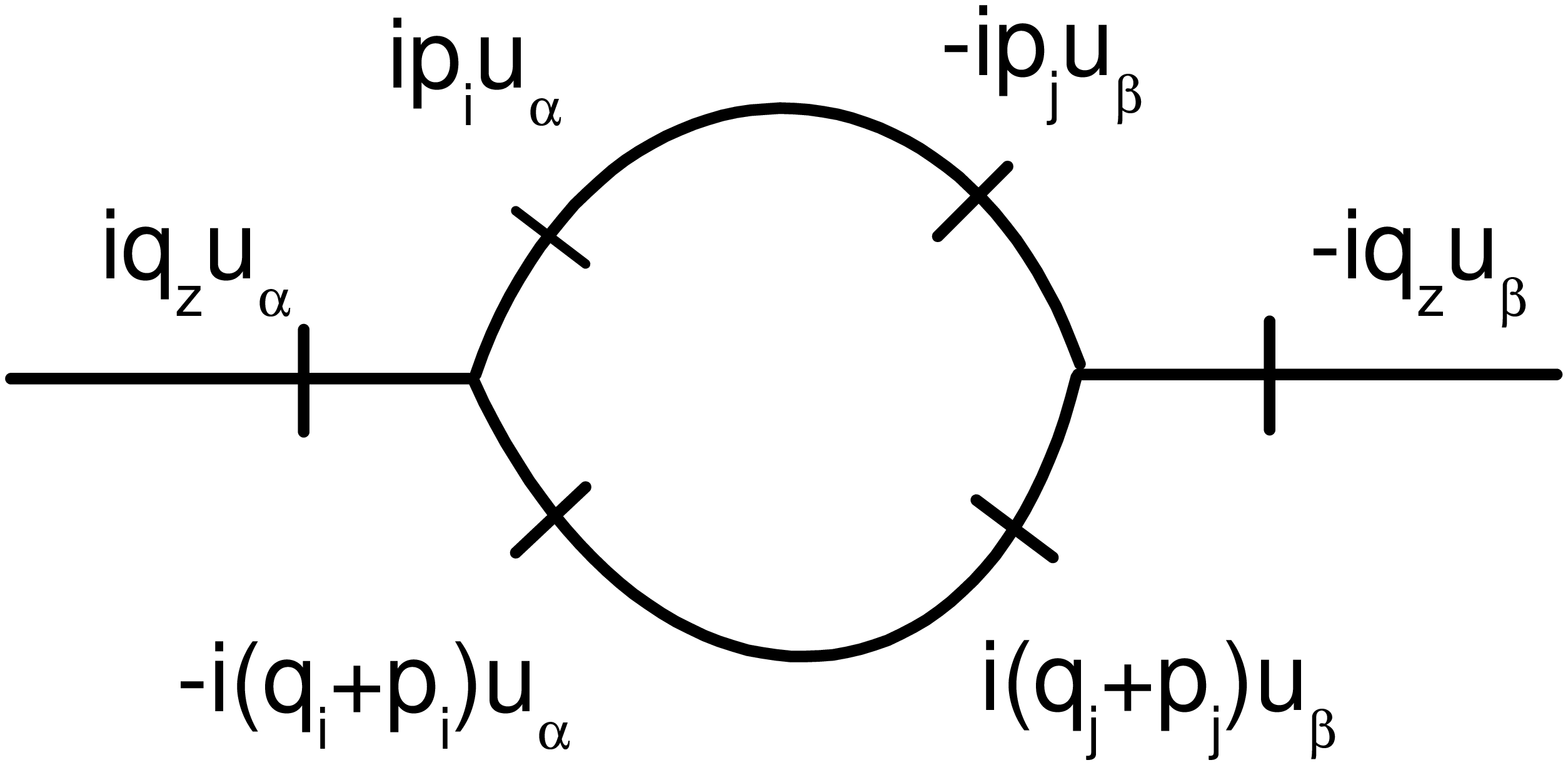}
   \caption{\label{fig: CorrectionB}The graphical correction to $B$ coming from the
   combination of the two cubic vertices.}
\end{figure}
We find the perturbation theory breaks down in $d<5$. This can be
seen by calculating the one-loop graphical correction to $B$, which
is represented by the Feynman diagram in Fig. \ref{fig:
CorrectionB}. An analysis of the Feynman diagram gives
\begin{eqnarray}
-{g^2\over 4T}\sum^n_{\alpha=1}\sum^n_{\beta=1}\sum_q
q^2_zu_{\alpha}(\vec{q})u_{\beta}
(-\vec{q})\int_{p}p_{\perp}^4G_{\alpha\beta}(\vec{p})^2,
 \label{PCorrectionToB}
\end{eqnarray}
in which
\begin{eqnarray}
G_{\alpha\beta}(\vec{p})^2&=&\left(T^2G(\vec{p})^2+2T\Delta_t
p_{\perp}^2G(\vec{p})^3\right)\delta_{\alpha\beta}\nonumber\\
&~&+\Delta_t^2p_{\perp}^4G(\vec{p})^4\, ,
 \label{propagator square}
\end{eqnarray}
where $\delta_{\alpha \beta}$ is a Kronecker delta function. This
diagram leads to two contributions. One of them is
\begin{eqnarray}
 -{g^2\Delta_t^2\over 4T}
 \sum^n_{\alpha=1}\sum^n_{\beta=1}
 \sum_q q^2_z u_{\alpha}(\vec{q}) u_{\beta} (-\vec{q})
 \int_{p} p_{\perp}^8 G(\vec{p})^4\, ,
 \label{}
\end{eqnarray}
which is actually a contribution to the random compression. We will
not discuss this contribution here since the random compression is
irrelevant. Later we will come back to it when we calculate the
wavevector-dependence of $\Delta_c$.
The other contribution is
\begin{eqnarray}
 &~&-{g^2\over 4} \sum^n_{\alpha=1}\sum_q
 q^2_z u_{\alpha}(\vec{q}) u_{\alpha} (-\vec{q})
 \nonumber\\
 &~&\times\int_{p}p_{\perp}^4[T
 G(\vec{p})^2+2\Delta_t p_{\perp}^2G(\vec{p})^3]\, ,
 \label{}
\end{eqnarray}
from which we obtain the correction to the compression modulus $B$
\begin{eqnarray}
\delta B &=&-{g^2\over2}\int_{p}p_{\perp}^4[T
G(\vec{p})^2+2\Delta_t
p_{\perp}^2G(\vec{p})^3]\nonumber\\
&\approx&-g^2\Delta_t\int_{p}{p_{\perp}^6\over
(Bp_z^2+Kp_{\perp}^4)^3}\nonumber\\
&\approx&-g^2\Delta_t \int^{\infty}_{-\infty}{dp_z\over
2\pi}\int^{\Lambda}_{1\over L}{d^{d-1}p_{\perp}\over (2\pi)^{d-1}}
{p_{\perp}^6\over (Bp_z^2+Kp_{\perp}^4)^3}\nonumber\\
&\approx&-{3\over 16}{S_{d-1}B\Delta_t\over
\left(2\pi\right)^{d-1}\left(5-d\right)} \left(g\over B\right)^2
\left(B\over K^5\right)^{1/2}L^{5-d},\nonumber\\
 \label{Pdelta B}
\end{eqnarray}
where, in the second equality we throw out the thermal fluctuation
piece because it is less divergent, $S_{d-1}$ is the surface area of
a $d-1$ dimensional sphere with unit radius. This result diverges as
a power law of the system size $L$ right at the critical point in
$d<5$, which implies that the upper critical dimension $d_{uc}$ for
this phase transition is 5, below which Gaussian theory breaks down.
Since the physical dimension $d=3$ is less than $d_{uc}$, the
standard momentum shell renormalization group (RG) transformation
has to be employed to study the phase transition.

Because a few features of our treatment are non-standard, we will
describe our approach in some detail.

The first unusual feature is that we use an infinite
hyper-cylindrical Brillouin zone: $|\vec{q}|<\Lambda$,
$-\infty<q_z<\infty$, where $\Lambda\sim{1/a}$ is an ultra-violet
(UV) cutoff. We separate the displacement field into high and low
wavevector components. $u_{\alpha}\left(\vec{r}\right) =
u_{\alpha}^< \left(\vec{r}\right) +u_{\alpha}^>
\left(\vec{r}\right)$ where $u_{\alpha}^> \left(\vec{r}\right)$ has
support in the hyper-cylindrical shell $\Lambda
e^{-\ell}<q_\perp<\Lambda$, $-\infty<q_z<\infty$, and $u_{\alpha}^<
\left(\vec{r}\right)$ has support in the remainder of the
hyper-cylinder (i.e., $q_\perp<\Lambda e^{-\ell}$,
$-\infty<q_z<\infty$). We then integrate the high wavevector part
$u_{\alpha}^> \left(\vec{r}\right)$, and rescale the length and long
wavelength part of the fields with $\vec{r}_{\perp} =
\vec{r}_{\perp}^{\,\prime} e^{\ell}$, $z = z^{\prime}e^{\omega
\ell}$, and $u_{\alpha}^> \left(\vec{r}\right) = e^{\chi
\ell}u^{\prime}_{\alpha}\left(\vec{r}\, ^{\prime}\right)$ so as to
restore the $UV$ cutoff back to $\Lambda$. This rescaling leads to
the zeroth order $RG$ flows of the effective couplings
\begin{eqnarray}
 K(\ell) &=& Ke^{(d+\omega+2\chi-5)\ell},\nonumber\\
 B(\ell) &=& Be^{(d-\omega+2\chi-1)\ell},\nonumber\\
 {\Delta_t\over T} \left(\ell\right) &=& \left(\Delta_t\over T\right)
 e^{(d+\omega+2\chi-3)\ell}.\nonumber
\end{eqnarray}
From these dimensional couplings one can construct four
dimensionless couplings:
\begin{eqnarray}
 g_1 &\equiv& C_{d-1}\left(B/K^3 \right)^{1 \over 2}
 \Lambda^{d-3}\, ,\\
 g_2 &\equiv& C_{d-1}\left(B/K^5 \right)^{1 \over 2}
 \Delta_t \Lambda^{d-5}\, ,\\
 g_3 &\equiv& C_{d-1}(g/B)^2\left(B/K^5 \right)^{1 \over 2}
 \Delta_t \Lambda^{d-5}\, ,\label{Defineg3}\\
 g_4 &\equiv& C_{d-1}(w/B)\left(B/K^5 \right)^{1\over 2}
 \Delta_t \Lambda^{d-5}\, ,\label{Defineg4}
\end{eqnarray}
where $C_{d-1}\equiv S_{d-1}/\left(2\pi\right)^{d-1}$. The RG flows
of the dimensionless couplings are independent of the arbitrary
rescaling exponents and given, ignoring graphical corrections, by
\begin{eqnarray}
 g_1(\ell) &=& g_1e^{(3-d)\ell}\, ,\\
 g_2(\ell) &=& g_2e^{(5-d)\ell}\, ,\\
 g_3(\ell) &=& g_3e^{(5-d)\ell}\, ,\\
 g_4(\ell) &=& g_4e^{(5-d)\ell}\, .
\end{eqnarray}
$g_1$ is just the dimensionless coupling found by Grinstein and
Pelcovits\cite{GP} in the pure smectic problem. It becomes
marginal and leads to very weak anomalous elasticity in $d=3$.
$g_2$ is actually the dimensionless coupling found by Toner and
Radsihovsky \cite{John} in the isotropic disordered smectic
problem, where it becomes relevant in $d<5$ and leads to much
stronger anomalous elasticity in $d=3$. $g_3$ and $g_4$, the truly
nontrivial dimensionless couplings in our problem, also become
relevant in $d<5$. As we will see below, they control the
structure of the RG flows and hence the critical behavior.
\begin{figure}
   \includegraphics[width=0.35\textwidth, angle=-90]{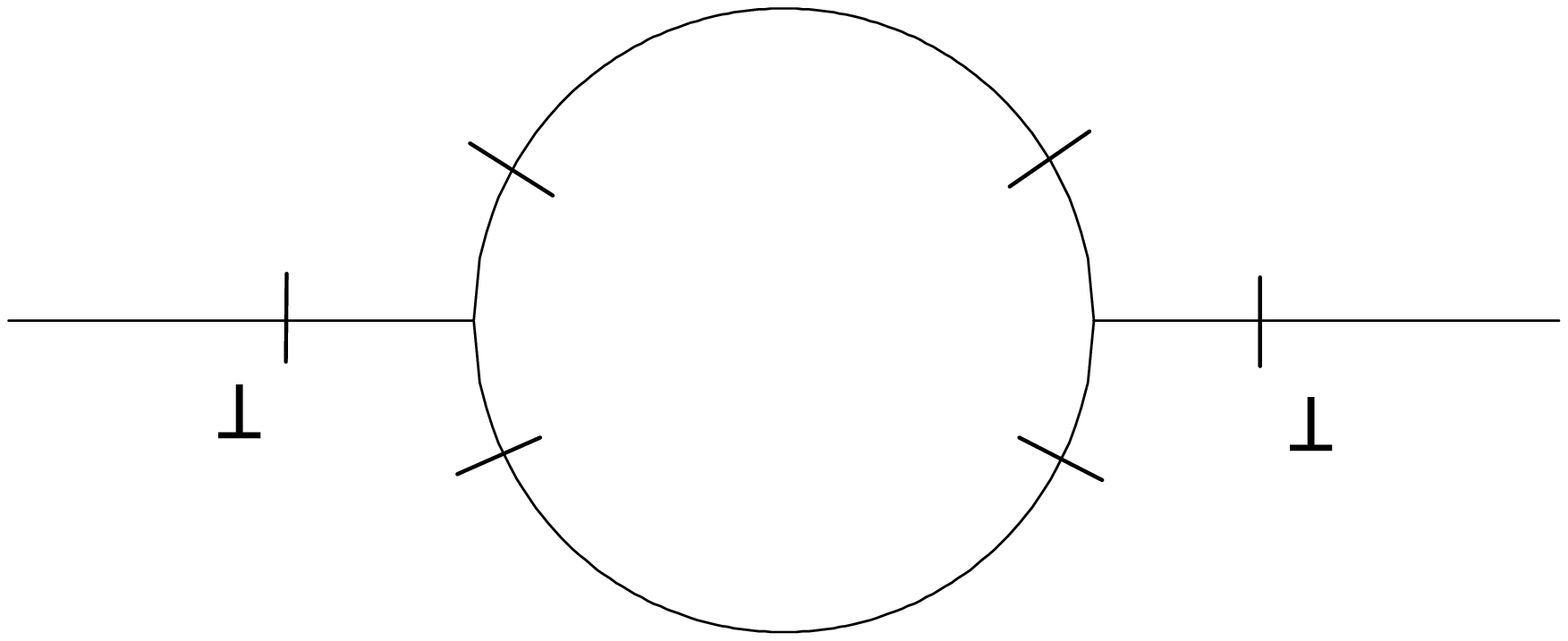}
   \caption{\label{fig: CorrectionD}Graphical corrections to the bend, tilt and random tilt terms.}
   \end{figure}
\begin{figure}
   \includegraphics[width=0.35\textwidth, angle=-90]{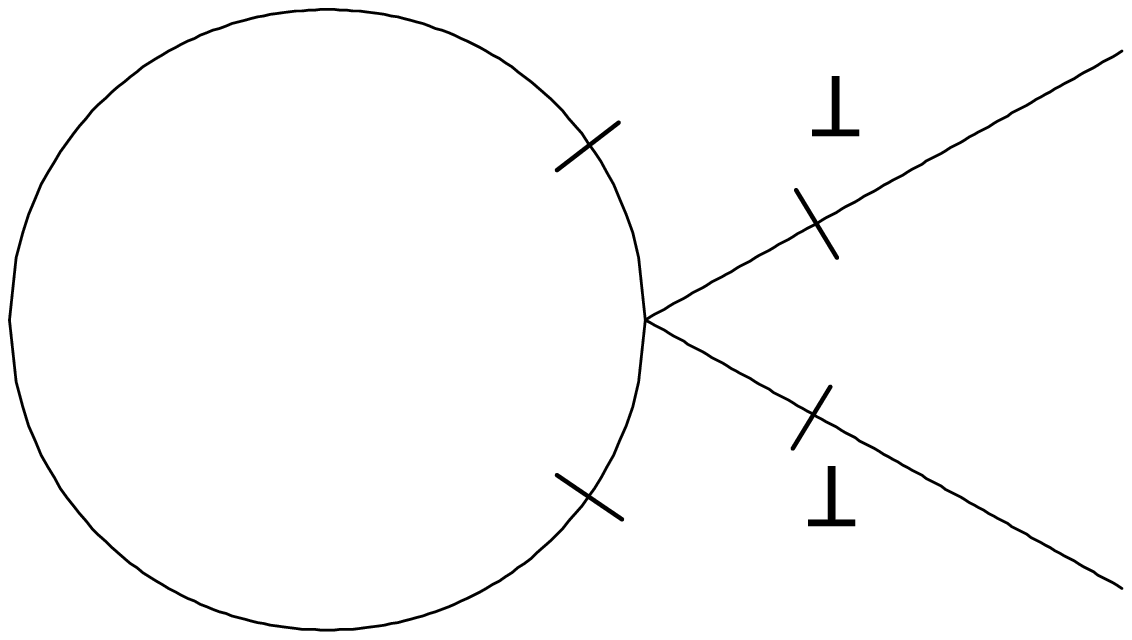}
   \caption{\label{fig: CorrectionD1}A graphical correction to the tilt term.}
   \end{figure}
\begin{figure}
   \includegraphics[width=0.35\textwidth, angle=-90]{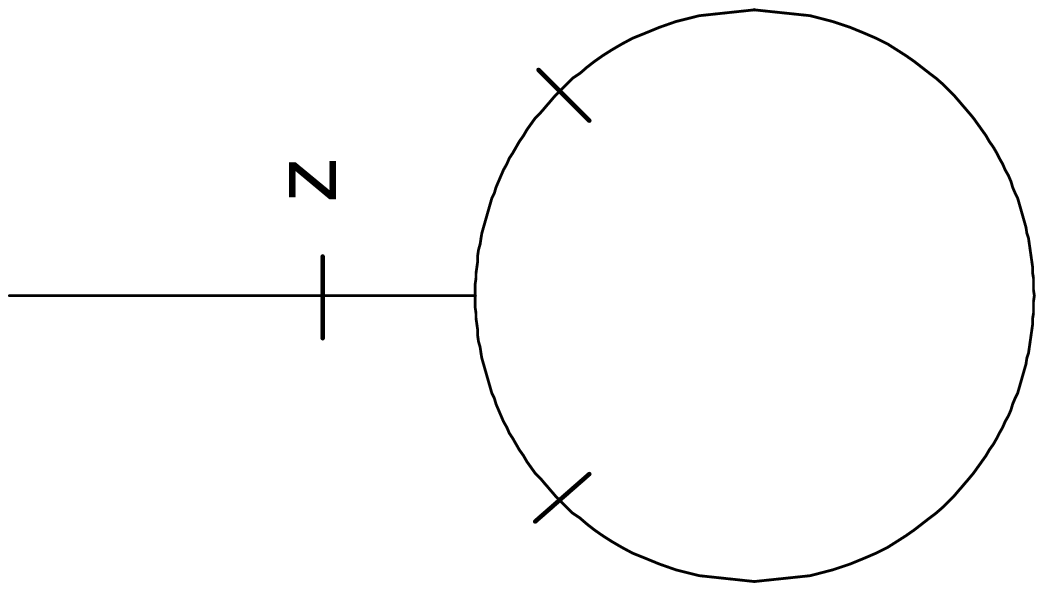}
   \caption{\label{fig: CorrectionLin}The graphical correction to the linear term $\displaystyle\sum_{n=1}^{\infty}
   \partial_z u$.}
   \end{figure}
\begin{figure}
   \includegraphics[width=0.35\textwidth, angle=-90]{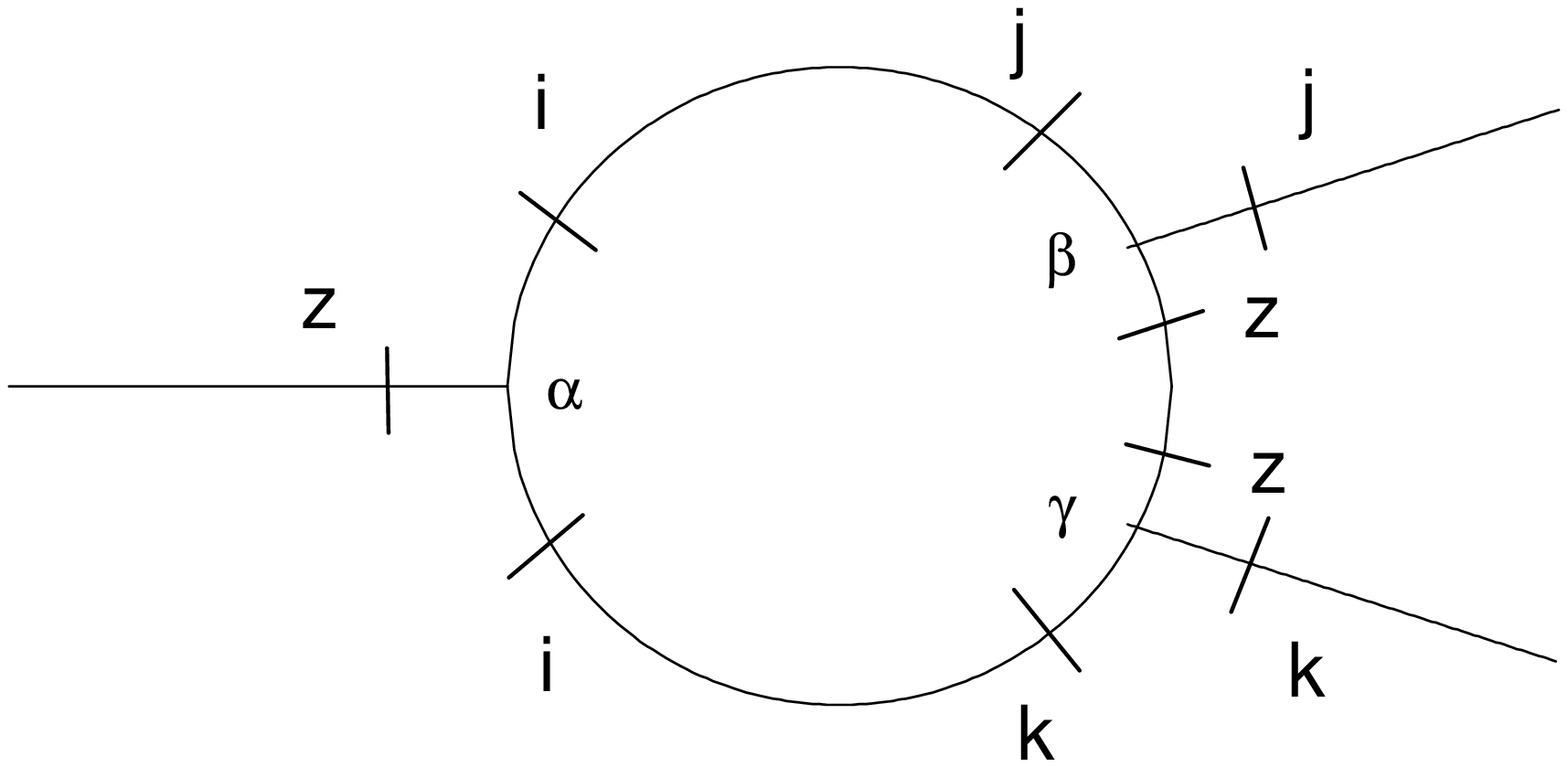}
   \caption{\label{fig: Correctiong1}A graphical correction to the cubic term.}
   \end{figure}
\begin{figure}
   \includegraphics[width=0.35\textwidth, angle=-90]{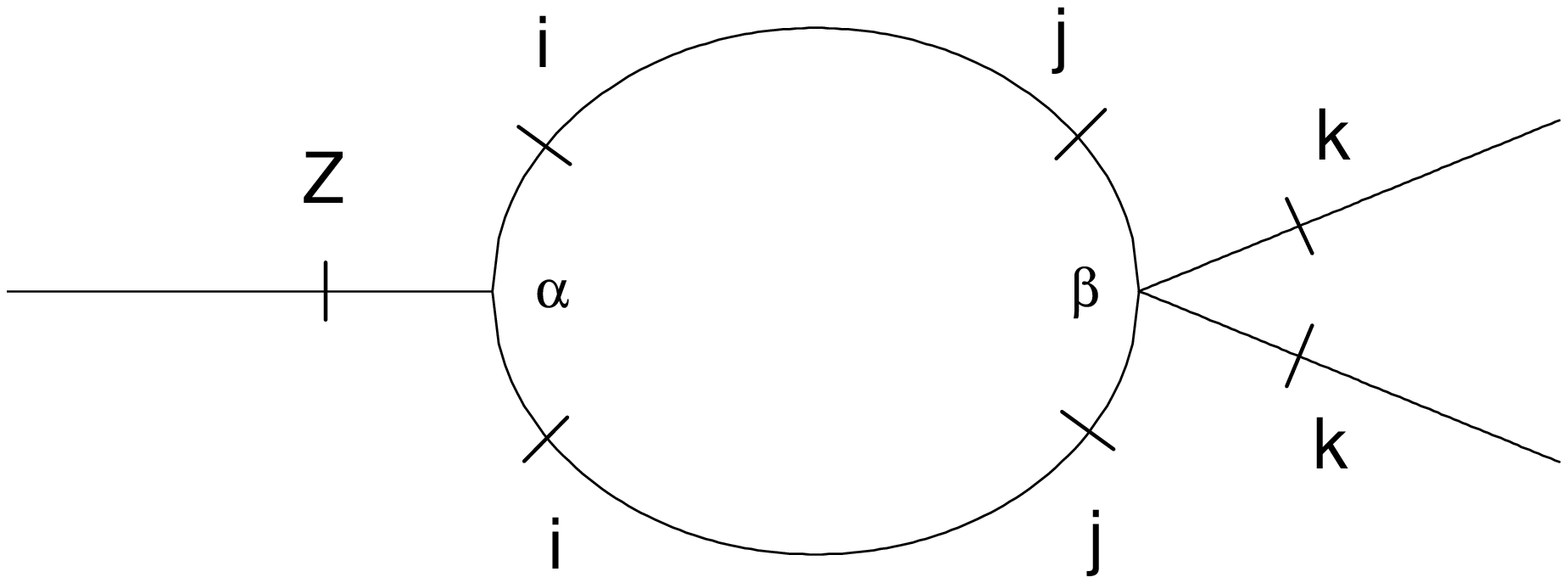}
   \caption{\label{fig: Correctiong2}Another graphical correction to the cubic term.}
   \end{figure}
\begin{figure}
   \includegraphics[width=0.35\textwidth, angle=-90]{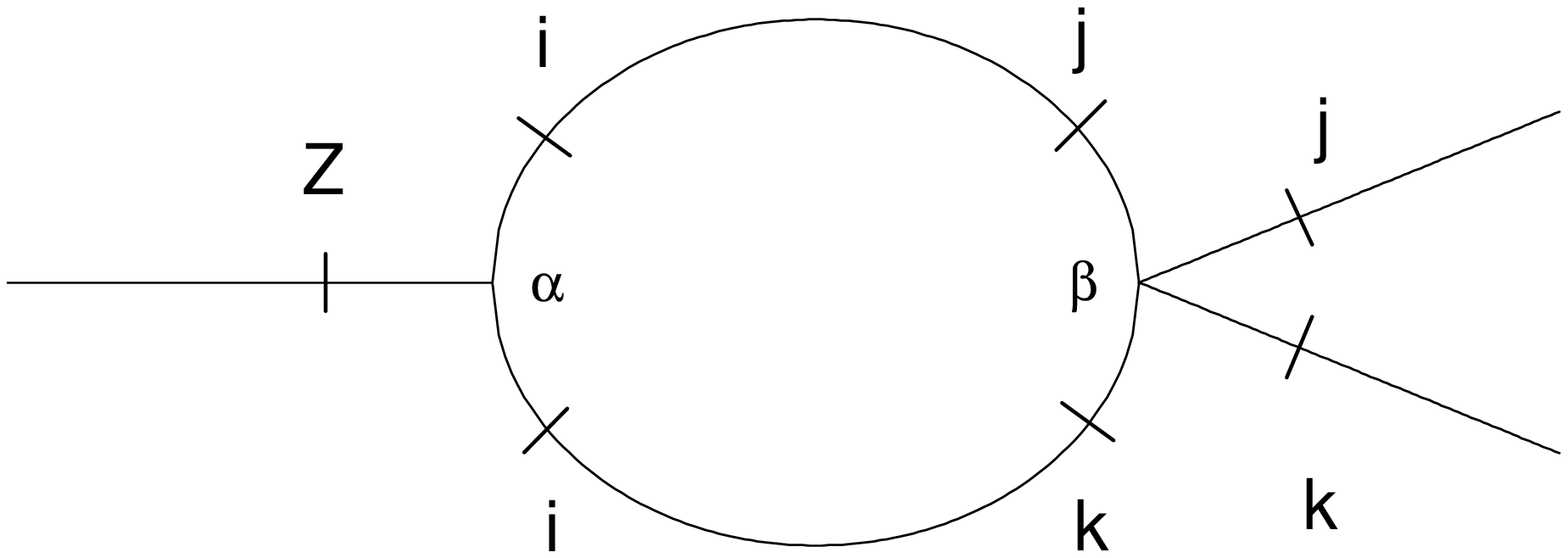}
   \caption{\label{fig: Correctiong3}One more graphical correction to the cubic term.}
   \end{figure}
\begin{figure}
   \includegraphics[width=0.35\textwidth, angle=-90]{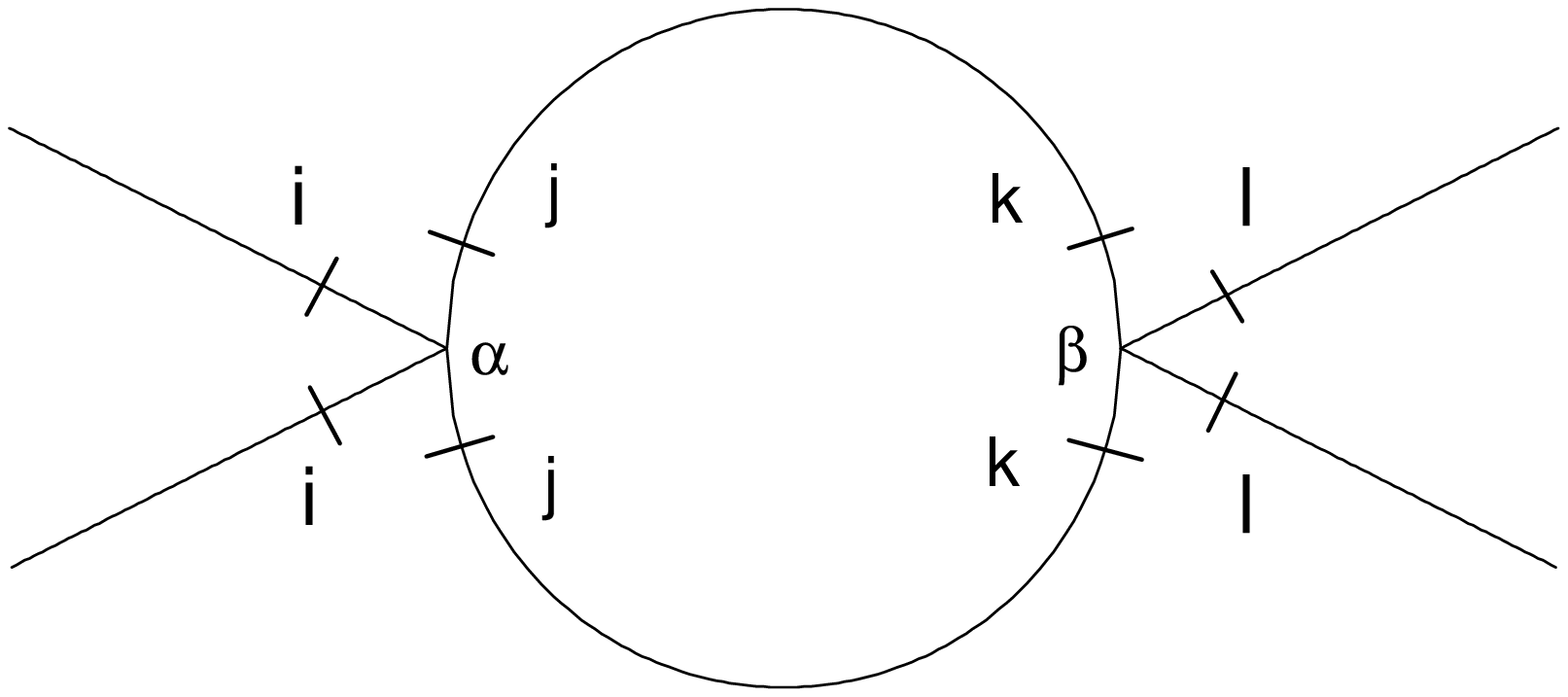}
   \caption{\label{fig: CorrectionW1}A graphical correction to the quartic term.}
   \end{figure}
\begin{figure}
   \includegraphics[width=0.35\textwidth, angle=-90]{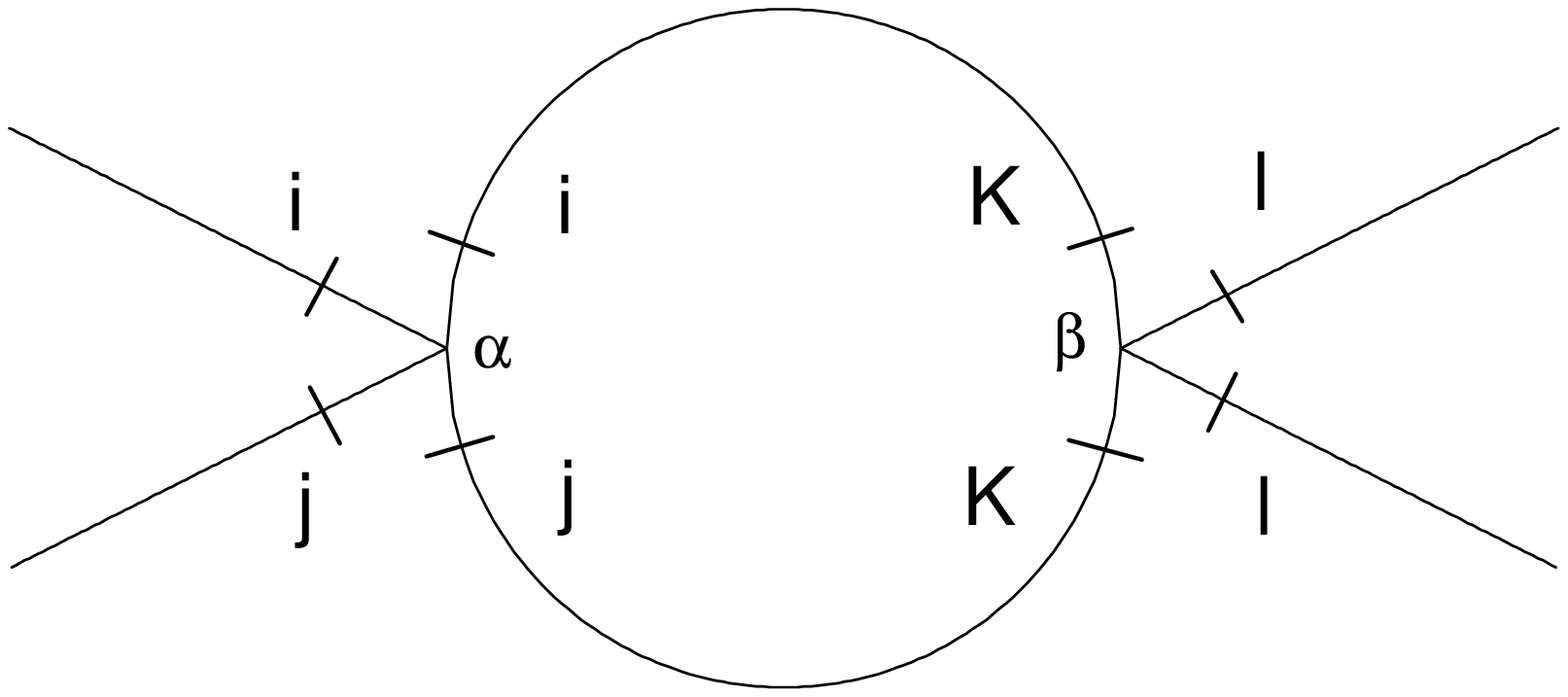}
   \caption{\label{fig: CorrectionW2}Another graphical correction to the quartic term.}
   \end{figure}
\begin{figure}
   \includegraphics[width=0.35\textwidth, angle=-90]{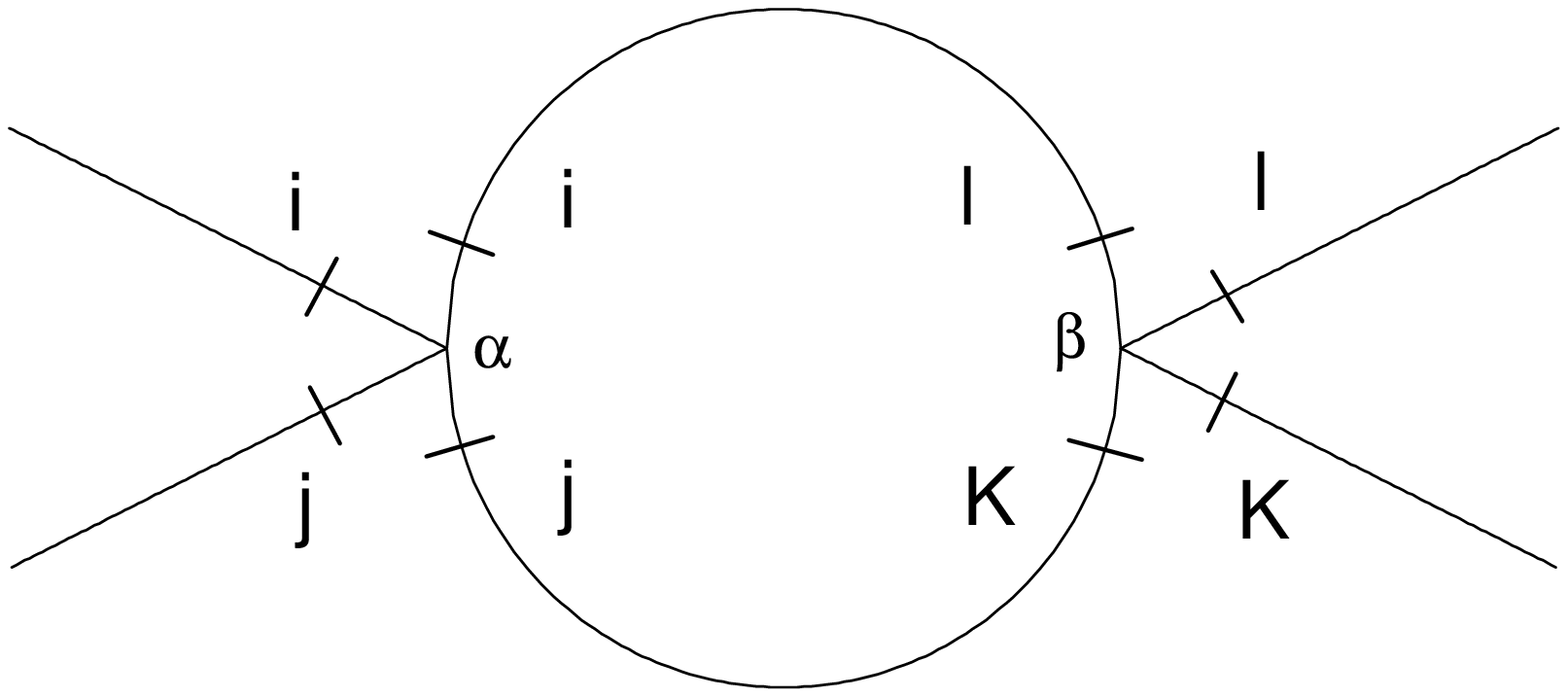}
   \caption{\label{fig: CorrectionW3}Another graphical correction to the quartic term.}
   \end{figure}
\begin{figure}
   \includegraphics[width=0.35\textwidth, angle=-90]{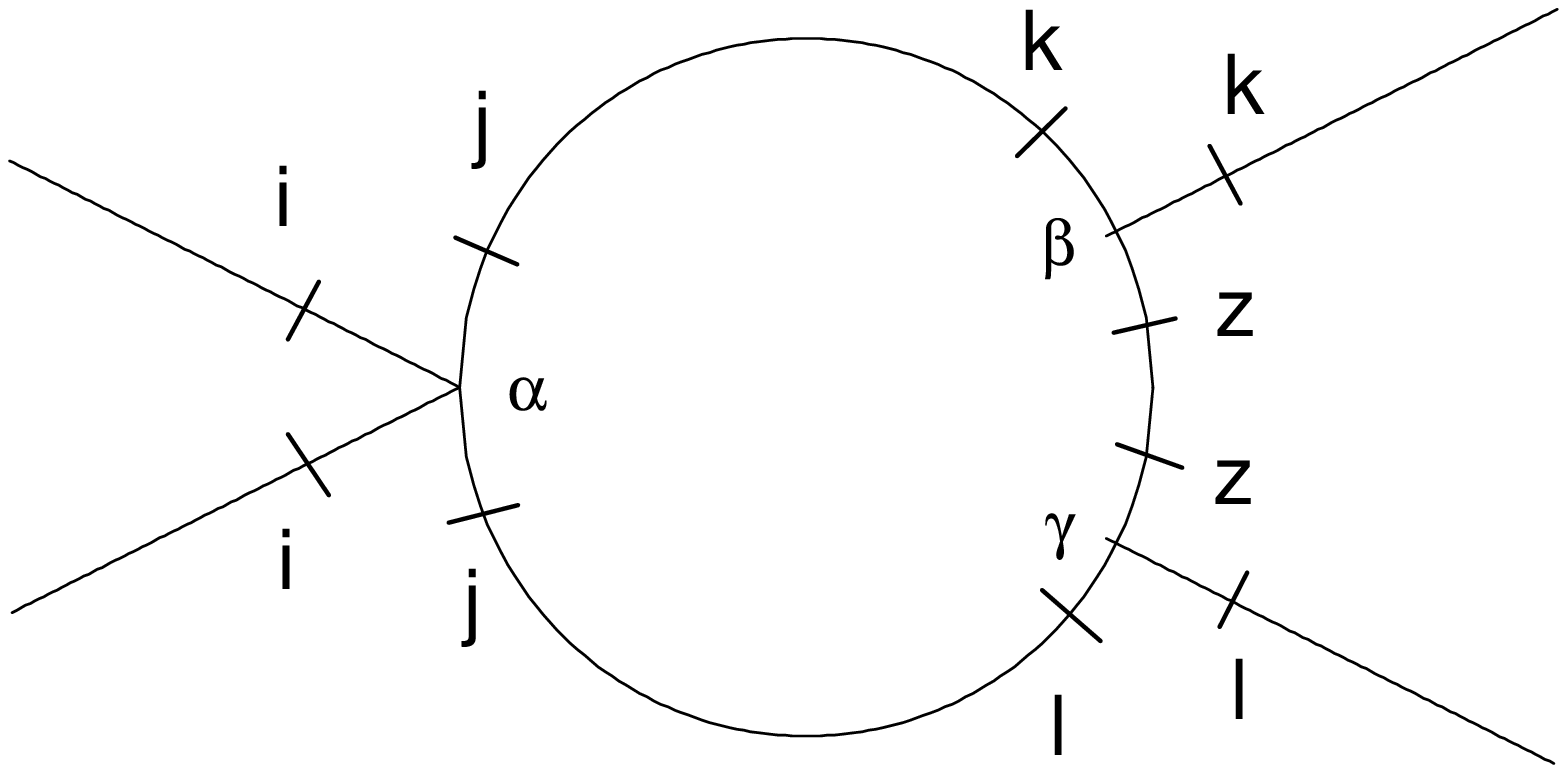}
   \caption{\label{fig: CorrectionW4}Yet another graphical correction to the quartic term.}
   \end{figure}
\begin{figure}
   \includegraphics[width=0.35\textwidth, angle=-90]{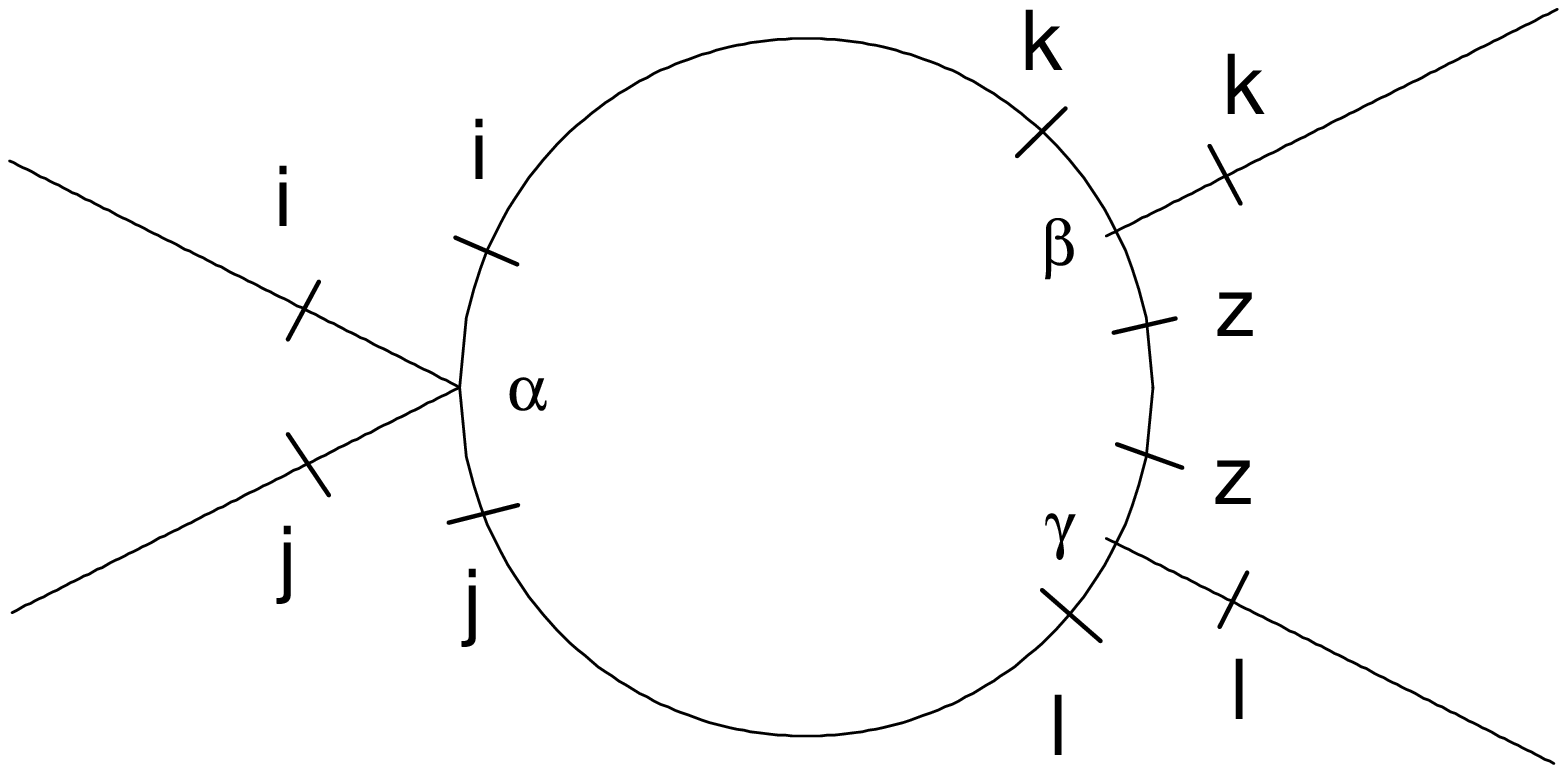}
   \caption{\label{fig: CorrectionW5}One more graphical correction to the quartic term.}
   \end{figure}
\begin{figure}
   \includegraphics[width=0.35\textwidth, angle=-90]{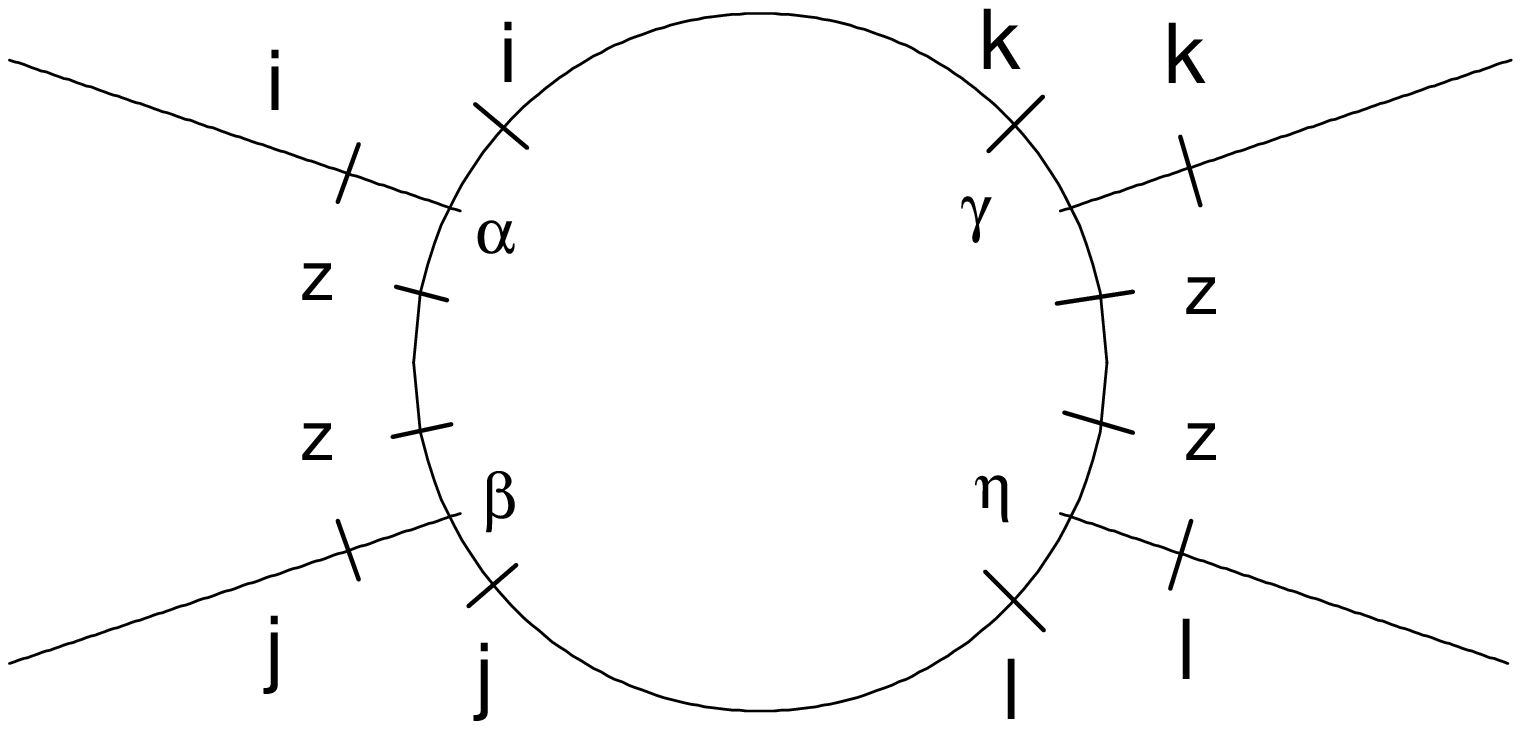}
   \caption{\label{fig: CorrectionW6}The final graphical correction to the quartic term.}
   \end{figure}

Since the integration over the high wavevector components of
$u_{\alpha}$ is always performed in the momentum shell
($-\infty<q_z<\infty$, $\Lambda e^{-d\ell}<q_{\perp}<\Lambda$), it
can be accomplished perturbatively in nonlinearities of $H[u]$. The
change of the effective Hamiltonian of the long length modes due to
the integration is
\begin{eqnarray}
\delta
H[u^<_{\alpha}]&=&\left<H_a[u_{\alpha}^<+u_{\alpha}^>]\right>_> -
{1\over
2T}\left<H_a^2[u_{\alpha}^<+u_{\alpha}^>]\right>_>\nonumber\\
               &~&+{1\over 6T^2}\left<H_a^3[u_{\alpha}^<+u_{\alpha}^>
               ]\right>_>\nonumber\\
               &~&-{1\over 24T^3}\left<H_a^4[u_{\alpha}^<+
               u_{\alpha}^>]\right>_>+......\, ,
               \label{RGperturb}
\end{eqnarray}
where $H_a$ represents the anharmonic terms which includes the cubic
and quartic terms, the average over the short length modes is done
by only using the harmonic part of the Hamiltonian.

Now we will discuss $\delta H[u^<]$ to lowest order (one loop
graphical correction) in detail.

Since as shown in (\ref{RGperturb}), the calculation of $\delta
H[u^<_{\alpha}]$ is just a modified perturbation theory, the
evaluation of the one-loop graphical correction to $B$ for long
wavelength modes can be accomplished by modifying (\ref{Pdelta B}),
restricting the integral range within the momentum shell
($-\infty<q_z<\infty$, $\Lambda e^{-d\ell}<q_{\perp}<\Lambda$).
Doing so we obtain
\begin{eqnarray}
\delta B &=&-{g^2\over2}\int^>_{p}p_{\perp}^4[T
G(\vec{p})^2+2\Delta_t
p_{\perp}^2G(\vec{p})^3]\nonumber\\
&\approx&-g^2\Delta_t\int^>_{p}{p_{\perp}^6\over
(Bp_z^2+Kp_{\perp}^4)^3}\nonumber\\
&\approx&-g^2\Delta_t \int^{\infty}_{-\infty}{dp_z\over
2\pi}\int^{\Lambda}_{\Lambda e^{-d\ell}}{d^{d-1}p_{\perp}\over
(2\pi)^{d-1}}
{p_{\perp}^6\over (Bp_z^2+Kp_{\perp}^4)^3}\nonumber\\
&\approx&-{3\over 16}C_{d-1}B\Delta_t \left(g\over B\right)^2
\left(B\over K^5\right)^{1/2}{\Lambda}^{d-5}d\ell\nonumber\\
&\approx&-{3\over 16}B g_3 d\ell. \label{delta B}
\end{eqnarray}

Now we discuss graphical corrections to the bend term (i.e., $K$-term),
the tilt term (i.e., $D$-term)
and the random tilt term (i.e, $\Delta_t$-term).
The Feynman diagram for these corrections are presented in Fig.
\ref{fig: CorrectionD} and \ref{fig: CorrectionD1}. Following
Feynman rules, from the diagram in Fig. \ref{fig: CorrectionD} we
obtain
\begin{eqnarray}
 &-&{g^2\over 2T}
 \sum^n_{\alpha=1}\sum^n_{\beta=1}\sum_{\vec{q}}\sum_{i j}
 q^{\perp}_i q^{\perp}_j
 u_{\alpha}(\vec{q})u_{\beta}(-\vec{q})\nonumber\\
 &\times& \int^>_{p} p_i^{\perp}(p_j^{\perp}+q_j^{\perp}) p^2_z
 G_{\alpha\beta}(\vec{p})
 G_{\alpha\beta}(\vec{p}+\vec{q}),
 \nonumber\\
 \label{noname1}\\
 &-&{g^2\over 2T}
 \sum^n_{\alpha=1}\sum^n_{\beta=1}\sum_{\vec{q}}\sum_{i j}
 q^{\perp}_i q^{\perp}_j
 u_{\alpha}(\vec{q})u_{\beta}(-\vec{q})\nonumber\\
 &\times& \int^>_{p} p_i^{\perp}p_j^{\perp} p^2_z
 G_{\alpha\beta}(\vec{p})
 G_{\alpha\beta}(\vec{p}+\vec{q}),\label{noname2}
\end{eqnarray}
where $G_{\alpha\beta}$ is given in (\ref{propagator square}). Note
that we have kept the $\vec{q}$-dependence of the internal legs
(i.e., the $\vec{q}$-dependence of the integrands in
(\ref{noname1}), (\ref{noname2})) since it is necessary for
obtaining the contributions to the bend term. Taylor expanding the
two integrands around $\vec{q}=0$, we find that, $O(1)$ terms lead
to the corrections to $\Delta_t$ and $D$; $O(q_{\perp})$ terms
vanish after the integration; $O(q_{\perp}^2)$ terms lead to the
corrections to $K$.

Let us first calculate the corrections to $\Delta_t$ and $D$. Setting
$\vec{q}$ inside the two integrands to $\vec{0}$, (\ref{noname1})
and (\ref{noname2}) become identical and can be added together to
get
\begin{eqnarray}
 -{g^2}
 \sum^n_{\alpha=1}\sum^n_{\beta=1}\sum_{\vec{q}}\sum_{i j}
 q_i^{\perp} q_j^{\perp}
 u_{\alpha}(\vec{q})u_{\beta}(-\vec{q})\nonumber\\
 \times \int^>_{p} p_i^{\perp} p_j^{\perp} p^2_z
 G_{\alpha\beta}(\vec{p})G_{\alpha\beta}(\vec{p})\, ,
 \label{noname3}
\end{eqnarray}
from which we obtain two distinct contributions. One of them is the
contribution to the random tilt term with the corresponding
correction to the random tilt variable $\Delta_t$:
\begin{eqnarray}
\delta \Delta_t &\approx&{g^2\over2}\int^>_{p}p^2_zp_{\perp}^6
\Delta^2_tG(\vec{p})^4\nonumber\\
&\approx&{g^2\over 2}\Delta_t^2\int^>_{p}{p_{\perp}^6p_z^2\over
(Bp_z^2+Kp_{\perp}^4)^4}\nonumber\\
&\approx&{g^2\over 2}\Delta_t^2 \int^{\infty}_{-\infty}{dp_z\over
2\pi}\int^{\Lambda}_{\Lambda e^{d\ell}}{d^{d-1}p_{\perp}\over
(2\pi)^{d-1}}
{p_{\perp}^6p_z^2\over (Bp_z^2+Kp_{\perp}^4)^3}\nonumber\\
&\approx&{1\over 64}C_{d-1}\Delta_t^2 \left(g\over
B\right)^2\left(B\over
K^5\right)^{1/2}{\Lambda}^{d-5}d\ell\nonumber\\
&\approx&{1\over 64} \Delta_t g_3 d\ell\, . \label{delta Delta_t}
\end{eqnarray}
The other one is the correction to the ordinary tilt term with the
corresponding correction to the tilt modulus $D$:
\begin{eqnarray}
(\delta D)_1 &\approx&-{4\over 3}g^2\Delta_t\int^>_{p}
p^2_zp_{\perp}^4 G(\vec{p})^3\nonumber\\
&\approx&-{4\over 3}g^2\Delta_t\int^>_{p}{p_{\perp}^4p_z^2\over
(Bp_z^2+Kp_{\perp}^4+Dp_{\perp}^2)^3}\nonumber\\
&\approx&(-{1\over 16}K\Lambda^2+{3\over 32}D)g_3d\ell\, ,
\label{deltaD1}
\end{eqnarray}
where we have used $(\delta D)_1$ instead of $(\delta D)$ since
there are more corrections to $D$ from other Feynman diagraphs,
which will be calculated later in this section. 

Now let us calculate the corrections to the bend term from
(\ref{noname1}) and (\ref{noname2}). As mentioned earlier, we have
to Taylor expand both of the integrands and use the $O(q_{\perp}^2)$
term. Fortunately, since (\ref{noname2}) is part of (\ref{noname1}),
we only need to do the calculation once for (\ref{noname1}), which,
however, is still very involved. In the following we will describe
the calculation in detail.

Only keeping the most divergent pieces which carry the factor
$\delta_{\alpha \beta}$ , the integrand in (\ref{noname1}) is
simplified as
\begin{eqnarray}
 &~~~&p_i^{\perp}(p_j^{\perp}+q_j^{\perp}) p^2_z
 \left[{\Delta_t p_{\perp}^2\delta_{\alpha\beta}
 \over (Kp_{\perp}^4+Bp_z^2)^2(K|\vec{p}+\vec{q}|^4+Bp_z^2)}
 \right.
 \nonumber\\
 &~~~&\left.+{\Delta_t |\vec{p}+\vec{q}|^2\delta_{\alpha\beta}
 \over (Kp_{\perp}^4+Bp_z^2)
 (K|\vec{p}+\vec{q}|^4+Bp_z^2)^2}\right].\nonumber\\
 \label{Integ function}
\end{eqnarray}
For convenience, we rewrite $K|\vec{p}+\vec{q}|^4$ as
\begin{eqnarray}
 K|\vec{p}+\vec{q}|^4=K(p_{\perp}^4+f(\vec{q}_{\perp}))
\end{eqnarray}
with $f(\vec{q}_{\perp})$ defined as
\begin{eqnarray}
 f(\vec{q}_{\perp})\equiv
 2p_{\perp}^2
 (2\vec{p}_{\perp}\cdot\vec{q}_{\perp}+q_{\perp}^2)
 +(2\vec{p}_{\perp}\cdot\vec{q}_{\perp}+q_{\perp}^2)^2.
 \label{qfunction}
\end{eqnarray}
Now let us do the following two Taylor expansions:
\begin{widetext}
\begin{eqnarray}
 {1\over (K|\vec{p}_{\perp}+\vec{q}_{\perp}|^4+Bp_z^2)}&=&
 {1\over Bp_z^2+Kp_{\perp}^4}
 \left[1-{K f(\vec{p}_{\perp})\over Bp_z^2+Kp_{\perp}^4}
 +{K^2 f^2(\vec{p}_{\perp})\over (Bp_z^2+Kp_{\perp}^4)^2}\right],
 \nonumber
 \\
 {1\over (K|\vec{p}_{\perp}+\vec{q}_{\perp}|^4+Bp_z^2)^2}&=&
 {1\over (Bp_z^2+Kp_{\perp}^4)^2}
 \left[1-{2 K f(\vec{p}_{\perp})\over Bp_z^2+Kp_{\perp}^4}
 +{3 K^2 f^2(\vec{p}_{\perp})\over
 (Bp_z^2+Kp_{\perp}^4)^2}\right].\nonumber
\end{eqnarray}
\end{widetext}
Plugging these two expressions into (\ref{Integ function}) makes
it extremely long and complicated. For better management, We
divided it into four pieces:
\begin{widetext}
\begin{eqnarray}
 {p_i^{\perp} p_j^{\perp} p^2_z
 \Delta_t p_{\perp}^2\delta_{\alpha\beta}
 \over (Kp_{\perp}^4+Bp_z^2)^3}
 \left[1-{K f(\vec{q}_{\perp})\over Bp_z^2+Kp_{\perp}^4}
 +{K^2 f^2(\vec{q}_{\perp})\over (Bp_z^2+Kp_{\perp}^4)^2}\right]\, ,
 \label{piece1}\\
 {p_i^{\perp} q_j^{\perp} p^2_z
 \Delta_t p_{\perp}^2\delta_{\alpha\beta}
 \over (Kp_{\perp}^4+Bp_z^2)^3}
 \left[1-{K f(\vec{q}_{\perp})\over Bp_z^2+Kp_{\perp}^4}
 +{K^2 f^2(\vec{q}_{\perp})\over (Bp_z^2+Kp_{\perp}^4)^2}\right]
 \label{piece2}\, ,\\
 {p_i^{\perp} p_j^{\perp} p^2_z
 \Delta_t |\vec{p}+\vec{q}|^2\delta_{\alpha\beta}
 \over (Kp_{\perp}^4+Bp_z^2)^3}
 \left[1-{2 K f(\vec{q}_{\perp})\over Bp_z^2+Kp_{\perp}^4}
 +{3 K^2 f^2(\vec{q}_{\perp})\over
 (Bp_z^2+Kp_{\perp}^4)^2}\right],
 \label{piece3}\\
 {p_i^{\perp} q_j^{\perp} p^2_z
 \Delta_t |\vec{p}+\vec{q}|^2\delta_{\alpha\beta}
 \over (Kp_{\perp}^4+Bp_z^2)^3}
 \left[1-{2 K f(\vec{q}_{\perp})\over Bp_z^2+Kp_{\perp}^4}
 +{3 K^2 f^2(\vec{q}_{\perp})\over
 (Bp_z^2+Kp_{\perp}^4)^2}\right].
 \label{piece4}
\end{eqnarray}
\end{widetext}
Inserting Eq. (\ref{qfunction}) into the above four pieces, we
obtain $O(q_{\perp}^2)$ terms:
\begin{widetext}
\begin{eqnarray}
  I_1(\vec{q}_{\perp})
  &\equiv& -2K \Delta_t q_{\perp}^2 \delta_{\alpha\beta}
 {p_z^2 p_{\perp}^4 p_i^{\perp} p_j^{\perp}
 \over (BP_z^2+Kp_{\perp}^4)^4}
 -4K \Delta_t \delta_{\alpha\beta} \sum_{k} \sum_{\ell} q_k^{\perp}
 q_{\ell}^{\perp}
 {p_z^2 p_{\perp}^2 p_i^{\perp} p_j^{\perp}
 p_k^{\perp} p_{\ell}^{\perp}
 \over (BP_z^2+Kp_{\perp}^4)^4}\nonumber\\
 &~&+16K^2 \Delta_t \delta_{\alpha\beta} \sum_{k} \sum_{\ell}
 q_k^{\perp} q_{\ell}^{\perp}
 {p_z^2 p_{\perp}^6 p_i^{\perp} p_j^{\perp}
 p_k^{\perp} p_{\ell}^{\perp}
 \over (BP_z^2+Kp_{\perp}^4)^5}
 \label{piece11}
\end{eqnarray}
from (\ref{piece1});
\begin{eqnarray}
 I_2(\vec{q}_{\perp})
 = -4K \Delta_t q_j^{\perp} \delta_{\alpha\beta}
 \sum_k q_k^{\perp}
 {p_z^2 p_{\perp}^4 p_i^{\perp} p_k^{\perp}
 \over (BP_z^2+Kp_{\perp}^4)^4}
 \label{piece21}
\end{eqnarray}
from (\ref{piece2});
\begin{eqnarray}
 I_3(\vec{q}_{\perp})
 &\equiv& \Delta_t q_{\perp}^2 \delta_{\alpha\beta}
 {p_z^2 p_i^{\perp} p_j^{\perp}
 \over (BP_z^2+Kp_{\perp}^4)^3}
 -16 K \Delta_t \delta_{\alpha\beta}
 \sum_{k} \sum_{\ell} q_k^{\perp}q_{\ell}^{\perp}
 {p_z^2 p_{\perp}^2 p_i^{\perp} p_j^{\perp}
 p_k^{\perp} p_{\ell}^{\perp}
 \over (BP_z^2+Kp_{\perp}^4)^4}\nonumber\\
 &~&-4K \Delta_t q_{\perp}^2 \delta_{\alpha\beta}
 {p_z^2 p_{\perp}^4 p_i^{\perp} p_j^{\perp}
 \over (BP_z^2+Kp_{\perp}^4)^4}
 -8 K \Delta_t \delta_{\alpha\beta}
 \sum_{k} \sum_{\ell} q_k^{\perp}q_{\ell}^{\perp}
 {p_z^2 p_{\perp}^2 p_i^{\perp} p_j^{\perp}
 p_k^{\perp} p_{\ell}^{\perp}
 \over (BP_z^2+Kp_{\perp}^4)^4}\nonumber\\
 &~&+48 K^2 \Delta_t \delta_{\alpha\beta}
 \sum_{k} \sum_{\ell} q_k^{\perp}q_{\ell}^{\perp}
 {p_z^2 p_{\perp}^6 p_i^{\perp} p_j^{\perp}
 p_k^{\perp} p_{\ell}^{\perp}
 \over (BP_z^2+Kp_{\perp}^4)^5}
 \label{piece31}
\end{eqnarray}
from (\ref{piece3});
\begin{eqnarray}
 I_4(\vec{q}_{\perp})
 = 2\Delta_t q_j^{\perp} \delta_{\alpha\beta}
 \sum_k q_k^{\perp}{p_z^2 p_i^{\perp} p_k^{\perp}
 \over (BP_z^2+Kp_{\perp}^4)^3}
 -8K \Delta_t q_j^{\perp} \delta_{\alpha\beta}
 \sum_k q_k^{\perp}{p_z^2 p_{\perp}^4 p_i^{\perp} p_k^{\perp}
 \over (BP_z^2+Kp_{\perp}^4)^4}
 \label{piece41}
\end{eqnarray}
\end{widetext}
from (\ref{piece4}). Finally the total correction to the bend term
can be written as
\begin{eqnarray}
 -{g^2\over 2T}
 \sum^n_{\alpha=1}\sum^n_{\beta=1}\sum_{\vec{q}}\sum_{i j k}
 q^{\perp}_i q^{\perp}_j
 u_{\alpha}(\vec{q})u_{\beta}(-\vec{q})
 \times \int^>_{p} I_k(\vec{q}_{\perp})\, ,\nonumber\\
 \label{BendCorrection}
\end{eqnarray}
where the subscript $k=1, 2, 3, 4$.

The main trick involving doing the integrals in
(\ref{BendCorrection}) can be summarized by the following two
integrals:
\begin{eqnarray}
 \int_{p'} p'_i p'_j \Gamma(\vec{p}\,')
 &=&{1\over d'}\delta_{ij} \int_p' (p')^2 \Gamma(\vec{p}\,')
 \label{Identity2}
 \\
 \int_{p'} p'_i p'_j p'_k p'_{\ell}\Gamma(\vec{p}\,')
 &=&{1\over d'(d'+1)}(\delta_{ij}\delta_{k\ell}+\delta_{ik}\delta_{j\ell}
 +\delta_{i\ell}\delta_{jk})\nonumber\\
 &~&\times\int_p' (p')^4 \Gamma(\vec{p}\,')\, ,
 \label{Identity2}
\end{eqnarray}
where both function $\Gamma(\vec{p}\,')$ and the integral region
are spherical symmetric, $d'$ is the dimension of the space in
which $\vec{p}\,'$ lies in.

Now we plug in the expression of $I_k$ into (\ref{BendCorrection})
and do the integrals. Let us start with $I_1$, which has three terms.
Integrating the first term gives
\begin{eqnarray}
 &~&-2K \Delta_t q_{\perp}^2 \delta_{\alpha\beta}
 \int_p{p_z^2 p_{\perp}^4 p_i^{\perp} p_j^{\perp}
 \over (BP_z^2+Kp_{\perp}^4)^4}
 \nonumber\\
 &=&-{2K \Delta_t \over d-1}q_{\perp}^2
 \delta_{\alpha\beta} \delta_{ij}
 \int_p^< {p_z^2 p_{\perp}^6 \over (BP_z^2+Kp_{\perp}^4)^4}
 \nonumber\\
 &=&-{2K \Delta_t \over d-1}q_{\perp}^2
 \delta_{\alpha\beta} \delta_{ij}
 \int_{-\infty}^{\infty} dp_z \int_{\Lambda e^{-d\ell}}^{\Lambda}
 dp_{\perp}
 {p_z^2 p_{\perp}^6 \over (BP_z^2+Kp_{\perp}^4)^4}
 \nonumber\\
 &=&-{1\over 64}\Delta_t\left(1\over B K\right)^{3\over 2} q_{\perp}^2
 \delta_{\alpha\beta} \delta_{ij} C_{d-1} \Lambda^{5-d} d\ell\, ;
\end{eqnarray}
integrating the second term gives
\begin{eqnarray}
 &~&-4K \Delta_t \delta_{\alpha\beta} \sum_{k} \sum_{\ell}
 q_k^{\perp} q_{\ell}^{\perp}
 \int_p {p_z^2 p_{\perp}^2 p_i^{\perp} p_j^{\perp}
 p_k^{\perp} p_{\ell}^{\perp}
 \over (BP_z^2+Kp_{\perp}^4)^4}
 \nonumber\\
 &=& -{4K \Delta_t \over d^2-1}
 \delta_{\alpha\beta}
 \sum_{k} \sum_{\ell}
 (\delta_{i j}\delta_{k \ell}+\delta_{i k}\delta_{j \ell}
 +\delta_{i \ell}\delta_{j k})
 q_k^{\perp} q_{\ell}^{\perp}
 \nonumber\\
 &~&\times\int_p^< {p_z^2 p_{\perp}^6
 \over (BP_z^2+Kp_{\perp}^4)^4}
 \nonumber\\
 &=& -{1\over 192}\Delta_t\left(1\over B K\right)^{3\over 2}
 \delta_{\alpha\beta}
 \nonumber\\
 &~&\times\sum_{k} \sum_{\ell}
 (\delta_{i j}\delta_{k \ell}+\delta_{i k}\delta_{j \ell}
 +\delta_{i \ell}\delta_{j k})
 q_k^{\perp} q_{\ell}^{\perp}
 C_{d-1} \Lambda^{5-d} d\ell
 \nonumber\\
 &=& -{1\over 192}\Delta_t\left(1\over B K\right)^{3\over 2}
 \delta_{\alpha\beta}
 (\delta_{i j}q_{\perp}^2 + 2q_i^{\perp}q_j^{\perp})
 C_{d-1} \Lambda^{5-d} d\ell\, ;\nonumber
\end{eqnarray}
integrating the third term gives
\begin{eqnarray}
 &~&16K^2 \Delta_t \delta_{\alpha\beta} \sum_{k} \sum_{\ell}
 q_k^{\perp} q_{\ell}^{\perp}
 \int_p {p_z^2 p_{\perp}^6 p_i^{\perp} p_j^{\perp}
 p_k^{\perp} p_{\ell}^{\perp}
 \over (BP_z^2+Kp_{\perp}^4)^5}
 \nonumber\\
 &=& {16K^2 \Delta_t \over d^2-1}
 \delta_{\alpha\beta}
 \sum_{k} \sum_{\ell}
 (\delta_{i j}\delta_{k \ell}+\delta_{i k}\delta_{j \ell}
 +\delta_{i \ell}\delta_{j k})
 q_k^{\perp} q_{\ell}^{\perp}
 \nonumber\\
 &~&\times\int_p^< {p_z^2 p_{\perp}^10
 \over (BP_z^2+Kp_{\perp}^4)^4}
 \nonumber\\
 &=& {5\over 384}\Delta_t\left(1\over B K\right)^{3\over 2}
 \delta_{\alpha\beta}
 \nonumber\\
 &~&\times\sum_{k} \sum_{\ell}
 (\delta_{i j}\delta_{k \ell}+\delta_{i k}\delta_{j \ell}
 +\delta_{i \ell}\delta_{j k})
 q_k^{\perp} q_{\ell}^{\perp}
 C_{d-1} \Lambda^{5-d} d\ell
 \nonumber\\
 &=& {5\over 384}\Delta_t\left(1\over B K\right)^{3\over 2}
 \delta_{\alpha\beta}
 (\delta_{i j}q_{\perp}^2 + 2q_i^{\perp}q_j^{\perp})
 C_{d-1} \Lambda^{5-d} d\ell\, .\nonumber
\end{eqnarray}
Adding the three results together we get
\begin{eqnarray}
 \int_p^< I_1(\vec{q}_{\perp})
 &=&\delta_{\alpha \beta}\Delta_t\left(1\over B K\right)^{3\over 2}
 C_{d-1} \Lambda^{5-d} d\ell
 \nonumber\\
 &~& \times \left[-{1\over 64}\delta_{i j}q_{\perp}^2-{1\over 192}
 (\delta_{i j}q_{\perp}^2 + 2q_i^{\perp}q_j^{\perp})\right.
 \nonumber\\
 &~&\left.+{5\over384}(\delta_{i j}q_{\perp}^2 +2q_i^{\perp}q_j^{\perp})\right]\, .
\end{eqnarray}
Plugging this result into (\ref{BendCorrection}) we find
corrections to the bend term:
\begin{eqnarray}
 &~&-{g^2\over 2}
 \sum^n_{\alpha=1}\sum^n_{\beta=1}\sum_{\vec{q}}
 \sum_{i j} q_i^{\perp} q_j^{\perp}
 u_{\alpha}(\vec{q})u_{\beta}(-\vec{q})\delta_{\alpha \beta}\Delta_t
 \nonumber\\
 &~&\times\left(1\over B K\right)^{3\over 2}
 C_{d-1} \Lambda^{5-d} d\ell
 \left[-{1\over 64}\delta_{i j}q_{\perp}^2-\right.
 \nonumber\\
 &~&\left.{1\over 192}
 (\delta_{i j}q_{\perp}^2 + 2q_i^{\perp}q_j^{\perp})+
 {5\over384}(\delta_{i j}q_{\perp}^2 + 2q_i^{\perp}q_j^{\perp})\right]
 \nonumber\\
 &=&-{g^2\over 256}
 \Delta_t\left(1\over B K\right)^{3\over 2}
 C_{d-1} \Lambda^{5-d} d\ell
 \nonumber\\
 &~&\times\sum_{\alpha} \sum_{\vec{q}} q_{\perp}^4
 u_{\alpha}(\vec{q})u_{\alpha}(-\vec{q})\, ,
\end{eqnarray}
from which the corresponding correction to the bend modulus $K$ is
\begin{eqnarray}
 (\delta K)_1
 &=&-{g^2\over 128}\Delta_t\left(1\over B K\right)^{3\over 2}
 C_{d-1}\Lambda^{5-d} d\ell
 \nonumber\\
 &=&-{1\over 128}K g_3 d\ell\, .
\end{eqnarray}
Similarly, inserting $I_2(\vec{q}_{\perp})$,
$I_3(\vec{q}_{\perp})$ and $I_4(\vec{q}_{\perp})$ into
(\ref{BendCorrection}), we obtain three more corrections to $K$:
\begin{eqnarray}
 (\delta K)_2&=&-{1\over 32}K g_3 d\ell\, ,
 \nonumber\\
 (\delta K)_3&=&{1\over 128}K g_3 d\ell\, ,
 \nonumber\\
 (\delta K)_4&=&-{1\over 32}K g_3 d\ell\, .
\end{eqnarray}
Now we have found all the correction to the bend modulus $K$ from
(\ref{noname1}).

As for (\ref{noname2}), since it is part of (\ref{noname1}), its
correction to $K$ can be easily shown to be $(\delta K)_1+ (\delta
K)_3$. Thus the total correction to the bend modulus $K$ is given by
\begin{eqnarray}
 \delta K
 &=& (\delta K)_1 \times 2 + (\delta K)_2
 +(\delta K)_3 \times 2 + (\delta K)_4
 \nonumber\\
 &=& -{1\over 32}K \left(B\over g\right)^2\Delta_t
 \left(B\over K^5\right)^5 C_{d-1} \Lambda^{5-d} d\ell
 \nonumber\\
 &=& -{1\over 32}K g_3 d\ell\, .
\end{eqnarray}

Now let us turn to the calculation of the graphical corrections to
$D$. The two Feynman diagrams which give the corrections to $D$ are
illustrated in Fig. \ref{fig: CorrectionD} and \ref{fig:
CorrectionD1}, respectively. The former has been discussed previously
and leads to $(\delta D)_1$ given in (\ref{deltaD1}).
A simple analysis of the latter gives
\begin{eqnarray}
(\delta D)_2 &\approx&-{3\over 4}\omega \Delta_t\int^>_{p}
p_{\perp}^4 G(\vec{p})^2\nonumber\\
&\approx&-{3\over 4}\omega\Delta_t\int^>_{p}{p_{\perp}^4\over
(Bp_z^2+Kp_{\perp}^4+Dp_{\perp}^2)^2}\nonumber\\
&\approx&({3\over 16}K\Lambda^2-{9\over 32}D)g_4d\ell\, .
\end{eqnarray}
Naively one would treat the sum of $(\delta D)_1$ and $(\delta D)_2$
as the total correction to $D$; however, a more careful analysis
shows that this treatment is {\it incorrect}. One has to take into
account that the disorder averaged ground state of the rescaled
system has been renormalized under the RG. The Feynman diagram in
Fig. \ref{fig: CorrectionLin} shows a contribution
\begin{eqnarray}
\delta L\sum_{\alpha=1}^n\partial_z u_{\alpha}\, ,
\end{eqnarray}
which is linear and does not exist in the original Hamiltonian. This
implies that the disorder averaged minimum of the Hamiltonian has
shifted from $\partial_z u_{\alpha}=0$ to $\partial_z
u_{\alpha}=-\delta L/B$. Therefore, after each cycle of the RG, the
Hamiltonian needs to be reexpanded around the new minimum, which is
accomplished by changing variable, defining $u'_{\alpha}\equiv
u_{\alpha}+(\delta L/B)z$. This treatment then makes the cubic term
$-g(\partial_z u) (\nabla_{\perp}u)^2/2$ to create another
contribution to the tilt term $(\delta D)_3|\nabla_{\perp}u'|^2/2$,
where $(\delta D)_3=g\delta L/B$. $\delta L$ can be easily
calculated from the Feynman diagram in Fig. \ref{fig: CorrectionLin}
\begin{eqnarray}
\delta L &\approx&-{1\over 2}g \Delta_t\int^>_{p}
p_{\perp}^4 G(\vec{p})^2\nonumber\\
&\approx&-{1\over 2}g \Delta_t\int^>_{p}{p_{\perp}^4\over
(Bp_z^2+Kp_{\perp}^4+Dp_{\perp}^2)^2}\nonumber\\
&\approx&(-{1\over 8}K\Lambda^2+{3\over 16}D)\left(B\over
g\right)g_3d\ell\, .
\end{eqnarray}
Taking $(\delta D)_3$ into account, the total correction to
$D$ should really be
\begin{eqnarray}
\delta D&=&(\delta D)_1+(\delta D)_2+(\delta D)_3\nonumber\\
        &=&{9\over 32}(g_3-g_4)+{3\over 16}K(g_4-g_3)\, .
        \label{delta D}
\end{eqnarray}
In the isotropic disordered smectic $A$ problem\cite{LC} the
rotation invariance prohibits the tilt term and this symmetry should be
preserved under the RG, which
provides an extra check of our calculation. Imposing $g_3=g_4$
(restoring the rotation invariance) in (\ref{delta D}), we
find indeed that $\delta D$ vanishes.

Now we calculate graphical corrections to anharmonic terms in the
Hamiltonian (\ref{H_AC}). This calculation has been avoided in
the isotropic disordered smectic $A$ problem\cite{LC} by using the
symmetry argument that the global rotation invariance requires
\begin{eqnarray}
 \delta B=\delta g=\delta w\, .
 \label{symmetry}
\end{eqnarray}
This argument does not apply in our problem due to the symmetry
breaking; however, it can be used to check our calculation. By
imposing $B=g=w$, our results should also satisfy
(\ref{symmetry}).

All the one-loop Feynman diagrams which give the corrections to the
cubic and quartic terms are shown through Fig. \ref{fig:
CorrectionW1} to \ref{fig: CorrectionW6}. The corresponding
graphical corrections to $g$ and $\omega$ are in order referred as
$(\delta g)_1$, $(\delta g)_2$, $(\delta g)_3$, $(\delta \omega)_1$,
$(\delta \omega)_2$, $(\delta \omega)_3$, $(\delta \omega)_4$,
$(\delta \omega)_5$, $(\delta \omega)_6$, respectively. Since the
evaluations of these diagrams are all similar, we choose to only
show in detail the evaluation of the Feynman diagram in Fig.
\ref{fig: CorrectionW6}, which is the most complicated, and for the
rest we only give results. An analysis of the Feynman diagram in
Fig. \ref{fig: CorrectionW6} gives
\begin{widetext}
\begin{eqnarray}
 &-&{2g^4\over T^3}{1\over V}
 \sum_{\alpha=1}^{\infty}\sum_{\beta=1}^{\infty}
 \sum_{\gamma=1}^{\infty}\sum_{\eta=1}^{\infty}
 \sum_{\vec{q}_1,\vec{q}_2,\vec{q}_3,\vec{q}_4}
 \sum_{i, j, k, l}
 \delta(\vec{q}_1+\vec{q}_2+ \vec{q}_3+\vec{q}_3)
 (iq_{1i}^{\perp})(iq_{2j}^{\perp})
 (iq_{3k}^{\perp})(iq_{4l}^{\perp})\nonumber\\
 &\times&
 u_{\alpha}(\vec{q}_1)u_{\beta}(\vec{q}_2)
 u_{\gamma}(\vec{q}_3)u_{\eta}(\vec{q}_4)\int_p^>
 p_i^{\perp} p_j^{\perp} p_k^{\perp} p_l^{\perp} p_z^4
 G_{\alpha\beta}G_{\beta\gamma}
 G_{\gamma\eta}G_{\alpha\eta}\, ,
 \label{correctionquartic6}
\end{eqnarray}
\end{widetext}
in which
\begin{eqnarray}
 &&G_{\alpha\beta}G_{\beta\gamma}G_{\gamma\eta}
   G_{\alpha\eta} \nonumber\\
 &&=(T G(\vec{q})\delta_{\alpha\beta}+\Delta_t
  q_{\perp}^2G(\vec{q})^2)
  (T G(\vec{q})\delta_{\beta\gamma}+\Delta_t
  q_{\perp}^2G(\vec{q})^2)\nonumber\\
  &&~~~(T G(\vec{q})\delta_{\gamma\eta}+\Delta_t
  q_{\perp}^2G(\vec{q})^2)
  (T G(\vec{q})\delta_{\alpha\eta}+\Delta_t
  q_{\perp}^2G(\vec{q})^2).\nonumber\\
  \label{propagatorquartic}
\end{eqnarray}
Note that the quartic term in the Hamiltonian (\ref{H_AC}) only
involves one sum over the replica index, while
(\ref{correctionquartic6}) has four. Thus only those terms in
$G_{\alpha\beta}G_{\beta\gamma}G_{\gamma\eta} G_{\alpha\eta}$ which
carry at least three Kronecker delta functions can lead to the
contributions to the quartic term. Plugging
(\ref{propagatorquartic}) into (\ref{correctionquartic6}) and
keeping the most relevant pieces, we obtain
\begin{eqnarray}
 (\delta \omega)_6
 &=&-{8g^4\Delta_t}
 \int_p^>p_{\perp}^6 p_z^4 G(\vec{q})^5\nonumber\\
 &=&-{3\over 32}{g_3^2\over g_4}\omega d\ell\, .
\end{eqnarray}
The corrections to $g$ and $w$ from other Feynman diagrams can be
calculated in a similar way and are
\begin{eqnarray}
 (\delta g)_1&=&{3\over 32} g_3 g d\ell\, ,~~~~
 (\delta g)_2=-{3\over 16} g_4 g d\ell\, ,\nonumber\\
 (\delta g)_3&=&-{3\over 32} g_4 g d\ell\, ,~~
 (\delta \omega)_1=-{3\over 16} g_4 \omega d\ell\, ,\nonumber\\
 (\delta \omega)_2&=&-{3\over 16} g_4 \omega d\ell\, , ~~
 (\delta \omega)_3=-{3\over 32} g_4 \omega d\ell\, ,\nonumber\\
 (\delta \omega)_4&=&{3\over 16} g_3 \omega d\ell\, ,~~~~
 (\delta \omega)_5={3\over 16} g_3 \omega d\ell\, .\nonumber
\end{eqnarray}
Adding them together we get
\begin{eqnarray}
 \delta g&=&({3\over 32} g_3-{9\over 32} g_4)d\ell\, ,\\
 \delta \omega&=&-{3\over 32}{g_3^2\over g_4}\omega
               -{15\over 32}g_4\omega
               +{3\over 8} g_3\omega\, .
\end{eqnarray}

Performing the rescalings described earlier and using the
graphical corrections we have calculated, we obtain the following
RG flow equations:
\begin{eqnarray}
{dB(\ell)  \over  d \ell} = \left(d - 1 - \omega +2\chi - {3 \over
16} g_3\right)B \label{10}\, ,
\end{eqnarray}
\begin{eqnarray}
{dK(\ell)  \over  d \ell} = \left(d - 5 + \omega + 2\chi + {1
\over 32} g_3\right)K \label{11}\, ,
\end{eqnarray}
\begin{eqnarray}
{d\left(\Delta_t/T \right)(\ell)  \over  d \ell} = \left(d - 3 +
\omega + 2\chi + {1 \over 64} g_3 \right) \label{12}\, ,
\end{eqnarray}
\begin{eqnarray}
{dg(\ell)  \over  d \ell} = \left(d - 3 + 3\chi + {3 \over
32}g_3-{9\over 32}g_4\right)g\label{15}\, ,
\end{eqnarray}
\begin{eqnarray}
{dD(\ell)  \over  d \ell} &=& \left(d - 3 + \omega + 2\chi + {9
\over 32}g_3 - {9 \over 32}g_4 \right) D \nonumber\\&+&{3 \over
16}K \left(g_4 - g_3\right) \label{13}\, ,
\end{eqnarray}
\begin{eqnarray}
{dw(\ell) \over  d \ell}&=& \left(d - 5 + \omega +4\chi -{3\over
32}{g_3^2\over g_4}\right)w\nonumber\\&+&\left({3\over
8}g_3-{15\over 32}g_4\right)w\label{14}\, .
\end{eqnarray}
These equations are not very transparent telling the physics since
they depend on the rescaling factor $\omega$ and $\chi$, which are
artificial. Instead, we should focus on the flow of the
dimensionless couplings. Combining the above flow equations with the
definition of $g_3$ and $g_4$,  we find
\begin{eqnarray}
{dg_3\over d\ell}&=&(\epsilon-{9\over 16}g_4+{13\over 32}g_3)g_3
\label{g3}\, ,\\
{dg_4\over d\ell}&=&(\epsilon-{15\over 32}g_4+{13\over
32}g_3-{3\over 32}{g_3^2\over g_4})g_4\, ,\label{g4}
\end{eqnarray}
where $\epsilon=5-d$.

Obviously the Gaussian fixed point ($g_3^*=0$, $g_4^*=0$) is one
of the fixed points of Eqs. (\ref{g3}) and (\ref{g4}). To check
it's stability, we linearize the two equations around it by
writing $g_3=0+\delta g_3$, $g_4=0+\delta g_4$, and get, to linear
order in $\delta g_{3,4}$,
\begin{eqnarray}
{d\delta g_3\over d\ell}&=&\epsilon \delta g_3\, ,\label{g3flow}\\
{d\delta g_4\over d\ell}&=&\epsilon \delta g_4\, .
\end{eqnarray}
These two linearized differential equations tell us that for
$\epsilon>0$ ($d<5$), given a small departure from the Gaussian
fixed point, both $g_3$ and $g_4$ will leave it exponentially, which
implies that the Gaussian fixed point is unstable in the physical
dimension $d=3$.


In addition to the Gaussian fixed point, these RG flow equations
(\ref{g3}, \ref{g4}) have other two fixed points. One of them
%
%
%
%
%
%
%
%
%
is exactly the one found in the isotropic disordered smectic $A$
problem \cite{LC}, for which the Hamiltonian is rotation
invariant. This is not a coincidence, because the Hamiltonian we
start with is more general in the sense that it has no symmetry
bound $(g/B)^2={w/B}$, so our solutions must include this specific
fixed point as a special case. To find this particular fixed point
, we can enforce this symmetry condition on the RG flow equations
(\ref{g3}, \ref{g4}) by setting $g_3=g_4$. After this enforcement
both of those two equations reduce to the recursion relation found
by Radzihovsky and Toner \cite{LC} for the isotropic problem:
\begin{eqnarray}
{dg_3\over d\ell}=(\epsilon-{5\over 32}g_3)g_3, \nonumber
\end{eqnarray}
which leads to $g_3^*=32\epsilon/5$. However, this fixed point is
not stable in our more general problem. We can examine this
instability by using a similar linearization analysis as for the
Gaussian fixed point. Writing $g_3=32\epsilon/5+\delta g_3$,
$g_4=32\epsilon/5+\delta g_4$ and plugging them into equations
(\ref{g3}) and (\ref{g4}), we obtain
\begin{eqnarray}
{d\delta g_3\over d\ell}&=&{13\epsilon\over 5}\delta
g_3-{18\epsilon\over 5}\delta
g_4\, ,\\
{d\delta g_4\over d\ell}&=&{7\epsilon\over 5}\delta
g_3-{12\epsilon\over 5}\delta g_4\, ,
\end{eqnarray}
which has the solution near the vicinity of the fixed point:
\begin{eqnarray}
\delta g_3&=&C_1e^{-\epsilon\ell}+18C_2e^{6\epsilon\ell/5}\, ,
\nonumber\\
\delta g_4&=&C_1e^{-\epsilon\ell}+7C_2e^{6\epsilon\ell/5}\, .
\nonumber
\end{eqnarray}
The $C_1$ and $C_2$ in the solution are constants determined by the
initial condition $(\delta g_3)_0$ and $(\delta g_4)_0$. This
solution shows that in general, this fixed point is also unstable,
although it is stable given that $(\delta g_3)_0=(\delta g_4)_0$,
which is just a verification of its stability in the isotropic
disordered smectic $A$ problem \cite{LC}.
%
%
%
%
%
%

The third fixed point can be obtained by assuming that $g_3$ flows
to zero. Under this assumption equation (\ref{g4}) reduces to
\begin{eqnarray}
{dg_4\over d\ell}=(\epsilon-{15\over32}g_4)g_4\, ,
\end{eqnarray}
which has a non-Gaussian fixed point $g_4^*=32\epsilon/15$. The linearized
differential equations around this fixed point are
\begin{eqnarray}
{d\delta g_3\over d\ell}&=&-{1\over 5}\epsilon\delta g_3\, ,\label{deltag3}\\
{d\delta g_4\over d\ell}&=&{13\over 15}\epsilon\delta
g_3-\epsilon\delta g_4\, ,\label{deltag4}
\end{eqnarray}
which have the solution:
\begin{eqnarray}
\delta g_3&=&c_1e^{-\epsilon\ell/5}\label{g31}\, ,\\
\delta g_4&=&{13\over
12}c_1e^{-\epsilon*\ell/5}+c_2e^{-\epsilon\ell}\, .\label{g41}
\end{eqnarray}
Note that both eigenvalues are negative for $d<5$, which implies
that this fixed point is stable and hence controls the second
order phase transition.
%
%
%
%
%
\begin{figure}[h]
\includegraphics[width=0.35\textwidth, angle=-90]{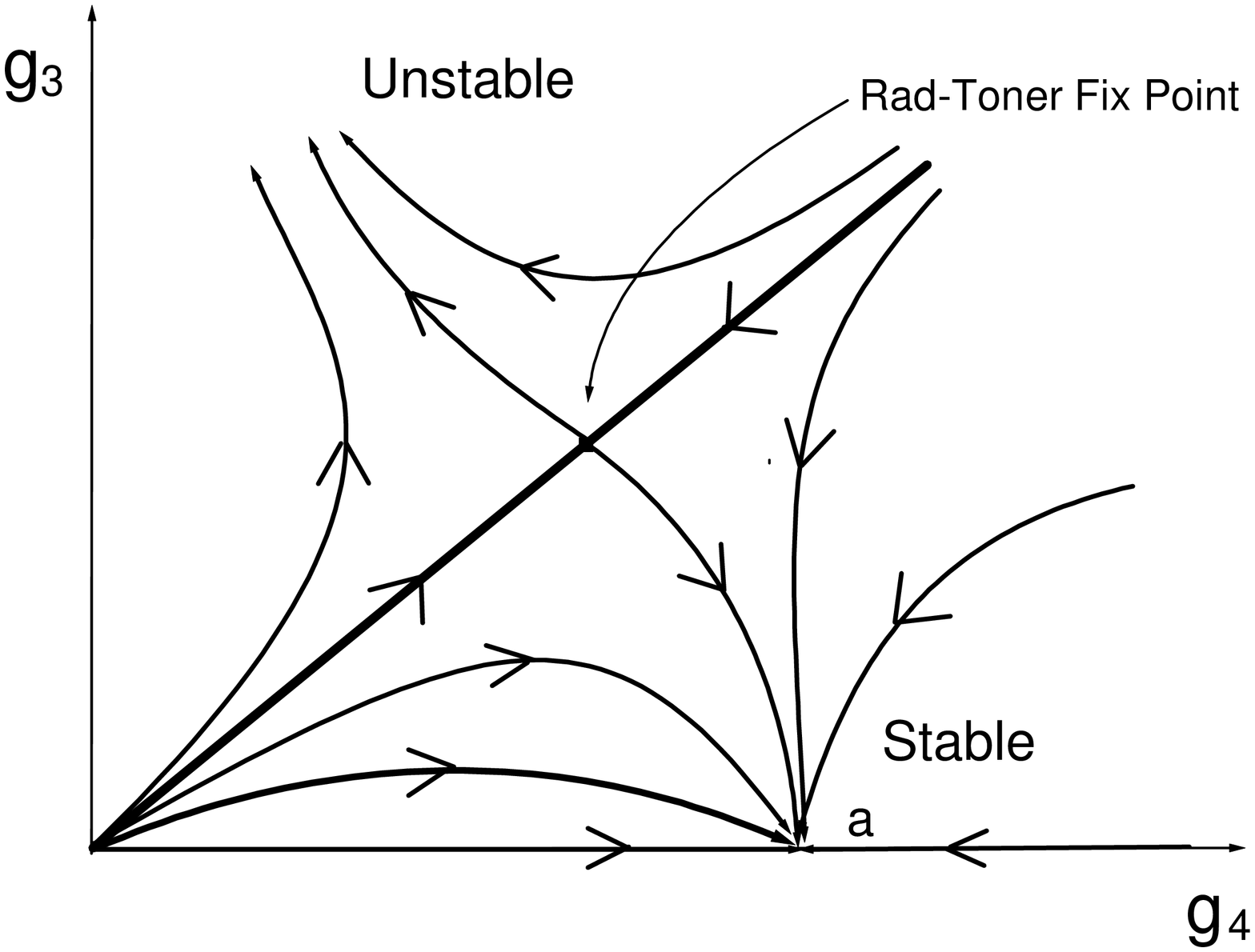}
\caption{\label{fig: RGflow}Graphical RG flow of the dimensionless
coupling $g_3$ and $g_4$. Point $a$ is the stable fixed point (0,
$32\epsilon/15$)}
\end{figure}

\section{\label{sec: CriticalExponents}Critical Exponents}
The RG we derived in the previous section is only valid in the
critical region, and hence named ``critical RG''. If the system is
right at the critical point, the critical RG holds for arbitrary
$\ell$, otherwise it breaks down at $\ell^*$ which is determined by
$D(\ell^*)=K(\ell^*)\Lambda^2$. In this section we first identify
the critical point, then study how fast the critical RG breaks down
if the system is slightly off the critical point, so as to calculate
the critical exponents. For convenience, we choose a special
rescaling
%
%
%
%
%
\begin{eqnarray}
\omega+2\chi=5-d-{1\over 32}g_3,\label{relation1}
\end{eqnarray}
which fixes $K$ at its initial value $K_0$.  The flow Eq. (\ref{13})
then becomes
\begin{eqnarray}
 {dD(\ell)  \over  d \ell}&=& \left(2 + {1\over 4}g_3- {9 \over 32}g_4\right) D
 \nonumber\\
 &~&+{3\over 16}K_0 (g_4-g_3)\, .
\label{Dflow}
\end{eqnarray}
Right at the critical point $D(\ell)$ also flows to a fixed value
$D^*$, which can be calculated by plugging $g_3^*=0$,
$g_4^*=32\epsilon/15$ into the Eq. (\ref{Dflow}) and requiring
$dD/d\ell=0$. Thus the fixed point controlling the phase transition
is identified in $D$-$g_3$-$g_4$-space as
\begin{eqnarray}
 g_3^*&=&0\, ,  ~~~~~~~g_4=32\epsilon/15\, ,\\\
 D^*&=&-\epsilon/5\, .
\end{eqnarray}
Linearizing the flow Eqs. (\ref{g3}), (\ref{g4}) and (\ref{Dflow}) around
this fixed point, we obtain (\ref{deltag3}), (\ref{deltag4})
and
%
%
%
%
%
%
%
%
%
%
%
%
%
%
%
%
%
%
%
\begin{eqnarray}
 {d\delta D(\ell) \over d \ell}&=&\lambda_D\delta D+\left({1\over 4}D^*-{3\over
 16}K_0\right)\delta g_3
 \nonumber \\
 &~& +\left({3\over 16}K_0-{9\over 32}D^*\right)\delta g_4
\label{21}\, ,
\end{eqnarray}
%
%
%
%
%
%
%
where
\begin{eqnarray}
 \lambda_D
          =2-{3 \epsilon \over5}+O(\epsilon^2).
\end{eqnarray}
The solutions of these flow equations are given by (\ref{g31}),
(\ref{g41}) and
\begin{eqnarray}
\delta D(\ell)=c_3e^{\lambda_D
\ell}-{K_0c_1\over128}e^{-\epsilon\ell/5} -{3K_0c_2\over
32}e^{-\epsilon\ell} \label{Linearsolution}\, ,
\end{eqnarray}
where $c_{1,2,3}$ are variables determined by the initial condition
via
\begin{eqnarray}
 c_1 &=& g_3^0-g_3^*, \label{g3initial}\\
 c_2 &=& g_4^0-g_4^* - {13\over 12}(g_3^0-g_3^*),\label{g4initial}\\
 c_3&=&D_0-D^*+{K_0c_1\over128}+{3K_0c_2\over32} \, ,
 \label{Dinitial}
\end{eqnarray}
where $D_0$, $g_3^0$ and $g_4^0$ are the ``complete bare'' values of
$D$, $g_3$ and $g_4$, respectively. According to the solution given
by (\ref{Linearsolution}), the fixed point is unstable with respect
to $D$ unless $c_3=0$, which defines exactly the critical point.
After some algebra, the critical point is identified as
\begin{eqnarray}
 D_0(T_{AC}) = {3K_0\over 32}g_3^0 - {3K_0\over  32}g_4^0,
\end{eqnarray}
which actually describes a plane in $D$-$g_3$-$g_4$ space. This
shows that the true critical point has dependence on $g_3^0$
and $g_4^0$, and is {\it not} exactly at $D_0(T)=0$, which is predicted by
the mean-field theory. $c_3$ measures the deviation from the critical point and
is often assumed to be
\begin{eqnarray}
 c_3\propto T-T_{AC}\, .
\end{eqnarray}
For initial values $D_0$, $g_3^0$ and $g_4^0$ off from the critical
plane, $D(\ell)$ departures exponentially under the RG and for large
$\ell$, behaves as
\begin{eqnarray}
 D(\ell)\propto c_3e^{\lambda_D\ell}
        \propto (T-T_{AC})e^{\lambda_D\ell}\, .
 \label{Dflow1}
\end{eqnarray}
The system is in the $A$ phase for $T>T_{AC}$ (i.e., $c_3>0$) and in
the $C$ phase for $T<T_{AC}$ (i.e., $c_3<0$). The universality
classes of both phases will be discussed in section \ref{sec:
ACPhases}.


Now we calculate the critical exponents $\nu_{\perp}$, $\nu_z$.
These two exponents are defined via the two correlation lengths
$\xi_{\perp}$ and $\xi_z$ by equations
\begin{eqnarray}
\xi_{\perp}&\propto& |T-T_{AC}|^{-\nu_{\perp}}\, ,\\
\xi_z&\propto&|T-T_{AC}|^{-\nu_z} \label{}\, .
\end{eqnarray}
These two correlation lengths are the characteristic lengths beyond
which the critical behavior crosses over to either the high
temperature (smectic $A$) or low temperature (smectic $C$) behavior.
More specifically, the correlation length $\xi_{\perp}$ satisfies
$K(\xi_{\perp})^{-2}=D$, which, under the RG, is mapped to
\begin{eqnarray}
K(\ell^*)\Lambda^2=D(\ell^*) \label{crossrelation}
\end{eqnarray}
with
\begin{eqnarray}
 \xi_{\perp} = e^{\ell^*}\Lambda^{-1}\, ,\label{trajectory}
\end{eqnarray}
where due to the special rescaling (\ref{relation1}), $K(\ell^*)$ is
fixed at $K_0$ and $D(\ell)$ is given by (\ref{Dflow1}). Thus we
obtain
\begin{eqnarray}
  \ell^*\propto-{1\over\lambda_D}\ln{|T-T_{AC}|}\, ,
\end{eqnarray}
which combined with Eq. (\ref{trajectory}), gives
\begin{eqnarray}
 \xi_{\perp}\propto |T-T_{AC}|^{1/\lambda_D}\, .
\end{eqnarray}
Therefore, the RG predicts
\begin{eqnarray}
\nu _{\perp} = {1\over \lambda} = {1 \over 2} + {3\epsilon \over 20
} +0\left(\epsilon^2 \right)\, . \label{}
\end{eqnarray}
The other correlation length $\xi_z$ can be obtained using the
scaling law $\xi_z \propto \xi_\perp^\zeta$; this leads to
\begin{eqnarray}
\nu_z = \zeta \nu_{\perp}=1+{3\epsilon\over10}+0\left(\epsilon^2
\right)\, . \label{}
\end{eqnarray}

In order to calculate the specific heat exponent $\alpha$, we need
to find the temperature-dependence of the free energy. Since the
free energy $F$ is invariant under the RG, we can write it as
\begin{eqnarray}
F &=& V_{r} f_r(K(\ell^*), B(\ell^*), D(\ell^*), \Delta(\ell^*),
g(\ell^*),
\omega(\ell^*))\nonumber\\
&=& V_0 e^{-(d-1)\ell^* - w\ell^*}f_{r},\nonumber
\end{eqnarray}
where $V_{r}$, $f_{r}$ are the renormalized volume and free energy
density, $V_0$ is the physical volume before renormalization. To
factor out the temperature dependence of $f_r$ as much as possible,
we make a special choice of $\ell^*$, $\omega$ and $\chi$:
\begin{eqnarray}
\ell^*&=&\ln(\xi_{\perp}\Lambda)\, ,\\ \omega &=& 2 - {\eta_B +
\eta_K \over 2}\, , \label{I3}\\ \chi &=& {1 \over 4} (6 -2d +\eta_B
- \eta_K) \label{I4}\, .
\end{eqnarray}
This choice fixes $K$ and $B$ so that they are $\ell$-independent
and hence temperature-independent. The choice of $\ell^*$ makes
$D(\ell^*)= K_0\Lambda^2$ and hence also temperature independent,
and for such large $D(\ell^*)$ fluctuations are small, and $f_r$ can
be evaluated by ignoring the anharmonic terms. In spite of all the
advantages just described, $f_r$ is a function of $\Delta(\ell^*)$,
which still has temperature-dependence. Fortunately by taking
$\partial f_r/\partial \Delta(\ell^*)$, which can be easily
calculated, we find $f_r$ is a linear function of $\Delta(\ell^*)$.
Finally the free energy can be nicely written as
\begin{eqnarray}
F&=&V_0 e^{-(d-1)\ell^* - w\ell^*}(\Delta(\ell^*) +
constant)\nonumber\\
&\propto&\left| T - T_c\right| ^{\nu _{\perp} \left(d+1-{\eta_K +
\eta_B \over 2}\right)}\times\nonumber\\
& &\left(\left| T - T_c\right| ^{\nu _{\perp} \left(-2+\eta_K
-\eta_t \right)}+constant\right)
   \label{I5}.
\end{eqnarray}
Now taking $\partial^2F/\partial T^2 $ and keeping the most
divergent part, we get the specific heat exponent
\begin{eqnarray}
\alpha = 2 - \nu_{\perp} \left(d - 1 +  {\eta_K -  \eta_B \over 2} -
\eta_t\right) \label{I7}\, .
\end{eqnarray}

\section{\label{sec: AnomalousElasticity}Anomalous Elasticity Right at the Critical Point}
In this section we show that the existence of the stable fixed point
leads to anomalous elasticity by using trajectory integral matching.
For the sake of simplicity we assume that the system is right at the
critical point, so that we do not need to worry about the crossover
from the critical region to the $A$ or $C$ phase, which will be
taken care of when we discuss correlations functions in section
\ref{sec: Cfunctions}. Let us calculate the disorder averaged
correlation function
$G(\vec{q})\equiv\overline{\left<|u(\vec{q})|^2\right>}-\overline
{\left<u(-\vec{q})\right>\left<u(\vec{q})\right>}$. The RG
establishes a connection between a correlation function at a small
wavevector (which is impossible to calculate in perturbation theory
due to the infrared divergence) and the same correlation function at
a large wavevector, which can be easily calculated in a controlled
perturbation theory. This connection for $G(\vec{q})$ is
\begin{eqnarray}
 &&G(\vec{q}, B, K, \Delta_t, g, w) = e^{(d-1+\omega+2\chi)\ell}\times\nonumber\\
 &&G(q_{\perp}e^{\ell},
 q_ze^{\omega\ell}, B(\ell), K(\ell), \Delta_t(\ell), g(\ell),
 w(\ell))\, ,\nonumber\\ \label{IntegralMatching}
\end{eqnarray}
where the prefactor on the right-hand side comes from the
dimensional and field rescaling. First let us consider the special
case $q_z=0$. The rescaling variable $\ell^*$ is chosen as
\begin{eqnarray}
 q_{\perp}e^{\ell^*}=\Lambda\, .
 \label{SpecialChoice}
\end{eqnarray}
We also choose $q_{\perp}$ sufficiently small such that
$g_3(\ell^*)$ and $g_4(\ell^*)$ has reached the stable fixed point
$g_3^*=0$, $g_4^*=32\epsilon/15$. Eliminating $\ell^*$ in favor of
$q_{\perp}$, we get
\begin{eqnarray}
 &&G(q_{\perp}, q_z=0, B, K, \Delta_t, g, w)=\left(\Lambda\over q_{\perp}\right)^{(d-1+\omega+2\chi)}
 \nonumber\\
 &&\times G(\Lambda, 0, B(\ell^*), K(\ell^*), \Delta_t(\ell^*), g(\ell^*),
 w(\ell^*)).\nonumber
\end{eqnarray}
Evaluating the right-hand side using a perturbation theory, we
obtain
\begin{eqnarray}
 &&G(q_{\perp}, q_z=0, B, K, \Delta_t, g, w)\nonumber\\
 &=&{1\over
 \Lambda^4 K(\ell^*)}\times\left(\Lambda\over
 q_{\perp}\right)^{(d-1+\omega+2\chi)}\nonumber\\
 &\equiv& {1\over K(q_{\perp}, q_z=0)q_{\perp}^4}\, ,
\end{eqnarray}
where
\begin{eqnarray}
 K(\ell^*)=
 K_0(\xi_{NL}^{\perp}\Lambda)^{d-5+\omega+2\chi}
 (\xi_{NL}^{\perp}q_{\perp})^{-(d-5+\omega+2\chi-g_3^*/32)}\nonumber
\end{eqnarray}
which is obtained by integrating the flow Eq. (\ref{11}), and the
anomalous bend modulus which diverges at long length scales is
defined as
\begin{eqnarray}
 K(q_{\perp}, q_z=0)=K_0(q_{\perp}\xi_{NL}^{\perp})^{-\eta_K},
\end{eqnarray}
with an anomalous exponent
\begin{eqnarray}
 \eta_K={1\over 32}g_3^*\, .
\end{eqnarray}
$\xi_{NL}^{\perp}$ is the nonlinear crossover length along $\perp$
directions, beyond which the perturbation theory breaks down.

The above calculations can be generalized for arbitrary direction of
$\vec{q}$ by using a more sophisticated choice of $\ell^*$, which
satisfies
\begin{eqnarray}
 K(\ell^*)\Lambda^2=K(\ell^*)(q_{\perp}e^{\ell^*})^4+B(\ell^*)
 (q_ze^{\omega\ell^*})^2\, .
 \label{GeneralChoice}
\end{eqnarray}
This choice ensures that the right-hand side of
(\ref{IntegralMatching}) can be evaluated in a controlled
perturbation theory and is consistent with (\ref{SpecialChoice}) in
the special case $q_z=0$. Solving (\ref{GeneralChoice}) we obtain
\begin{eqnarray}
 e^{\ell^*}=\left(\Lambda\over q_{\perp}\right)f_{\ell}\left[\left(q_{\perp}\xi_{NL}\right)^{\zeta}
 \over q_z\xi_{NL}^z\right]\, ,
 \label{GeneralSolution}
\end{eqnarray}
where the anisotropy exponent $\zeta$ is given by
\begin{eqnarray}
\zeta=2-{\left(\eta_K+\eta_B\right)\over 2}\, ,
\end{eqnarray}
$\eta_B$ is defined as the anomalous exponent of the compression
modulus $B$
\begin{eqnarray}
 B(q_{\perp}, q_z=0)=B_0(q_{\perp}\xi_{NL}^{\perp})^{\eta_B}\, ,
\end{eqnarray}
and given by
\begin{eqnarray}
\eta_B &=& {3\over 16}g_3^*\, ,
\end{eqnarray}
$f_{\ell}(x)$ is a scaling function
\begin{eqnarray}
 f_{\ell}(x)=\left\{
 \begin{array}{ll}
 1, &x\ll 1\\
 x^{1/\zeta}, &x\gg 1
 \end{array}\right.
 \label{}.
\end{eqnarray}
Plugging (\ref{GeneralSolution}) into (\ref{IntegralMatching}) leads
to the general $\vec{q}$-dependence of both $K$ and $B$, which is
summarized in (\ref{K}, \ref{B}).

The $\vec{q}$-dependence of $\Delta_t$ can not be calculated by
using the correlation function
$G(\vec{q})\equiv\overline{\left<|u(\vec{q})|^2\right>}-\overline
{\left<u(-\vec{q})\right>\left<u(\vec{q})\right>}$ since it is
independent of $\Delta_t$. Instead we perform trajectory integral
matching on the correlation function $\overline{\langle
u(\vec{q})\rangle\langle u(-\vec{q})\rangle}=\Delta_t
q_{\perp}^2G(\vec{q})^2$. After a similar calculation for the
special case $q_z=0$, we find $\Delta_t$ also diverges at small
$q_{\perp}$ as
\begin{eqnarray}
 \Delta_t(q_{\perp}, q_z=0)=\Delta_t^0(q_{\perp}\xi_{NL}^{\perp})^{-\eta_t}\, ,
\end{eqnarray}
where the anomalous exponent $\eta_t$ is given by
\begin{eqnarray}
 \eta_t &=& {1\over 64}g_3^*\, .
\end{eqnarray}
Generalizing the calculation for arbitrary $\vec{q}$, we get the
result summarized in (\ref{Deltacrit}).

Since $g_3^*$ is zero, $\eta_{K, t}$ are zero to $O(\epsilon)$, and
$\eta_B$ is zero {\it exactly}. To calculate $\eta_{K, t}$ to
$O(\epsilon^2)$, we performed a two-loop RG calculation showing that
\begin{eqnarray}
 \eta_K &=& C_K\epsilon^2 + O(\epsilon^3)\, ,\nonumber\\
 \eta_t &=& C_{\Delta}\epsilon^2 + O(\epsilon^3)\, ,
\end{eqnarray}
where $C_K=(32\ln\left(4/3\right)-10)/225$,
$C_{\Delta}=(12\ln\left(4/3\right)-1/3)/225$.

\section{\label{sec: ACPhases}$A$ and $C$ Phases}
In this section we will identify the universality classes of both
the high temperature ($A$) and low temperature ($C$) phases. For now
let us assume that the system is far away from the critical point so
that we do not need to worry about the crossovers between the
critical region and the two phases, which will be discussed in
section \ref{sec: Cfunctions}. First we discuss the $A$ phase. The
model we start with is the Hamiltonian (\ref{H}). Since at long
wavelength the tilt term $D_0 |\vec{\nabla}_{\perp}u|^2/2$ dominates
the bend term $K(\nabla^2_{\perp}u)^2/2$, the latter can be
neglected. The model reduces to
\begin{eqnarray}
H &=& \int d^dr  \left[{B \over  2}(\partial_zu)^2+{D\over 2}
\left|\vec{\nabla}_{\perp}u \right|^2-{g \over 2}(\partial_zu)
|\vec{\nabla}_{\perp}u|^2\right.\nonumber\\&+&\left.{w \over 8}
\left|\vec{\nabla}_{\perp}u \right|^4
+\vec{h}\left(\vec{r}\right)\cdot\vec{\nabla}u +
V_p(u-\phi(\vec{r}))\right] \label{}\, .
\end{eqnarray}
For this model a simple power counting shows that both the
anharmonic terms and the random tilt disorder are irrelevant in
$d>2$. Thus in $d=3$ the effective model for the $A$ phase is simply
\begin{eqnarray}
H &=& \int d^dr  \left[{B \over  2}(\partial_zu)^2+{D\over 2}
\left|\vec{\nabla}_{\perp}u \right|^2+ V_p(u-\phi(\vec{r}))\right]\, ,\nonumber\\
\label{Aphase}
\end{eqnarray}
which is just the random field $XY$ model with anisotropic
stiffness. This model has been studied extensively, and the
correlation $\overline{\langle (u(\vec{r})-u(\vec{0}))^2\rangle}$
has a logarithmic divergence as $r\to\infty$. As a result the
density correlation function for the $A$ phase decays as a power
law, which implies that the $A$ phase only has quasi-long-range
translational order. However, unlike the conventional smectic $A$
phase, in which the power law exponent depends on the temperature,
here the exponent is {\it universal}. These correlation functions
will be calculated in detail in section \ref{sec: Cfunctions}.

Deep in the $C$ phase, $D|\vec{\nabla}_{\perp}u|^2/2$ is negative.
Let us first use a mean field theory to find the new {\it disorder
averaged} ground state of the {\it disordered} Hamiltonian
(\ref{H}). Since the two disordered interactions in the Hamiltonian
are totally random and isotropic, they should not affect the mean
field theory. Therefore we ignore them for the moment and work on
the {\it clean} Hamiltonian
\begin{eqnarray}
  H &=& \int d^dr  \left[{B \over  2}(\partial_zu)^2+{D_0\over 2}
 \left|\vec{\nabla}_{\perp}u \right|^2+{K\over 2}(\nabla^2 u)^2
 \right.\nonumber\\
 &~&\left.-{g \over 2}(\partial_zu)
 |\vec{\nabla}_{\perp}u|^2+{w \over 8}
 \left|\vec{\nabla}_{\perp}u \right|^4\right]\, , \label{clean1}
\end{eqnarray}
which can be rewritten as
\begin{eqnarray}
 H &=& \int d^dr \left[{B\over 2}\left(\partial_z u-
 {g\over 2B}|\vec{\nabla}_{\perp}u|^2\right)^2
 +{K\over 2}(\nabla^2 u)^2
 \right.\nonumber\\
 &~&\left.+{w'\over 8}\left(|\vec{\nabla}_{\perp}u|^2+{2D\over w'}\right)^2
 -{D^2\over 2w'}\right]\label{clean2}\, ,
\end{eqnarray}
where
\begin{eqnarray}
w'\equiv w-(g^2/B)\, .\label{Definitionw'}
\end{eqnarray} $w'=0$ is the tricritical point
separating the first order ($w'<0$) and second order ($w'>0$) phase
transitions. Assuming $w'>0$ since we are interested in the
second order phase transition, the ground state of the Hamiltonian
(\ref{clean2}) is given by
\begin{eqnarray}
 |\vec{\nabla}_{\perp}u| &=& \sqrt{-{2D\over w'}}\, ,\\
 \partial_z u &=& {gD\over B w'}\, .
\end{eqnarray}
Now we go back to the disordered Hamiltonian (\ref{H}) and expand it
around its {\it disorder averaged} minimum by making the
substitution
\begin{eqnarray}
u=\sqrt{-{2D\over w'}}x+{gD\over B w'}z + u'\label{},
\end{eqnarray}
where we have assumed that in the $C$ phase the system spontaneously
breaks azimuthal rotation symmetry and the layer normal tilts along
$\hat{x}$, which is an arbitrary direction within $\perp$ plane.
Also for simplicity we throw away the random field term by using a
{\it posterior} reasoning; that is, if the $C$ phase is ``$m=1$
Bragg glass'', then the random field disorder is irrelevant
\cite{Karl}. For brevity, we define $\theta_0\equiv
\sqrt{-{2D/ w'}}$. 
After the substitution and simplification, the Hamiltonian (\ref{H})
becomes
\begin{widetext}
\begin{eqnarray}
H[u'] &=& {1 \over 2} \int d^dr \left[K\left(\nabla_s^2
u'\right)^2 + B\left(\partial _zu'\right)^2
+w\theta_0^2\left(\partial_x u'\right)^2
-2g\theta_0(\partial_x u')(\partial_zu') - g(\partial_z
u')\left|\vec{\nabla}_s u'\right|^2\right.\nonumber\\
&+&\left.w\theta_0(\partial_x u')\left|\vec{\nabla}_s
u'\right|^2 +{w \over 4} \left|\vec{\nabla}_s
u'\right|^4+\vec{h}\left(\vec{r}\right)\cdot\vec{\nabla}u'\right]
\label{},
\end{eqnarray}
\end{widetext}
where $\hat{s}$ denotes the direction which is perpendicular to both $\hat{x}$ and $\hat{z}$.
Treating the tilt disorder by using the replica trick, we get
\begin{widetext}
\begin{eqnarray}
H[u'_{\alpha}] &=& {1 \over 2} \int d^dr \sum^n_{\alpha = 1}
\left[K\left(\nabla_s^2 u'_{\alpha}\right)^2 + B\left(\partial
_zu'_{\alpha}\right)^2 +w\theta_0^2\left(\partial_x
u'_{\alpha}\right)^2 -2g\theta_0(\partial_x
u'_{\alpha})(\partial_zu'_{\alpha}) - g(\partial_z
u'_{\alpha})\left|\vec{\nabla}_s u'_{\alpha}\right|^2\right.\nonumber\\
&+&\left.w\theta_0(\partial_x
u'_{\alpha})\left|\vec{\nabla}_s u'_{\alpha}\right|^2 +{w
\over 4} \left|\vec{\nabla}_s u'_{\alpha}\right|^4
\right]\nonumber\\&-&{\Delta_t\over 2T}\int d^dr \sum^n_{\alpha,
\beta = 1} \nabla_{\perp}u'_{\alpha} \cdot
\nabla_{\perp}u'_{\beta}-{\Delta_c\over 2T}\int d^dr
\sum^n_{\alpha, \beta = 1} \partial_zu'_{\alpha} \cdot
\partial_zu'_{\beta}\, , \label{H'}
\end{eqnarray}
\end{widetext}
whose universality class is still not clear. Then we make the
following coordinate transformation:
\begin{eqnarray}
r_{z'}&=&(g/{w\theta_0})r_x+r_z,\nonumber\\
r_{x'}&=&r_x,\nonumber\\
r_{s'}&=&r_s,\label{Ctransformation}
\end{eqnarray}
which, in momentum space, corresponds to the transformation
\begin{eqnarray}
 q_{z'} &=& q_z\, ,\nonumber\\
 q_{x'} &=& q_x - \Gamma q_z\, ,\nonumber\\
 q_{s'} &=& q_s\, ,
\end{eqnarray}
with
\begin{eqnarray}
 \Gamma = g/w \theta_0={g\over w}\sqrt{w'\over{-2D}}\, .
 \label{Gamma}
\end{eqnarray}
After this coordinate transformation the
Hamiltonian (\ref{H'}) becomes
\begin{widetext}
\begin{eqnarray}
H[u'_{\alpha}] &=& {1 \over 2} \int d^dr' \sum^n_{\alpha = 1}
\left[\tilde{K}\left(\nabla_{s'}^2 u'_{\alpha}\right)^2 +
\tilde{B}\left(\partial _{z'}u'_{\alpha}\right)^2
+\gamma\left(\partial_{x'} u'_{\alpha}\right)^2 +
\tilde{g}(\partial_{x'} u'_{\alpha})\left|\vec{\nabla}_{s'}
u'_{\alpha}\right|^2 +{w \over 4} \left|\vec{\nabla}_{s'}
u'_{\alpha}\right|^4 \right]\nonumber\\&-&{1\over 2T}\int d^dr'
\sum^n_{\alpha, \beta =
1}\left[\Delta_{s'}\left(\vec{\nabla}_{s'}u'_{\alpha} \cdot
\vec{\nabla}_{s'}u'_{\beta}\right)+\Delta_{x'}
(\partial_{x'}u'_{\alpha})
(\partial_{x'}u'_{\beta})+\Delta_{z'}(\partial_{z'}u'_{\alpha})(\partial_{z'}u'_{\beta})
\right]\nonumber\\
&+&{1\over 2T}\int d^dr' \sum^n_{\alpha, \beta =
1}\Delta_{x'z'}\left(\vec{\nabla}_{x'}u'_{\alpha} \cdot
\vec{\nabla}_{z'}u'_{\beta}\right) \label{H_C}\, ,
\end{eqnarray}
\end{widetext}
with
\begin{eqnarray}
\tilde{B}&=&B-{g^2\over w}\, ,\\
\tilde{K}&=&K\, ,\\
\gamma&=&w\theta_0^2\, ,\\
\tilde{g}&=&w\theta_0\, ,\\
\Delta_{s'}&=&\Delta_t\, ,\\
\Delta_{x'}&=&\Delta_t\, ,\\
\Delta_{x'z'}&=&{2g\Delta_t\over{w\theta_0}}\, ,\\
\Delta_{z'}&=&\Delta_t\left(g\over w\theta_0\right)^2+\Delta_c\, .
\end{eqnarray}
It is easy to check that $\gamma$, $\tilde{g}$ and $w$ satisfy the
magic relation $\tilde{g}=\sqrt{w\gamma}$. This is not a coincidence
but due to the symmetry, that is, the Hamiltonian must be invariant
under the rotation about $\hat{z}$-axis since the environment (i.e.,
the aerogel stretched along $\hat{z}$) is azimuthally symmetric.
Because of this symmetry the model (\ref{H_C}) belongs to the
universality class of ``$m=1$ smectic Bragg glass''.

\section{\label{sec: AnomalousElasticityIrrelevant}Wavevector-dependences of Irrelevant Disorder Variances}
In our RG calculations in section \ref{sec: ACTransition} we have not
included in the Hamiltonian (\ref{H_AC}) the random compression disorder
\begin{eqnarray}
-{\Delta_c\over 2T}\int d^dr \sum^n_{\alpha, \beta = 1}
\partial_zu_{\alpha} \cdot \partial_zu_{\beta}\, .
\end{eqnarray}
A simple power counting shows that this term is irrelevant and thus
has no effect on the critical behavior and anomalous elasticity.
However, while it does not affect the anomalous elasticity, the
disorder invariance $\Delta_c$ itself develops strong power-law
dependence on the wavevector at long wavelength. Now we calculate
the power-law exponent, which we denote as $\eta_c$, to
$O(\epsilon)$. This is necessary since the random compression
disorder has non-trivial contribution to the correlation function
$\overline{\left<|u(\vec{q})|^2\right>}$, and therefore affects the
light scattering predictions.

The most important one-loop graphical correction to $\Delta_c$ comes
from the Feynman diagram in Fig. \ref{fig: CorrectionB}, from which
we obtain
\begin{eqnarray}
\delta\Delta_c  &=& {g^2 \over 2} \int ^>_q \Delta_t^2
q^8_{\perp}G\left(\vec{q}\right)^4\nonumber \\
&=& {g^2\Delta_t^2\over 2} \int^{\infty}_{-\infty} {dq_z \over
2\pi}\int ^> {d^{d - 1}q_{\perp} \over  (2\pi)^{d-1}}{q^8_{\perp}
\over \left(Kq^4_{\perp} + Bq^2_z
\right)^4}\nonumber\\
&=& {5 \over 64} C_{d-1}\Delta^2_t \left({g^4 \over K^7B} \right)^{1
\over 2} \Lambda ^{d-7} \label{}.
\end{eqnarray}
Thus using the dimension and field rescaling we used in section
\ref{sec: ACTransition}, the RG flow equation of $\Delta_c$ to one-loop order is
given by
%
\begin{eqnarray}
{d\Delta_c(\ell)  \over  d \ell} = \left(d + 1 - \omega +2\chi+
{5\over 64}g_5\right)\Delta_c \label{Delta_c}\, ,
\end{eqnarray}
where the dimensionless coupling $g_5$ is defined as
\begin{eqnarray}
g_5\equiv B\Delta_t g_3/K\Delta_c\, .\label{g5}
\end{eqnarray}
Combining this RG flow equation with the flow Eqs.
(\ref{10}, \ref{11}, \ref{12}, \ref{g3} \ref{g4}), we get
\begin{eqnarray}
 {d g_5\over d \ell} = \left( 2 - {1\over 5}\epsilon -
 {5\over 64}g_5 \right)g_5 \label{},
\end{eqnarray}
which flows to a stable nontrivial fixed point $g_5^* = 64\left(10-
\epsilon\right)/25$. Then the wavevector-dependence of $\Delta_c$
can be calculated by performing trajectory integral matching on the
correlation function $\overline{\langle u(-\vec{q})\rangle\langle
u(\vec{q})\rangle}=(\Delta_t q_{\perp}^2+\Delta_c q_z^2
)G(\vec{q})^2$. The calculations are essentially the same as those
in section \ref{sec: AnomalousElasticity} and will not be repeated
here again. The result is given in (\ref{Deltacrit}) with the
anomalous exponent
\begin{eqnarray}
 \eta_c={5\over 64}g_5^*=2-{1\over 5}\epsilon+O(\epsilon^2)\, .
 \label{etac}
\end{eqnarray}
Note that $\eta_c$ is nonzero even to the {\it zeroth} order of $\epsilon$,
which is quite common for irrelevant variables.

It is also interesting to point out that $\eta_c$ is not fully independent,
but are related to other anomalous exponents by an {\it exact} scaling
relation, which is implied by the fact that $g_5$ flows to a non-zero
stable fixed point. For large enough $\ell$, $g_5$ reaches the
fixed point and thus
\begin{eqnarray}
{d \ln {g_5}\over d \ell}=0\, .
\end{eqnarray}
This equation, after decomposing $g_5$ into small pieces by its definition
(\ref{g5}), leads to
\begin{eqnarray}
 {d \ln {B}\over d \ell}+{d \ln {\Delta_t}\over d \ell}+{d \ln {g_3}\over d \ell}
 -{d \ln {K}\over d \ell}-{d \ln {\Delta_c}\over d \ell}=0\,
 .\nonumber\\
\end{eqnarray}
After plugging the flow Eqs. (\ref{10}, \ref{11}, \ref{12},
\ref{eta_3}, \ref{Delta_c}) into the above equation, the rescaling
factors $\chi$ and $\omega$ vanish and we are left with
\begin{eqnarray}
\eta_c = 2 + \eta_t - \eta_B - \eta_K - \eta_3 \label{Scalingeta_c}\, .
\end{eqnarray}
Since we already know the value of $\eta_t$, $\eta_B$, $\eta_K$ and
$\eta_3$ to $O(\epsilon)$, using (\ref{Scalingeta_c}) also leads to
(\ref{etac}).

Likewise in the model of the $C$ phase, which is given by
(\ref{H_C}), there are also disordered terms which are irrelevant in
the RG calculation but important for making light scattering
predictions. These terms are
\begin{eqnarray}
-{\Delta_{x'}\over 2T}\int d^dr \sum^n_{\alpha, \beta = 1}
\partial_{x'}u'_{\alpha} \cdot \partial_{x'}u'_{\beta},\nonumber\\
{\Delta_{x'z'}\over 2T}\int d^dr \sum^n_{\alpha, \beta = 1}
\partial_{z'}u'_{\alpha} \cdot \partial_{x'}u'_{\beta},\nonumber\\
-{\Delta_{z'}\over 2T}\int d^dr \sum^n_{\alpha, \beta = 1}
\partial_{z'}u'_{\alpha} \cdot \partial_{z'}u'_{\beta}\nonumber.
\end{eqnarray}

$\Delta{z'}$ and $\Delta_{x'z'}$ have no wavevector-dependences
since there are no graphical corrections to them.

As for $\Delta_{x'}$, we will use a different RG, namely the RG
calculation for ``$m=1$ Brag glass'', to show the power-law
wavevector-dependence of $\Delta_{x'}$ and calculate the power-law
exponent, which we denote as $\tilde{\eta}_{x'}$.
For convenience, we transform the Hamiltonian (\ref{H_C}) into the
exact form studied in \cite{Karl}. This transformation is
accomplished by doing the field rescaling
\begin{eqnarray}
 u'=\theta_0 u''.\label{FieldRescaling}
\end{eqnarray}
In terms of $u''$ the Hamiltonian (\ref{H_C}) can be written as
\begin{widetext}
\begin{eqnarray}
H[u''_{\alpha}] &=& {1 \over 2} \int d^dr' \sum^n_{\alpha = 1}
\left[\tilde{K}'\left(\nabla_{s'}^2
u^{\prime\prime}_{\alpha}\right)^2 + \tilde{B}'\left(\partial
_{z'}u^{\prime\prime}_{\alpha}\right)^2 +\gamma'\left(\partial_{x'}
u^{\prime\prime}_{\alpha}\right)^2 + \gamma'(\partial_{x'}
u^{\prime\prime}_{\alpha})\left|\vec{\nabla}_{s'}
u^{\prime\prime}_{\alpha}\right|^2 +{\gamma' \over 4}
\left|\vec{\nabla}_{s'}
u^{\prime\prime}_{\alpha}\right|^4 \right]\nonumber\\
 &~&-{1\over 2T}\int d^dr'
\sum^n_{\alpha, \beta =
1}\left[\Delta_{s'}'\left(\vec{\nabla}_{s'}u^{\prime\prime}_{\alpha}
\cdot \vec{\nabla}_{s'}u^{\prime\prime}_{\beta}\right)+\Delta_{x'}'
(\partial_{x'}u^{\prime\prime}_{\alpha})
(\partial_{x'}u^{\prime\prime}_{\beta}) \right] \label{H_C'}\, ,
\end{eqnarray}
\end{widetext}
where the coefficients of the quadratic terms are related to the
original ones by
\begin{eqnarray}
 \tilde{K}', \tilde{B}', \gamma', \Delta'_{s', x'}=\theta_0^2(\tilde{K}, \tilde{B}, \gamma, \Delta_{s',
 x'})\, .\label{Relation}
\end{eqnarray}
For this model, the detailed RG calculation can be found in
\cite{Karl}. Here we describe the calculation very briefly. A
hyper-cylindrical Brillouin zone is used: $-\Lambda<q_{s'}<\Lambda$,
$-\infty<q_{x'}<\infty$, $0<|\vec{q}_{z'}|<\infty$, where $\Lambda$
is an ultra-violet cutoff. Note that based on the arguments in the
previous section, $\Lambda$ has to be replaced by $\xi_{\perp}^{-1}$
when the system is within the critical region. The displacement
field is separated into high and low wavevector components,
$u^{\prime \prime}_{\alpha}\left(\vec{r}\right) = u^{\prime \prime
<}_{\alpha} \left(\vec{r}\right) +u^{\prime \prime <}_{\alpha}
\left(\vec{r}\right)$ where $u^{\prime \prime
>}_{\alpha} \left(\vec{r}\right)$ has support in the
hyper-cylindrical shell $\Lambda e^{-\ell}<|q_{s'}|<\Lambda$,
$-\infty<q_{x'}<\infty$, $0<|\vec{q}_{z'}|<\infty$, and
$u^{\prime\prime <}_{\alpha}\left(\vec{r}\right)$ has support in the
remainder of the hyper-cylinder (i.e., $0<|q_{s'}|<\Lambda
e^{-\ell}$, $-\infty<q_{x'}<\infty$, $0<|\vec{q}_{z'}|<\infty$).
Then integrate out the high wavevector part $u^{\prime\prime
>}_{\alpha}\left(\vec{r}\right)$, and rescale the length and long
wavelength part of the fields with
$r_{s'} = r'_{s'}e^{\ell}$, $r_{x'} = r'_{x'}e^{\omega_{x'}\ell}$,
$r_{z'} = r'_{z'}e^{\omega_{z'}\ell}$,
$u_{\alpha}^{\prime\prime}(\vec{r}) =
e^{\chi'\ell}u_{\alpha}^{\prime\prime\prime}(\vec{r}^{\,\prime})$ so
as to restore the $UV$ cutoff back to $\Lambda$.

The underlying symmetry of the model (\ref{H_C'}) guarantees that
the unit
$(\partial_{x'}u^{\prime\prime}_{\alpha}-(\nabla_{s'}u^{\prime\prime}_{\alpha})^2/2)$
is graphically renormalized as a whole. Therefore it is convenient
to choose the rescaling which also preserve the unit; the
appropriate choice is $\chi'=2-\omega_{x'}$.

Using the hard continuation (namely, keeping the number of the {\it
soft} directions fixed at 1 for all spacial dimension d), the RG to
one-loop order gives the following flow equations:
\begin{eqnarray}
 {d\gamma'(\ell)\over d\ell} &=&
 \left[5+(d-2)\omega_{z'}-3\omega_{x'}-{3g_6\over
 32\sqrt{2}}\right]\gamma',\nonumber\\
 \label{flowgamma'}\\
 {d \tilde{K}'(\ell)\over d\ell} &=& \left[1 + (d-2)\omega_{z'} -
 \omega_{x'} + {g_6\over 8\sqrt{2}}\right]K'\label{flowK'}\\
 {d(\Delta'_{s'}/T)(\ell)\over d\ell} &=& \left[3
 +(d-2)\omega_{z'}-\omega_{x'}+{g_6\over 32\sqrt{2}}\right](\Delta'_{s'}/T)
 \nonumber\\
 \label{flowDeltas'}
\end{eqnarray}
where $g_6 \equiv
\Delta'_{s'}(\gamma'/({K'}^{7-d}{B'}^{d-2}))^{1/2}C_{d-1}\Lambda^{2d-7}$
is a dimensionless measure of disorder, whose flow equation is
\begin{eqnarray}
 {dg_6(\ell)\over d\ell} = 2\tilde{\epsilon} g_6 - {15\over
 64\sqrt{2}}g_6^2\, ,
\end{eqnarray}
where $\tilde{\epsilon}=7/2-d$. This equation has a stable fixed
point
\begin{eqnarray}
 g_6={128\sqrt{2}\over 15}\tilde\epsilon\, .
\end{eqnarray}
Using trajectory integral matching we can calculate the
wavevector-dependences of $\gamma'$, $\tilde{K}'$ and
$\Delta'_{s'}$, which, combined with (\ref{Relation}), leads to the
wavevector-dependences of $K$, $\Delta_{s'}$ and $\gamma$ summarized
by (\ref{KanomC}, \ref{gammaanomC}, \ref{DeltaanomC}), with the
exponents
\begin{eqnarray}
\tilde{\zeta}_{x'}&=&2-{{\tilde{\eta}_{\gamma}+\tilde{\eta}_K}\over
2}\, ,\\
\tilde{\zeta}_{z'}&=&2-{\tilde{\eta}_K\over 2}\, ,\\
\tilde{\eta}_K&=&{g_6^*\over 8\sqrt{2}}={16\over 15}\tilde{\epsilon}+O({\tilde{\epsilon}}^2)\, ,\\
\tilde{\eta}_{s'}&=&{g_6^*\over 32\sqrt{2}}={4\over 15}\tilde{\epsilon}+O({\tilde{\epsilon}}^2)\, ,\\
\tilde{\eta}_{\gamma}&=&{3\over 32\sqrt{2}}g_6^*={4\over
5}\tilde{\epsilon}+O({\tilde{\epsilon}}^2)\, .
\end{eqnarray}

\begin{figure}
   \includegraphics[width=0.35\textwidth, angle=-90]{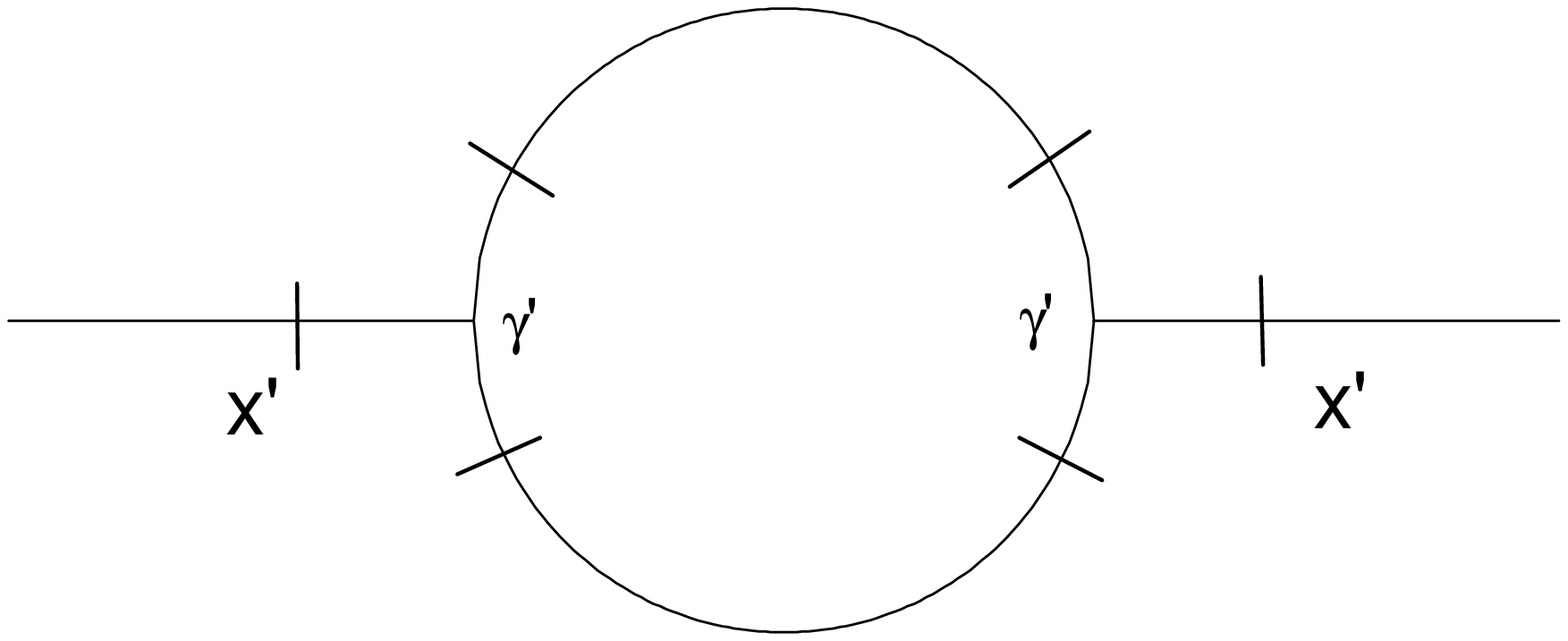}
   \caption{\label{fig: CorrectionBm=1}The graphical correction to $\Delta'_{x'}$.}
\end{figure}

So far we have briefly reviewed the RG calculation for ``$m=1$ Brag
glass'', now we use it to calculate the wavevector-dependence of the
irrelevant disorder invariance $\Delta'_{x'}$. The Feynman diagram
for the one-loop graphical correction to $\Delta'_{x'}$ is presented
in Fig. \ref{fig: CorrectionBm=1}. An analysis of the diagram gives
\begin{eqnarray}
\delta\Delta'_{x'} &=& {{\gamma'}^2{\Delta'_{s'}}^2\over 2}
\int^{\infty}_{-\infty} {dq_{z'} \over 2\pi}
 \int^{\infty}_{-\infty} {d^{d-2}q_{x'} \over (2\pi)^{d-2}}
 \nonumber\\
 &~&\times
 \int ^> {{dq_{s'} \over (2\pi)}{q_{s'}^8 \over
\left(K'q^4_{s'} + \tilde{B}'q^2_{z'}+\gamma' q^2_{x'}
\right)^4}}\nonumber\\
&=& {21 \over 128\sqrt{2}}g_7\, , \label{}
\end{eqnarray}
where $g_7$ is another dimensionless coupling defined as $g_7\equiv
g_6\gamma'\Delta'_{s'} \Lambda ^{-2}/(K'\Delta'_{x'})$. Thus the RG
flow of $\Delta'_{x'}$ to one-loop order is given by
\begin{eqnarray}
 {d\Delta'_{x'}(\ell)\over d\ell} &=&
 \left[5+(d-2)\omega_{z'}-3\omega_{x'}+{21\over
 128\sqrt{2}}g_7\right]\Delta'_{x'}\, .
 \nonumber\\
 \label{flowDeltax'}
\end{eqnarray}
 Combining this flow equation with (\ref{flowgamma'}),
(\ref{flowK'}) and (\ref{flowDeltas'}), we
obtain the flow equation of the dimensionless coupling $g_7$:
\begin{eqnarray}
 {dg_7\over d\ell} = \left(2-{8\over 5}\tilde{\epsilon}\right)g_7
 -{21\over 128\sqrt{2}}g_7^2\, ,
\end{eqnarray}
which flows to the fixed point $g_7^* =
256\sqrt{2}(5-4\tilde{\epsilon})/105$. At this point, our experience
from the previous calculations immediately gives us the
wavevector-dependence of $\Delta'_{x'}$, which, combined with
(\ref{Relation}), leads to the wavevector-dependence of
$\Delta_{x'}$ summarized by (\ref{DeltaanomC}). The power-law
exponent $\tilde{\eta}_{x'}$ is given by
\begin{eqnarray}
 \tilde{\eta}_{x'} = {21\over 128\sqrt{2}}g_7^*=2-{8\over 5}\tilde{\epsilon}+O({\tilde{\epsilon}}^2)\, .
\end{eqnarray}

That both $g_6$ and $g_7$ flows to non-zero fixed points implies the
{\it exact} scaling relations (\ref{scaling1}, \ref{scaling3},). For
the derivation, please refer to how we derive (\ref{Scalingeta_c}).

In addition, using soft continuation (i.e.; keeping the number of
{\it hard} directions fixed at 2) obtains
\begin{eqnarray}
 \tilde{\eta}_{x'} = 2 + \tilde{\tilde{\epsilon}} +
 O({\tilde{\tilde{\epsilon}}}^2)
\end{eqnarray}
and the same {\it exact} scaling relation given by (\ref{scaling3}),
where $\tilde{\tilde{\epsilon}}= 4-d$. The numerical estimate and
error bar for $\eta_{x'}$ can be obtained using the approach
described in Appendix \ref{sec: Average}.

\section{\label{sec: Cfunctions}Correlation Functions}
In this section we discuss correlation functions and make X-ray and
light scattering predictions. We start with the calculation of the
disorder averaged fluctuations in momentum space
$\overline{<|u(\vec{q})|^2>}$. In our problem this quantity has
three main distinct types of behavior. They are characterized by the
fluctuations in the $A$ phase, the fluctuations in the $C$ phase and
the fluctuations right at the critical point, respectively. The
first two types have been well studied and known, and the third type
is unique to this problem. Our main interest is the rich crossovers
between these distinct behaviors near the critical point, which
leads to amazing experimental consequences. We warn the reader of
heavy algebra in this section.

As we pointed out earlier, in some cases while certain type of disorder
is irrelevant in the RG sense, it can have important contributions to
correlation functions. Thus we prefer to use the complete
model given by (\ref{CompleteH_AC}) to calculate
$\overline{<|u(\vec{q})|^2>}$.

For large $\vec{q}$s (i.e., $q_{\perp}\ll \xi_{\perp}^{-1}$ or
$q_z\ll \xi_z^{-1}$), the model can be treated by the critical RG,
and we get
\begin{eqnarray}
 \overline{<|u(\vec{q})|^2>}&=&
 \overline{<|u(\vec{q})|^2>_{AC}}\nonumber\\
 &\equiv&{\Delta_t\left(\vec{q}\right)q_{\perp}^2 \over \left(Bq^2_z
 +D(T, \vec{q})q_{\perp}^2+K\left(\vec{q}\right)q^4_{\perp}\right)^2}+\nonumber\\
 &~&{\Delta_c\left(\vec{q}\right)q_z^2
 \over\left(Bq^2_z
 +D(T, \vec{q})q_{\perp}^2+K\left(\vec{q}\right)q^4_{\perp}\right)^2}\,
 ,\nonumber\\
 \label{CfunctionCrit}
\end{eqnarray}
where $K$ and $\Delta_{t,c}$ are all $\vec{q}$-dependent and given
by (\ref{K}, \ref{Deltacrit}), $B$ is $\vec{q}$-independent, $D$ has
dependences on both $\vec{q}$ and the temperature:
\begin{eqnarray}
 &~&D\left(\vec{q}, T=T_{AC}\right)\nonumber\\
 &\approx&\left\{
\begin{array} {ll}
(T-T_{AC})(\xpnl q_{\perp})^{\eta_D}, &\xznl q_z \ll (\xpnl q_{\perp})^{\zeta}\\
(T-T_{AC})(\xznl q_{z})^{{\eta_D\over \zeta}}, &\xznl q_z \gg
(\xpnl q_{\perp})^{\zeta}
\end{array}\right.\, ,
\nonumber\\
\label{D}
\end{eqnarray}
where $\eta_D=2-\eta_K-1/\nu_{\perp}$. The first and the second
terms in the Eq. (\ref{CfunctionCrit}) come from the random tilt and
random compression disorders, respectively. In this case the random
field disorder is not only irrelevant in the RG calculation, but
also has negligible contributions to $\overline{<|u(\vec{q})|^2>}$
in any region in $\vec{q}$-space, and is thus neglected.

For small $\vec{q}$s (i.e., $q_{\perp}\ll \xi_{\perp}^{-1}$ or
$q_z\ll \xi_z^{-1}$), the tilt term in the model dominates the bend
term, and the critical RG is no longer valid. A different RG needs
to be used to treat the model. First let us derive the effective
model for small $\vec{q}$s. The derivation has two steps. The first
step is to run the critical RG to the crossover point
$\ell^*=\ln(\Lambda\xi_{\perp})$ to obtain a model for the rescaled
system. The second step is to undo the dimension and field rescaling
we performed during the RG. implementing this procedure we obtain
\begin{widetext}
\begin{eqnarray}
H[u_{\alpha}] &=& {1 \over 2} \int d^dr \left(\sum^n_{\alpha = 1}
\left[K(T)\left(\nabla_{\perp}^2u_{\alpha}\right)^2 + B\left(\partial
_zu_{\alpha}\right)^2 - g(T)(\partial
_zu_{\alpha})\left(\nabla_{\perp}u_{\alpha}\right)^2 + {w(T) \over 4}
\left|\vec{\nabla}_{\perp}u_{\alpha}\right|^4 +
D(T)\left|\nabla_{\perp}u_{\alpha}\right|^2
\right]\right.\nonumber\\&-&\left. \sum^n_{\alpha, \beta =
1}\left[{\Delta_t(T)\over 2T}  \nabla_{\perp}u_{\alpha} \cdot
\nabla_{\perp}u_{\beta} +{\Delta_c(T)\over 2T}
\partial_zu_{\alpha} \cdot \partial_zu_{\beta}+{1\over
T}\Delta_p(u_{\alpha}-u_{\beta})\right]\right),
\label{SmallCompleteH_AC}
\end{eqnarray}
\end{widetext}
where the coefficients are renormalized by the critical fluctuations and hence
temperature-dependent.

In the $A$-side critical region, based on the argument we give in
section \ref{sec: ACPhases}, the model (\ref{SmallCompleteH_AC})
belongs to the universality class of ``random field $XY$ model''. We
treat it using the functional RG \cite{XY} and obtain
\begin{eqnarray}
\overline{<|u(\vec{q})|^2>}&=&\overline{<|u(\vec{q})|^2>_A}\nonumber\\
 &\equiv& {\Delta_t \left(\vec{q}, T
\right)q_{\perp}^2 \over \left(Bq^2_z+D(T)q^2_{\perp}
+K(T)q_{\perp}^4\right)^2}+\nonumber\\
&~&{\Delta_c\left(\vec{q}, T \right)q_z^2
\over\left(Bq^2_z+D(T)q^2_{\perp}+K(T)q_{\perp}^4\right)^2}
+\nonumber\\
 &~& {C{B^{1\over 2}D}\over
\left(Bq^2_z+D(T)q^2_{\perp}+K(T)q_{\perp}^4\right) ^{3\over2}q_0^2}\, ,\nonumber\\
\label{CfunctionA}
\end{eqnarray}
where the third piece comes from the random field disorder. The
random tilt and compression disorder are irrelevant in the RG
calculation; however, their contributions to
$\overline{<|u(\vec{q})|^2>}$ are important in some region in
$\vec{q}$-space and hence kept.
The constants $K$, $D$, $\Delta_{t, c}$ in (\ref{CfunctionA}) are
{\it not} wavevector-dependent and given by
\begin{eqnarray}
 B &=& B_0(\xi_{NL}^{\perp}/\xi_{\perp})^{\eta_B}=B_0\, ,\nonumber\\
 K &=& K_0(\xi_{NL}^{\perp}/\xi_{\perp})^{-\eta_K}\propto
 (T-T_{AC})^{-\eta_K\nu_{\perp}}\, ,\nonumber\\
 D &=& D_0(\xi_{NL}^{\perp}/\xi_{\perp})^{\eta_D}\propto
 (T-T_{AC})^{(2-\eta_K)\nu_{\perp}}\, ,\nonumber\\
 \Delta_{t,c} &=&
 \Delta_{t,c}^0(\xi_{NL}^{\perp}/\xi_{\perp})^{-\eta_{t,c}}\propto
 (T-T_{AC})^{-\eta_{t,c}\nu_{\perp}}\, .\nonumber
\end{eqnarray}

In the $C$-side critical region, we have to expand the Hamiltonian
(\ref{SmallCompleteH_AC}) around the disorder-averaged minimum.
Following the procedure in section \ref{sec: ACPhases} leads to the
model (\ref{H_C}), where for convenience a special coordinate system
other than the lab one has been used. The relations between these
two coordinate systems are controlled by the parameter $\Gamma$ and, in
Fourier space, given in Eqs. (\ref{qx}, \ref{qs}, \ref{qz}).
$\Gamma$ is now renormalized by critical fluctuations and hence
temperature-dependent, given by
\begin{eqnarray}
\Gamma\sim \left(\xi_{\perp}\Lambda\right)^{1-{{\eta_K+\eta_3}\over
2}}\, .
\end{eqnarray}
The exponent is equal to $4/5$ in $d=3$ to $O(\epsilon)$; therefore,
$\Gamma$ is expected to be very large as $T\to T_{AC}^-$. The
derivation of this result is given in appendix \ref{Sec: Gamma}. The
temperature-dependences of the coefficients in the model (\ref{H_C})
can be derived in a similar way and are given by
\begin{eqnarray}
 \tilde{K}_c &=& K_0(\xi_{NL}^{\perp}/\xi_{\perp})^{-\eta_K}\propto
 (T_{AC}-T)^{-\eta_K\nu_{\perp}}\, ,\label{K_c}\\
 \gamma_c &=& D_0(\xi_{NL}^{\perp}/\xi_{\perp})^{\eta_D}\propto
 (T_{AC}-T)^{(2-\eta_K)\nu_{\perp}}\, ,\nonumber\\ \label{gamma_c}\\
 \Delta_{s',x'}^c &=&
 \Delta_t^0(\xi_{NL}^{\perp}/\xi_{\perp})^{-\eta_t}\propto(T_{AC}-T)^{-\eta_t\nu_{\perp}}\, ,\label{Deltasx_c}\\
 \Delta_{z'}^c &=& \left[\left(\xi_{NL}^{\perp}\over\lambda\right)^2\Delta_t^0+\Delta_c^0\right]
 \left(\xi_{\perp}\over\xi_{NL}^{\perp}\right)^{\eta_c}\nonumber\\
 &\propto& (T-T_{AC})^{-\eta_c\nu_{\perp}}\, . \label{Deltaz_c}
\end{eqnarray}
These coefficients have been named ''half-dressed'' values (i.e.
unrenormalized by the $C$ fluctuations) in Eqs. (\ref{KanomC},
\ref{gammaanomC}, \ref{DeltaanomC}). Treating the model (\ref{H_C})
using the RG for ``$m=1$ smectic Bragg glass''\cite{Karl}, we obtain
\begin{eqnarray}
\overline{<|u(\vec{q})|^2>}&=&
\overline{<|u(\vec{q})|^2>_C}\nonumber\\ &\equiv& {{\Delta_{s'}
\left(\vec{q}\,', T \right)q_{s'}^2}\over \left(\gamma(\vec{q}\,',
T)q_{x'}^2 + \tilde{B}q^2_{z'}+\tilde{K}\left (\vec{q}\,',
T\right)q^4_{s'}\right)^2}\nonumber\\
&~& +{\Delta_{x'}\left(\vec{q}\,', T \right)q_{x'}^2\over
\left(\gamma(\vec{q}\,', T)q_{x'}^2 +
\tilde{B}q^2_{z'}+\tilde{K}\left(\vec{q}\,', T
\right)q^4_{s'}\right)^2}
\nonumber\\
&~&+{\Delta_{z'}\left( T\right)q_{z'}^2\over
\left(\gamma(\vec{q}\,', T)q_{x'}^2 +
Bq^2_{z'}+\tilde{K}\left(\vec{q}\,', T \right)q^4_{s'}\right)^2}
\nonumber\\
\label{CfunctionC}
\end{eqnarray}
where the $\vec{q}\, '$-dependence's of $\tilde{K}$, $\gamma$,
$\Delta_{s',x'}$ are given in Eqs. (\ref{KanomC}, \ref{gammaanomC},
\ref{DeltaanomC}), $\tilde{B}$ and $\Delta_{z'}$ are $\vec{q}\,
'$-independent. The first and the second terms correspond to the
contributions from the random tilt disorder along the soft and the
hard directions in the original ``$m=1$ smectic'' problem,
respectively, and the third term corresponds to the contribution
from the random compression disorder in the same problem. The random
field disorder is neglected since its contribution to
$\overline{<|u(\vec{q})|^2>}$ is always subdominant to the others.

Thus, near the critical point the behavior of
$\overline{<|u(\vec{q})|^2>}$ can be summarized in the following
compact way:
\begin{widetext}
\begin{eqnarray}
\overline{<|u(\vec{q})|^2>} = \left\{
\begin{array} {ll}
 \overline{<|u(\vec{q})|^2>_A}\, , &q_{\perp}\ll \xi_{\perp}^{-1},\, q_z \ll \xi_z^{-1},~~~\mbox{for $T>T_{AC}$}\\
 \overline{<|u(\vec{q})|^2>_{AC}}\, , &q_{\perp}\gg \xi_{\perp}^{-1}\, $or$ \,\,q_z \gg \xi_z^{-1},
\\
 \overline{<|u(\vec{q})|^2>_C}\, , &q_{\perp}\ll \xi_{\perp}^{-1},\, q_z \ll \xi_z^{-1},~~~\mbox{for $T< T_{AC}$}\\
\end{array}\right.\, .
 \label{CfunctionAC}
\end{eqnarray}
\end{widetext}

%

The X-ray Scattering pattern for both $A$ and $C$ phases has been
studied before. It is quasi sharp, isotropic for the $A$ phase,
broad, anisotropic for the $C$ phase. In the following we will
discuss the crossover between the two distinct patterns in the
critical region. The X-ray scattering intensity is given by the
Fourier transform of the thermal and disorder averaged $\rho-\rho$
correlation function:
\begin{eqnarray}
I(\vec{q})\propto\int{d^dr'd^dr''}\quad{\overline{\left<\rho(\vec{r}\,')
\rho(\vec{r}\,'')\right>}e^{-i\vec{q}\cdot\left(\vec{r}\,'-\vec{r}\,''\right
)}} \label{xray},
\end{eqnarray}
where $\rho(\vec{r})$ is the molecule density. $\rho(\vec{r})$ can
be expanded in a Fourier series with period $a$, the distance
between nearest layers, via
\begin{eqnarray}
\rho(\vec{r})=\sum_{n=-\infty}^{\infty}\rho_n
e^{inq_0(Z+u\left(\vec{r})\right)}\, , \label{density}
\end{eqnarray}
where $\rho_n$ is the (complex) amplitude of the $n$'th harmonic
of the smectic density wave, $u(\vec{r})$ is layer displacement at
the point $\vec{r}$. Inserting the decomposition (\ref{density})
into (\ref{xray}) gives
\begin{widetext}
\begin{eqnarray}
I(\vec{q})\propto\sum_{n,n'}\rho_n\rho_{n'}\int{d^dr'd^dr''}\quad{e^{-
i\vec{q}\cdot(\vec{r}\,'-\vec{r}\,'')}e^{iq_0(nZ'+n'Z'')} \times
\overline{\left<e^{iq_0(nu(\vec{r}\,')+n'u(\vec{r}\,''))}\right>}}\,.
\label{}
\end{eqnarray}
\end{widetext}
Changing variables of integration from $\vec{r}\,'$ to the
different variable $\vec{r}\equiv\vec{r}\,'-\vec{r}\,''$ and using
the fact that, in a homogeneous system
\begin{eqnarray}
\overline{\left<e^{iq_0(nu(\vec{r}+\vec{r}\,'')+n'u(\vec{r}\,''))}\right>}
=\overline{\left<e^{iq_0(nu(\vec{r})+n'u(\vec{0}))}\right>}
\label{}
\end{eqnarray}
the integral over $\vec{r}\,''$ now just gives $V\delta_{n+n'}$,
where $V$ is the volume of the system and $\delta_{n+n'}$ is a
Kronecker delta which can be used to collapse the sum on $n'$ to
the single term $n'=-n$. Doing so, we obtain
\begin{eqnarray}
I\left(\vec{q}\right)\propto
V\sum_n|\rho_n|^2\int{d^dr}\quad{{e^{i(nq_0\hat{Z}-\vec{q})\cdot
\vec{r}}}F_n(\vec{r}\,')}\, , \label{113}
\end{eqnarray}
where
\begin{eqnarray}
F_n(\vec{r})\equiv\overline{\left<e^{iq_0(nu(\vec{r})+n'u(\vec{0}))}
\right>}\, . \label{}
\end{eqnarray}
Now approximating the $u$ - fluctuation as Gaussian, we can write
\begin{eqnarray}
F_n(\vec{r})=\exp\left({{-}{{{n^2}{q^2_0}}\over
2}C(\vec{r})}\right) \label{115}
\end{eqnarray}
with
\begin{eqnarray} C(\vec{r}) &\equiv&  \overline{
\left<\left[u(\vec{r}) - u(0) \right]^2
\right>}\nonumber\\
&=&\int {d^dk \over (2 \pi)^d} \quad {2\left[1 -
\cos(\vec{r}\cdot\vec{q})\right]\overline{\left<u\left(\vec{q}\right)u
\left(-\vec{q}\right) \right>}}\, .\nonumber\\
 \label{CfunctionReal}
\end{eqnarray}
Putting (\ref{115}) into (\ref{113}) we get
\begin{eqnarray}
I\left(\vec{q}\right)\propto\sum_{n=1}^{\infty}\int{d^dr}\quad{{e^{i(n
q_0\hat{Z}-\vec{q})\cdot\vec{r}}}\exp\left({{-}{{{n^2}{q^2_0}}\over
2}C(\vec{r})}\right)}\, . \nonumber
\\
\label{inti}
\end{eqnarray}
For $\vec{q}$ near the Bragg peaks at the $mq_o\hat{Z}$, the sum
is readily seen to be dominated by the $n=m$ term; thus, the
scattering near the Bragg peaks is essentially the Fourier
transform of $\exp({-n^2q_0C(\vec{r})/2})$.

Right at the critical point ($T=T_{AC}$), in spite of the symmetry
broken we mentioned earlier, the fluctuations of the system is
still qualitatively like those of ``Glassy smectic $A$'', hence
the X-ray scattering pattern is also broad and anisotropic.

For $T \to T_{AC}^+$, plugging Eq. (\ref{CfunctionAC})
into Eq. (\ref{CfunctionReal}) and performing an asymptotic
analysis, we obtain, in 3D,
\begin{widetext}
\begin{eqnarray}
C(\vec{r}) =
\left\{\begin{array}{ll}{\lambda^2(r_{\perp}/\xi_{NL}^{\perp})^{\varGamma}f_{\varGamma}\left(\left(r_z/
\xi^z_N\right)/{\left(r_{\perp}/\xi^{\perp}_N\right)^{\zeta}}\right)},
&r_{\perp}\ll \xi_{\perp},\ \ r_z\ll \xi_z\\
\\
{\lambda^2(\xi_{\perp}/\xi_{NL}^{\perp})^{\varGamma}+
 {\lambda^5(\xi_{\perp}/\xi_{NL}^{\perp})^{\eta_t}\over {\varrho^3 (\xi_{NL}^{\perp})^2}}
 \left({1\over\xi_{\perp}}-{1\over \sqrt{r_{\perp}^2
 +(\varrho r_z)^2}}
\right)}\nonumber\\
{+{\varsigma\over q^2_0}{\ln\left(\sqrt{r_{\perp}^2+(\varrho
r_z)^2}\over\xi_{\perp}\right)}},&r_{\perp}\gg \xi_{\perp}\ \
$or$\ \ r_z\gg \xi_z
\end{array}\right.
\label{real}
\end{eqnarray}
\end{widetext}
where $\varGamma\equiv{2+\eta_t-({3\eta_K}/2)}$, $f_{\varGamma}$
is another universal scaling function, $\varsigma\approx 1.10$,
$\varrho=(\lambda/\xi_{\perp})(\xi_{\perp}/\xi_{NL}^{\perp})^{\eta_K/2}$.
For small $r$, i.e., $r_{\perp}\ll\xi_{\perp}, r_z\ll\xi_z$,
$C(\vec{r})$ increases as a power law, while for big $r$, i.e.,
$r_{\perp}\gg \xi_{\perp}$ or $r_z\gg\xi_z$, $C(\vec{r})$
increases logarithmically. This indicates that $F_n(\vec{r})$
varies faster (exponentially) for small $r$
($r_{\perp}\ll\xi_{\perp}, r_z\ll\xi_z$), slower (as a power law)
for big $r$ ($r_{\perp}\gg \xi_{\perp}$ or $r_z\gg\xi_z$). So the
X-ray scattering intensity $I(\delta\vec{q})$ for large $|\delta
\vec{q}|=|\vec{G}-\vec{q}|$, which is dominated by
$I^f(\delta\vec{q})$, defined as the contribution coming from the
fast varying part of $F_n(\vec{r})$, should be broad and {\it
anisotropic}, qualitatively like a Lorentzian squared. The line
widths $\delta q_z^x$, $\delta q_{\perp}^x$ of this Lorentzian
squared are defined as $\delta q_z^x=({\xi^x_z})^{-1}$ and $\delta
q_{\perp}^x=({\xi^x_{\perp}})^{-1}$ with $\xi^x_z$ and
$\xi^x_{\perp}$ satisfying
\begin{eqnarray}
C(0,\xi^x_z)=a^2, \label{xwith}\\
C(\xi^x_{\perp},0)=a^2\label{zwith}.
\end{eqnarray}
Solving these two equations we get
\begin{eqnarray}
\xi^x_\perp=\xi^\perp_{NL}\left(a\over\lambda\right)^{2/
\varGamma}\, , \label{fxwid}\\
\xi^x_z=\xi^z_{NL}\left(a\over\lambda\right)^{{2\zeta}/
\varGamma}\,.
 \label{fzwid}
\end{eqnarray}
The temperature dependence of $\xi^x_{\perp}$ and $\xi^x_z$ could
be used to determine the exponents $2/ \varGamma$, ${2\zeta}/
\varGamma$, $\eta_k$, and $\eta_\Delta$ since the bulk $K(T)$ and
$B(T)$ that implicitly appear in Eqs. (\ref{fxwid}, \ref{fzwid})
have temperature dependence that can be extracted from
measurements on bulk materials.

For small $\delta q$, $I(\delta\vec{q})$ is dominated by
$I^s(\delta\vec{q})$, which is defined as the contribution coming
from the slow varying part of $F_n(\vec{r})$. Thus,
$I(\delta\vec{q})$ is quasi-sharp and {\it isotropic}, and diverges
as a power law as $\delta{q}$ approaches $0$.

The crossover between these two distinct scattering patterns is
defined as the point where the two contributions become comparable,
that is, $I^s(\delta \vec{q})=I^f(\delta \vec{q})$. Assuming that
this happens at very small $|\delta \vec{q}|$, so that
$I^f(\delta\vec{q})$ can be treated as a constant
$\left(\xi^x_{\perp}\right)^2{\xi^x_z}$.  We will verify this
assumption as a posteria. Let us first calculate the crossover in
$\perp$ directions
\begin{eqnarray}
I\left(\delta q_{\perp}=\delta q^c_{\perp}, \delta
q_z=0\right)&\propto& \left(\delta
q^c_{\perp}\right)^{-3+0.55n^2}\xi_{\perp}^{-0.55n^2}\nonumber\\
&~&\times e^{-n^2\lambda^2q_0^2\left(\xi_{\perp}/\xi^{\perp}_{NL}\right)^{\varGamma}/2}\nonumber\\
&\propto&\left(\xi^x_{\perp}\right)^2{\xi^x_z}, \label{}
\end{eqnarray}
which gives
\begin{eqnarray}
\delta q^c_{\perp}&\propto&\left|T-T_{AC}\right|^
{0.55n^2\Omega/(3-0.55n^2)}\nonumber\\
&~&\times\exp(-{n^2\over
{3-0.55n^2}}\left|T-T_{AC}\right|^{-\Omega})\, .
\end{eqnarray}
where $\Omega\equiv \varGamma\nu_{\perp}$. Likewise the crossover
in $z$ direction is found to be
\begin{eqnarray}
\delta q^c_z&\propto&\left|T-T_{AC}\right|^{
\left(2-\eta_k\right)\nu_{\perp}/2}\delta q_{\perp}^c\, .
\end{eqnarray}
Both $\delta q_{\perp}^c$ and $\delta q_z^c$ become extremely small
when $T\to T_{AC}^+$, which is consistent with our earlier
assumption.

As for $T\to T_{AC}^-$, since the
X-ray scattering pattern is always broad, there is no significant
crossover which can be detected by experiments.

Now we investigate light scattering behavior near the critical
region. The light scattering intensity is proportional to a linear
sum of the fluctuations of the nematic director $\sum_{ij}A_{ij}\overline{\langle
\delta n_i^{\perp}(-\vec{q})\delta n_j^{\perp}(\vec{q})\rangle}$ \cite{Lubensky},
where the values of $A_{ij}$ are determined by the
electric polarization directions of the incident and transmitted
light. Since in Sec. \ref{sec: Model} we have shown
that the fluctuations of $\hat{n}$ and the layer normal $\hat{N}$
are bound together, the light scattering intensity is thus
proportional to a linear sum of $C_{ij}(\vec{q})$, which is defined
as
\begin{eqnarray}
 C_{ij}(\vec{q}) \equiv \overline{\left<N_i^{\perp}(-\vec{q})N_j^{\perp}(\vec{q})\right>}
 =
 L_{ij}^{\perp}(\vec{q})q_{\perp}^2\overline{<\left|u(\vec{q})\right|^2>}\,
 .\nonumber\\
 \label{light1}
\end{eqnarray}

We now derive the eight distinct regions with qualitatively
different wavevector-dependences of $C_{ij}(\vec{q})$ in the
$A$-side critical region, which are illustrated in Fig. \ref{fig: 8regions}.
For big $q$'s (i.e., $q_{\perp}\gg \xi_{\perp}^{-1}$ or $q_z\gg
\xi_z^{-1}$, which corresponds to the regions outside the
rectangle OJFI ), $C_{ij}(\vec{q})$ is
given by
\begin{eqnarray}
 C_{ij}(\vec{q}) &=& L_{ij}^{\perp}\left({\Delta_t \left(\vec{q}\right)q_{\perp}^2 \over \left(Bq^2_z+D(T)q^2_{\perp}
+K\left(\vec{q}\right)q^4_{\perp}\right)^2}\right.\nonumber\\
&+&\left.{\Delta_c\left(\vec{q}\right)q_z^2
\over\left(Bq^2_z+D(T)q^2_{\perp}
+K\left(\vec{q}\right)q^4_{\perp}\right)^2}\right)\, ,
\label{light2}
\end{eqnarray}
where in the denominator $Dq_{\perp}^2$ is negligible compared to
$Kq_{\perp}^4$. The function for the crossover (i.e., locus EG) between the random
compression and tilt can be obtained by doing
\begin{eqnarray}
&~&{\Delta_t \left(\vec{q}\right)q_{\perp}^4 \over
\left(D(T)q_{\perp}^2 + Bq^2_z +K\left(\vec{q}, T
\right)q^4_{\perp}\right)^2}\nonumber\\
&=&{\Delta_c\left(\vec{q}\right)q_z^2 q_{\perp}^2
\over\left(D(T)q_{\perp}^2 + Bq^2_z +K\left (\vec{q},
T\right)q^4_{\perp}\right)^2},\label{RTRC}
\end{eqnarray}
which leads to
\begin{eqnarray}
\Delta_t^0(q_z\xi_{NL}^z)^{-\eta_t/\zeta}q_{\perp}^2=
\Delta_c^0(q_z\xi_{NL}^z)^{-\eta_c/\zeta}q_z^2.
\end{eqnarray}
Using the scaling relation (\ref{Scalingeta_c}), this equation can
be further transformed into
\begin{eqnarray}
 q_z\xi_{NL}^z =
 \left(q_{\perp}\xi_{NL}^z\sqrt{\Delta_t^0/\Delta_c^0}\right)^A\, ,
\end{eqnarray}
where $A\equiv \zeta/(1+\eta_3/2)$. Above EG (i.e., in region 8),
$C_{ij}(\vec{q})$ is dominated by the random compression fluctuations and given by
\begin{eqnarray}
 C_{ij}(\vec{q})&\approx& L_{ij}^{\perp}(\vec{q})\left(\Delta_c q_z^2q_{\perp}^2\over B^2 q_z^4\right)\nonumber\\
                &\approx& L_{ij}^{\perp}(\vec{q})\left(\left(\Delta_c^0\over B_0^2\right)
                \left(q_{\perp}\over q_z\right)^2(q_z\xi_{NL}^z)^{-\eta_c/\zeta}\right)\nonumber\\
                &\sim& L_{ij}^{\perp}(\vec{q})\left({\lambda^5\Delta_c^0\over \Delta_t^0(\xi_{NL}^{\perp})^2}\left(q_{\perp}\over q_z\right)^2
                   (q_z\xi_{NL}^z)^{-\eta_c/\zeta}\right)\,
                   ,\nonumber\\
\end{eqnarray}
where we remind reader that $\lambda$ is the smectic penetration
length defined by $\lambda=\sqrt{K_0/B_0}$, and $\xi_{NL}^{\perp,
z}$ are the nonlinear crossover lengths defined in Eqs.
(\ref{NonlinearPerp}, \ref{NonlinearZ}). Below EG, $C_{ij}(\vec{q})$
is dominated by the random tilt fluctuations but still has two
different wavevector-dependences due to the crossover (locus FH)
between $Kq_{\perp}^4$ and $Bq_z^2$ in the common denominator, which
satisfies
\begin{eqnarray}
 q_z\xi_{NL}^z=(q_{\perp}\xi_{NL}^{\perp})^{\zeta}
\end{eqnarray}
and is below locus EG since $\zeta>A$. $C_{ij}(\vec{q})$is given by
\begin{eqnarray}
 C_{ij}(\vec{q})&\approx& L_{ij}^{\perp}(\vec{q})\left(\Delta_t q_{\perp}^4\over B^2 q_z^4\right)\nonumber\\
                &\approx& L_{ij}^{\perp}(\vec{q})\left(\left(\Delta_t^0\over B_0^2\right)\left(q_{\perp}\over q_z\right)^4
                (q_z\xi_{NL}^z)^{-\eta_t/\zeta}\right)\nonumber\\
                &\sim& L_{ij}^{\perp}(\vec{q})\left({\lambda^5\Delta_c^0\over (\xi_{NL}^{\perp})^2}\left(q_{\perp}\over
                q_z\right)^4(q_z\xi_{NL}^z)^{-\eta_t/\zeta}\right)\,
                \nonumber\\
\end{eqnarray}
above locus FH (i.e., in
region 7) and
\begin{eqnarray}
 C_{ij}(\vec{q}) &\approx& L_{ij}^{\perp}(\vec{q})\left(\Delta_t q_{\perp}^4\over K^2 q_{\perp}^8\right)\nonumber\\
                 &\approx& L_{ij}^{\perp}(\vec{q})\left(\left(\Delta_t^0\over K_0^2\right)\left(1\over
                 q_{\perp}\right)^4(q_{\perp}\xi_{NL}^{\perp})^{2\eta_K-\eta_t}\right)\nonumber\\
                 &\sim& L_{ij}^{\perp}(\vec{q})\left({\lambda\over (\xi_{NL}^{\perp})^2}\left(1\over
                 q_{\perp}\right)^4(q_{\perp}\xi_{NL}^{\perp})^{2\eta_K-\eta_t}\right)\,
                 \nonumber\\
\end{eqnarray}
below locus FH (i.e., in region 6).

For small $q$'s (i.e., $q_{\perp}\ll \xi_{\perp}^{-1}$, $q_z\ll
\xi_z^{-1}$, which corresponds to the regions within the rectangle
OJFI in Fig. \ref{fig: 8regions}),
$\overline{<|u(\vec{q})|^2>}$ is given by (\ref{CfunctionA}),
and $C_{ij}(\vec{q})$ is thus
\begin{eqnarray}
 C_{ij}(\vec{q}) &=& L_{ij}^{\perp}(\vec{q})\left({\Delta_t \left(\vec{q}, T
 \right)q_{\perp}^4 \over \left(Bq^2_z+D(T)q^2_{\perp}
 +K\left(\vec{q}, T
 \right)q^4_{\perp}\right)^2}\right.\nonumber\\
 &~&\left.+{\Delta_c\left(\vec{q}, T
 \right)q_z^2 q_{\perp}^2\over\left(Bq^2_z+D(T)q^2_{\perp} +K\left (\vec{q},
 T\right)q^4_{\perp}\right)^2}\right.\nonumber\\
 &~&\left.+ {C{B^{1\over 2}D}q_{\perp}^2\over
 \left(Bq^2_z+D(T)q^2_{\perp}+K\left(\vec{q},
 T\right)q_{\perp}^4\right) ^{3\over2}q_0^2}\right)\, .\nonumber\\
 \label{light3}
\end{eqnarray}
where in the denominator $Kq_{\perp}^4$ is negligible compared to $Dq_{\perp}^2$. Now since
$C_{ij}(\vec{q})$ has three pieces, the crossovers
are quite complicated. For simplicity, let us divide the rectangle OJEI
into two subregions separated by locus OF, which satisfies
$Bq_z^2=Dq_{\perp}^2$ and hence
\begin{eqnarray}
 q_z = {\lambda\over
 \xi_{NL}^{\perp}}(\xi_{NL}^{\perp}/\xi_{\perp})^{1-\eta_K/2}q_{\perp}\,
 .
\end{eqnarray}

First we consider the subregion above locus OF (i.e., triangle IOF),
in which $Bq_z^2$ dominates $Dq_{\perp}^2$ in the common denominator
in the Eq. (\ref{light3}). In this region, the three pieces in the Eq. (\ref{light3}),
which correspond to the contributions from the the random tilt, the random
compression and the random field respectively, are given respectively by
\begin{eqnarray}
 &~& L_{ij}^{\perp}(\vec{q})\left(\Delta_t q_{\perp}^4\over B^2 q_z^4\right)\nonumber\\
                &=& L_{ij}^{\perp}(\vec{q})\left(\left(\Delta_t^0\over B_0^2\right)\left(q_{\perp}\over q_z\right)^4
                (\xi_{NL}^{\perp}/\xi_{\perp})^{-\eta_t}\right)\nonumber\\
                &\sim& L_{ij}^{\perp}(\vec{q})\left({\lambda^5\over (\xi_{NL}^{\perp})^2}\left(q_{\perp}\over
                q_z\right)^4(\xi_{NL}^{\perp}/\xi_{\perp})^{-\eta_t}\right)\, ,
                \nonumber\\
                \label{lightRT1}
\end{eqnarray}
\begin{eqnarray}
 &~& L_{ij}^{\perp}(\vec{q})\left(\Delta_c q_z^2q_{\perp}^2\over B^2 q_z^4\right)\nonumber\\
                &=& L_{ij}^{\perp}(\vec{q})\left(\left(\Delta_c^0\over B_0^2\right)
                \left(q_{\perp}\over q_z\right)^2(\xi_{NL}^{\perp}/\xi_{\perp})^{-\eta_c}\right)\nonumber\\
                &\sim& L_{ij}^{\perp}(\vec{q})\left({\lambda^5\Delta_c^0\over \Delta_t^0(\xi_{NL}^{\perp})^2}\left(q_{\perp}\over q_z\right)^2
                   (\xi_{NL}^{\perp}/\xi_{\perp})^{-\eta_c}\right)\, ,
                   \nonumber\\
                \label{lightRC}
\end{eqnarray}
\begin{eqnarray}
 &~& L_{ij}^{\perp}(\vec{q})\left({CD\over Bq_0^2}{q_{\perp}^2\over
                 q_z^3}\right)\nonumber\\
                 &=& L_{ij}^{\perp}(\vec{q})\left({CK_0\over B_0 q_0^2}{q_{\perp}^2\over q_z^3}(\xi_{NL}^{\perp}/\xi_{\perp})^{-\eta_K}\xi_{\perp}^{-2}
                 \right)\nonumber\\
                 &\sim& L_{ij}^{\perp}(\vec{q})\left({\lambda^2\over (\xi_{NL}^{\perp})^2 q_0^2}{q_{\perp}^2\over
                 q_z^3}(\xi_{NL}^{\perp}/
                 \xi_{\perp})^{2-\eta_K}\right)\, .\nonumber\\
                 \label{lightRF1}
\end{eqnarray}
The crossovers between them is
obtained by equating them to each other. The function for the crossover (i.e., locus DC) between the
random compression and random field is given by
\begin{eqnarray}
 q_z &=& {\Delta_t^0\over \lambda^3 q_0^2
 \Delta_c^0}(\xi_{NL}^{\perp}/\xi_{\perp})^{2-\eta_K+\eta_c}\nonumber\\
 &=& {\Delta_t^0\over \lambda^3 q_0^2
 \Delta_c^0}(\xi_{NL}^{\perp}/\xi_{\perp})^{4+\eta_t-2\eta_K-\eta_3}\, ,
\end{eqnarray}
where we have used the {\it exact} scaling relation given in the Eq.
(\ref{Scalingeta_c}); the function for the crossover (i.e., locus CE)
between the random compression and random tilt is given by
\begin{eqnarray}
 q_z &=& \sqrt{\Delta_c^0/\Delta_t^0}(\xi_{NL}^{\perp}/\xi_{\perp})^{(\eta_c-\eta_t)/2}q_{\perp}\nonumber\\
 &=& \sqrt{\Delta_c^0/\Delta_t^0}(\xi_{NL}^{\perp}/\xi_{\perp})^{1-{{\eta_K+\eta_3}\over
 2}}q_{\perp}\, ,
 \label{CTCross}
\end{eqnarray}
where again we have used the same {\it exact} scaling relation given in Eq.
(\ref{Scalingeta_c}); the function for the crossover (i.e., locus BC)
between the random tilt and random field is given by
\begin{eqnarray}
 q_z = \lambda^3 q_0^2
 (\xi_{NL}^{\perp}/\xi_{\perp})^{\eta_K-2-\eta_t}q_{\perp}^2\, .
\end{eqnarray}
These three loci intersect at point C, at which
\begin{eqnarray}
 q_{\perp} &\propto & \xi_{\perp}^{{\eta_3\over 2}+{3\over 2}\eta_K-\eta_t-3}\, ,\nonumber\\
 q_z &\propto & \xi_{\perp}^{\eta_3+2\eta_K-\eta_t-4}\, .
\end{eqnarray}
It is easy to show that this point is within the triangle IOF, as
illustrated in Fig. \ref{fig: 8regions}. Thus, these three loci
divide the triangle IOF into three distinct regions with different
wavevector-dependence of $C_{ij}(\vec{q})$ for each region.
Specifically, $C_{ij}(\vec{q})$ is dominated by the random
compression and given by the Eq. (\ref{lightRC}) in region 5, and
dominated by the random tilt and given by Eq. (\ref{lightRT1}) in
region 4, and dominated by the random field and given by the Eq.
(\ref{lightRF1}) in region 2.

Now we consider the subregion (i.e., triangle OJF) below the locus OF, in which
$Bq_z^2$ is dominated by $Dq_{\perp}^2$ in the common denominator in
the Eq. (\ref{light3}). It is easy to show that the contribution to $C_{ij}(\vec{q})$
from the random compression never dominates in this region. The contributions from the random
tilt and the random field disorder are given by
\begin{eqnarray}
 &~& L_{ij}^{\perp}(\vec{q})\left(\Delta_t q_{\perp}^4\over D^2 q_{\perp}^4\right)\nonumber\\
                &=& L_{ij}^{\perp}(\vec{q})\left({\Delta_t^0\over K_0^2}\left(1\over q_{\perp}\right)^4(\xi_{NL}^{\perp}/\xi_{\perp})
                ^{2\eta_K-\eta_t}\right)\nonumber\\
                &\sim& L_{ij}^{\perp}(\vec{q})\left({\lambda\over (\xi_{NL}^{\perp})^2}\left(1\over
                 q_{\perp}\right)^4(\xi_{NL}^{\perp}/\xi_{\perp})^{2\eta_K-\eta_t}\right)\,
                 \nonumber\\
                 \label{lightRT2}
\end{eqnarray}
and
\begin{eqnarray}
 &~& L_{ij}^{\perp}(\vec{q})\left({CB^{1/2}\over D^{1/2}}{1\over
                     q_{\perp}}\right)\nonumber\\
                 &=& L_{ij}^{\perp}(\vec{q})\left({CB_0^{1/2}\over K_0^{1/2}\xi_{\perp}^{-1}}
                     {1\over q_{\perp}}(\xi_{NL}^{\perp}/\xi_{\perp})^{\eta_K/2}\right)\nonumber\\
                 &\sim& L_{ij}^{\perp}(\vec{q})\left({\xi_{NL}^{\perp}\over \lambda q_0^2}
                        {1\over q_{\perp}}(\xi_{NL}^{\perp}/\xi_{\perp})^{{\eta_K\over
                        2}-1}\right)\, ,
                 \label{lightRF2}
\end{eqnarray}
respectively. The crossover (i.e., locus AB)
between them is
\begin{eqnarray}
 q_{\perp} = q_{\perp}^R\propto\xi_{\perp}^{{3\over
 2}\eta_K-\eta_t-3}\, .
\end{eqnarray}
$C_{ij}(\vec{q})$ is dominated by the random field and given by
the Eq. (\ref{lightRF2}) in region 1, and dominated by the random tilt and
given by the Eq. (\ref{lightRT2}) in region 2.

Now we discuss the wavevector-dependence of
$C_{ij}\left(\vec{q}\right)$ in the $C$-side critical region. For
big $q$'s (i.e., $q_{\perp}\gg \xi_{\perp}^{-1}$ or
$q_z\gg\xi_z^{-1}$), the wavevector-dependence of
$C_{ij}\left(\vec{q}\right)$ is the same as that in the $A$-side
critical region. For small $q$'s (i.e.,
$q_{\perp}\ll\xi_{\perp}^{-1}$, $q_z\ll\xi_z^{-1}$),
$C_{ij}(\vec{q})$ is given by
\begin{eqnarray}
C_{ij}(\vec{q}) &=& L^{\perp}_{ij}(\hat{q})\left({{\Delta_{s'}
\left(\vec{q}\,', T \right)q_{s'}^2q_ {\perp}^2}\over
\left(\gamma(\vec{q}\,', T)q_{x'}^2 +
\tilde{B}q^2_{z'}+\tilde{K}\left (\vec{q}\,',
T\right)q^4_{s'}\right)^2}\right.\nonumber\\
&~&\left.+{\Delta_{z'}\left( T\right)q_{z'}^2q_{\perp}^2 \over
\left(\gamma(\vec{q}\,', T)q_{x'}^2 +
\tilde{B}q^2_{z'}+\tilde{K}\left(\vec{q}\,',T\right)q^4_{s'}\right)^2}\right.\nonumber\\
&~&\left.+{\Delta_{x'}\left(\vec{q}\,', T \right)q_{x'}^2q_{\perp}^2
\over \left(\gamma(\vec{q}\,', T)q_{x'}^2 +
\tilde{B}q^2_{z'}+\tilde{K}\left(\vec{q}\,', T
\right)q^4_{s'}\right)^2}\right)\, .\nonumber\\
\label{light4}
\end{eqnarray}
where we have neglected the contribution from the $\Delta_{x'z'}$ disorder (see
Hamiltonian (\ref{H_C})) since it is not important compared to the contributions
from the other disorders. For convenience, we name the three pieces in the above
equation as
``$\Delta_{s'}$-piece'', ``$\Delta_{z'}$-piece'' and ``$\Delta_{x'}$-piece'',
respectively. We find that there are five distinct regions
in the wavevector space with qualitatively different wavevector-dependences of
$C_{ij}\left(\vec{q}\right)$. Since the
azimuthal symmetry about $\hat{z}$-axis is now broken
due to the tilting of the layers, the five regions have to be illustrated
in a three-dimensional picture, which is shown in Fig. \ref{fig: 5regions}.
In the following we will show how we derive the five regions and the
wavevector-dependence of $C_{ij}\left(\vec{q}\right)$ for each region.

First we calculate the various crossovers within the three planes
(i.e., $q_{z'}$-$q_{x'}$, $q_{z'}$-$q_{s'}$, $q_{s'}$-$q_{x'}$). Let
us start with $q_{z'}$-$q_{x'}$ plane, in which case
$\Delta_{s'}$-piece vanishes. One crossover in this plane is the
scaling locus $OG$, which divides the rectangle $OHGI$ into two
distinct regions with different wavevector-dependences of
$\Delta_{s', x'}$, $K$ and $\gamma$. The function for $OG$ is given
by
\begin{eqnarray}
 q_{z'}\xi_z=\left(q_{x'}\xi_{\perp}\right)^{\tilde{\zeta}_{z'}/
 \tilde{\zeta}_{x'}}\, .\label{CrossOG}
\end{eqnarray}
The other one is the crossover (i.e., locus $OB$) between
$\Delta_{z'}$-piece and $\Delta_{x'}$-piece, which satisfies
\begin{eqnarray}
\Delta_{z'}q_{z'}^2=\Delta_{x'}q_{x'}^2\, , \label{z'x'}
\end{eqnarray}
where $\Delta_{z'}$ is wavevector-independent but
temperature-dependent and given in (\ref{Deltaz_c}), $\Delta_{x'}$
has dependences on both the temperature and wavevector, which is
summarized in (\ref{DeltaanomC}). Assuming that $OB$ is in one of
the two regions separated by $OG$ which is closer to $q_{z'}$-axis
(this assumption will be verified as a posterior), from (\ref{z'x'})
we obtain
\begin{eqnarray}
\left[\left(\xi_{NL}^{\perp}\over\lambda\right)^2\Delta_t^0+\Delta_c^0\right]
\left(\xi_{\perp}\over\xi_{NL}^{\perp}\right)^{\eta_c}q_{z'}^2\nonumber\\
=\Delta_t^0
\left(\xi_{\perp}\over\xi_{NL}^{\perp}\right)^{\eta_t}\left(
\xi_zq_{z'}\right)^{-{\tilde{\eta}_{x'}\over\tilde{\zeta}_{z'}}}
q_{x'}^2\, .\nonumber
\end{eqnarray}
A further reorganization leads to
\begin{eqnarray}
q_{z'}\xi_z=\left[\sqrt{\Delta_t^0\over\Delta_n^0}\left(\xi_{\perp}\over
\xi_{NL}^{\perp}\right)^{\eta_3/2}\left(\xi_{NL}^{\perp}\over\lambda\right)
 \left( q_{x'}\xi_{\perp}\right)\right]^{\tilde{\phi}_{x'z'}}\label{CrossOB}
\end{eqnarray}
where
\begin{eqnarray}
 \Delta_n^0 &\equiv& \left(\xi_{NL}^{\perp}\over\lambda\right)^2\Delta_t^0+\Delta_c^0
 \, ,\nonumber\\
 \phi_{x'z'}&\equiv&
 2\tilde{\zeta}_{z'}/(2\tilde{\zeta}_{z'}+\tilde{\eta}_{x'})\,
 .\nonumber
\end{eqnarray}
It is easy to verify that the function for the locus $OB$ given by
the Eq. (\ref{CrossOB}) is indeed consistent with the assumption we
just made.

The crossovers in the other two planes can be obtained through
similar calculations, which will not be repeated. We only give
the results. In $q_{z'}$-$q_{s'}$ plane, in which case
$\Delta_{x'}$-piece vanishes, the scaling locus (i.e., $OF$) and the
crossover (i.e., locus $OA$) between $\Delta_{z'}$-piece and
$\Delta_{s'}$- piece are given respectively by
\begin{eqnarray}
 q_{z'}\xi_z=\left(q_{s'}\xi_{\perp}\right)^{\tilde{\zeta}_{z'}}\,
 \label{CrossOF}
\end{eqnarray}
and
\begin{eqnarray}
 q_{z'}\xi_z=\left[\sqrt{\Delta_t^0\over\Delta_n^0}\left(\xi_{\perp}\over
 \xi_{NL}^{\perp}\right)^{\eta_3/2}\left(\xi_{NL}^{\perp}\over\lambda\right)\left(
 q_{s'}\xi_{\perp}\right)\right]^{\tilde{\phi}_{s'z'}}\,
 ,\nonumber\\
 \label{CrossOA}
\end{eqnarray}
where
\begin{eqnarray}
 \tilde{\phi}_{s'z'} \equiv
 2\tilde{\zeta}_{z'}/(2\tilde{\zeta}_{z'}+\tilde{\eta}_{s'})\, .
\end{eqnarray}
In $q_{x'}$-$q_{s'}$ plane, in which case $\Delta_{z'}$-piece vanishes,
coincidentally, the scaling locus and the crossover between
$\Delta_{x'}$-piece and $\Delta_{s'}$-piece merge into one (i.e.,
$OD$), and both are given by the same function:
\begin{eqnarray}
 q_{x'}\xi_{\perp}=\left(q_{s'}\xi_{\perp}\right)^{\tilde{\zeta}_{x'}}\,
 .\label{CrossOD}
\end{eqnarray}

Having found the crossovers in the three planes, now we can use them
to obtain the crossovers in 3D wavevector space. Moving the loci
$OA$ and $OB$ along $\hat{q}_{x'}$ and $\hat{q}_{s'}$, respectively,
we obtain the two boundaries $OAC$ and $OBC$ as the trajectories of
$OA$ and $OB$, which meet at the locus $OC$. Thus, $OAC$ and $OBC$
are the crossovers between $\Delta_{z'}$-piece and
$\Delta_{s'}$-piece, and $\Delta_{z'}$-piece and
$\Delta_{x'}$-piece, respectively, and they are also defined by the
functions (\ref{CrossOA}) and (\ref{CrossOB}), respectively, and
$OC$ is hence defined by the the combination of (\ref{CrossOA}) and
(\ref{CrossOB}). Likewise, $OEF$, $OEG$ and $ODE$ are the scaling
crossovers, which separate distinct regions with different
wavevector-dependence of $\Delta_{x', s'}$, $\gamma$ and $K$, and
are defined by the functions (\ref{CrossOF}), (\ref{CrossOG}) and
(\ref{CrossOD}), respectively. It is easy to show that these three
crossovers meet at the same locus $OE$, whose function is thus the
combination of any two of (\ref{CrossOF}), (\ref{CrossOG}) and
(\ref{CrossOD}). In addition, $ODE$ also serves as the crossover
between $\Delta_{x'}$-piece and $\Delta_{s'}$-piece due to the
duality of the locus OD. $OCE$ is another crossover between
$\Delta_{x'}$-piece and $\Delta_{s'}$-piece; however, since $OCE$
and $ODE$ are in different regions with different
wavevector-dependences of $\Delta_{x'}$ and $\Delta_{s'}$, $OCE$ is
defined by a function which is different from (\ref{CrossOD}):
\begin{eqnarray}
q_{x'}=\left(q_{z'}\xi_z\right)^{\tilde{\phi}_{s'x'}}q_{s'}
\end{eqnarray}
with
$\tilde{\phi}_{s'x'}=\left(2-\tilde{\eta}_K-\tilde{\eta}_{\gamma}\right)
/2\tilde{\zeta}_{z'}$.


After the definitions of the crossovers separating the five regions
in Fig. \ref{fig: 5regions} become known, the calculation of the
wavevector-dependence of $C_{ij}(\vec{q})$ in each region is
straightforward. In the region $O-ABCH$, $C_{ij}(\vec{q})$ is
dominated by $\Delta_{z'}$-piece and given by
\begin{eqnarray}
 &~&L^{\perp}_{ij}(\hat{q})\left({\Delta_{z'}\left( T\right)q_{z'}^2q_{\perp}^2 \over
 \left(\gamma(\vec{q}\,', T)q_{x'}^2 +
 \tilde{B}q^2_{z'}+\tilde{K}\left(\vec{q}\,',T\right)q^4_{s'}\right)^2}\right)\nonumber\\
 &=& L^{\perp}_{ij}(\hat{q})\left({\Delta_{z'}\left( T\right)q_{\perp}^2 \over
 \tilde{B}^2q^2_{z'}}\right)\nonumber\\
 &=& L^{\perp}_{ij}(\hat{q}) \left({\Delta_n^0\over B_0^2}
 {q_{\perp}^2\over
 q_{z'}^2}(\xi_{\perp}/\xi_{NL}^{\perp})^{\eta_c}\right)\nonumber\\
 &\sim& L^{\perp}_{ij}(\hat{q}) \left({\Delta_n^0\over \Delta_t^0}
 {\lambda^5\over (\xi_{NL}^{\perp})^2}{q_{\perp}^2\over q_{z'}^2}
 (\xi_{\perp}/\xi_{NL}^{\perp})^{\eta_c}\right)\, .
\label{}
\end{eqnarray}
In both of the regions $O-AFEC$ and $O-FJDE$, $C_{ij}(\vec{q})$ is
dominated by $\Delta_{s'}$-piece but with different
wavevector-dependences, since these two regions are separated by the
scaling crossover $OEF$.
In $O-AFEC$, $C_{ij}(\vec{q})$ is given by
\begin{eqnarray}
 &~&L^{\perp}_{ij}(\hat{q})\left({\Delta_{s'} \left(\vec{q}\,', T
 \right)q_{s'}^2q_ {\perp}^2}\over \tilde{B}^2q^4_{z'}\right)\nonumber\\
 &=& L^{\perp}_{ij}(\hat{q})\left({\Delta_t^0\over B_0^2}
 {q_{s'}^2q_{\perp}^2\over q_z^4}(\xi_{\perp}/\xi_{NL}^{\perp})^{\eta_t}
 (q_{z'}\xi_z)^{-\tilde{\eta}_{s'}/\tilde{\zeta}_{z'}}\right)\nonumber\\
 &\sim& L^{\perp}_{ij}(\hat{q})\left({\lambda^5\over (\xi_{NL}^{\perp})^2}
 {q_{s'}^2q_{\perp}^2\over q_z^4}(\xi_{\perp}/\xi_{NL}^{\perp})^{\eta_t}
 (q_{z'}\xi_z)^{-\tilde{\eta}_{s'}/\tilde{\zeta}_{z'}}\right)\, ,\nonumber\\
\end{eqnarray}
while in O-FJDE, it is given by
\begin{eqnarray}
 &~&L^{\perp}_{ij}(\hat{q})\left({\Delta_{s'} \left(\vec{q}\,', T
 \right)q_ {\perp}^2}\over K^2 q_{s'}^6\right)\nonumber\\
 &=&L^{\perp}_{ij}(\hat{q})\left({\Delta_t^0\over K_0^2}{q_{\perp}^2\over q_{s'}^6}
 (\xi_{\perp}/\xi_{NL}^{\perp})^{\eta_t-2\eta_K}(q_{s'}\xi_{\perp})^{2\eta_K-\eta_{s'}}
 \right)\nonumber\\
 &\sim& L^{\perp}_{ij}(\hat{q})\left({\lambda\over(\xi_{NL}^{\perp})^2}{q_{\perp}^2\over q_{s'}^6}
 (\xi_{\perp}/\xi_{NL}^{\perp})^{\eta_t-2\eta_K}(q_{s'}\xi_{\perp})^{2\eta_K-\eta_{s'}}
 \right)\, .\nonumber\\
\end{eqnarray}
Likewise, in both of the regions $O-CEBG$ and $O-EDIG$,
$C_{ij}(\vec{q})$ is dominated by $\Delta_{x'}$-piece but with
different wavevector-dependences.
In $O-CEGB$, it is given by
\begin{eqnarray}
 &~& L^{\perp}_{ij}(\hat{q})\left(\Delta_{x'}\left(\vec{q}\,', T
\right)q_{x'}^2q_{\perp}^2 \over \tilde{B}^2q^4_{z'}\right)\nonumber\\
 &=& L^{\perp}_{ij}(\hat{q})\left({\Delta_t^0\over B_0^2}
 {q_{s'}^2q_{\perp}^2\over q_z^4}(\xi_{\perp}/\xi_{NL}^{\perp})^{\eta_t}
 (q_{z'}\xi_z)^{-\tilde{\eta}_{x'}/\tilde{\zeta}_{z'}}\right)\nonumber\\
 &\sim& L^{\perp}_{ij}(\hat{q})\left({\lambda^5\over (\xi_{NL}^{\perp})^2}
 {q_{s'}^2q_{\perp}^2\over q_z^4}(\xi_{\perp}/\xi_{NL}^{\perp})^{\eta_t}
 (q_{z'}\xi_z)^{-\tilde{\eta}_{x'}/\tilde{\zeta}_{z'}}\right)\, ,\nonumber\\
\label{}
\end{eqnarray}
while in $O-EDIG$, it is
\begin{eqnarray}
 &~&L^{\perp}_{ij}(\hat{q})\left(\Delta_{x'}\left(\vec{q}\,', T
 \right)q_{\perp}^2 \over \gamma^2q_{x'}^2 \right)\nonumber\\
 &=&L^{\perp}_{ij}(\hat{q})\left({\Delta_t^0 (\xi_{NL}^{\perp})^4\over K_0^2}
 \left(q_{\perp}\over
 q_{x'}\right)^2(\xi_{\perp}/\xi_{NL}^{\perp})^{\eta_t+4-2\eta_K}\right.
 \nonumber\\
 &~&\left.\times(q_{x'}\xi_{\perp})^{-(2\tilde{\eta}_{\gamma}+\tilde{\eta}_K)/\tilde{\zeta}_{x'}}\right)
 \nonumber\\
 &\sim&L^{\perp}_{ij}(\hat{q})\left({\lambda (\xi_{NL}^{\perp})^2}
 \left(q_{\perp}\over
 q_{x'}\right)^2(\xi_{\perp}/\xi_{NL}^{\perp})^{\eta_t+4-2\eta_K}\right.
 \nonumber\\
 &~&\left.\times(q_{x'}\xi_{\perp})^{-(2\tilde{\eta}_{\gamma}+\tilde{\eta}_K)/\tilde{\zeta}_{x'}}\right)
 \, .
\end{eqnarray}

So far we have discussed $C_{ij}(\vec{q})$ for small $q$'s (i.e.,
$q_{\perp}\ll\xi_{\perp}^{-1}$, $q_z\ll\xi_z^{-1}$) in the $C$-side
critical region in the transformed reciprocal coordinate system.
However, what is needed for experiments is the behavior of
$C_{ij}(\vec{q})$ in the lab reciprocal coordinate system. Since the
relation between these two coordinate systems is known, the
calculation is straightforward but somewhat tedious. Instead of
listing $C_{ij}(\vec{q})$ for all $\vec{q}$s, which is quite
complicated and far more than necessary, we now discuss
$C_{ij}(\vec{q})$ for special $\vec{q}$s, which is
simple and adequate for making useful experiment predictions. First we
consider the case in which $\vec{q}$ is restricted in $q_z$-$q_x$
plane. In this case $\Delta_{s'}$-piece vanishes, and
$C_{ij}(\vec{q})$ reduces to
\begin{eqnarray}
 C_{xx}(\vec{q}) = {\Delta_{z'}\left( T\right)q_z^2q_x^2 \over
 \left(\gamma(\vec{q}\,', T)q_{x'}^2 +
 \tilde{B}q^2_z\right)^2}+{\Delta_{x'}\left(\vec{q}\,', T \right)q_{x'}^2q_x^2 \over
 \left(\gamma(\vec{q}\,', T)q_{x'}^2 + \tilde{B}q^2_z\right)^2}\,.
 \nonumber\\
 \label{light5}
\end{eqnarray}
The crossover between $\Delta_{z'}$ and $\Delta_{x'}$ has been
calculated previously and given by Eq. (\ref{CrossOB}), which, in
terms of $\vec{q}$, becomes
\begin{eqnarray}
 &~&q_z\xi_z\nonumber\\
 &=&\left[\sqrt{\Delta_t^0\over\Delta_n^0}\left(\xi_{\perp}\over
 \xi_{NL}^{\perp}\right)^{\eta_3/2}\left(\xi_{NL}^{\perp}\over\lambda\right)
 \left( q_x-\Gamma
 q_z\right)\xi_{\perp}\right]^{\tilde{\phi}_{x'z'}}\nonumber\\
 &=&\left[\sqrt{\Delta_t^0\over\Delta_n^0}\left(\xi_z\over \Gamma\right)
 \left( q_x-\Gamma q_z\right)\right]^{\tilde{\phi}_{x'z'}}
 \, .
 \label{}
\end{eqnarray}
This function defines the loci $OE$ and $OE'$ in Fig. \ref{fig:
10regions}. The coordinates of $E$ and $E'$ are given respectively
by
\begin{eqnarray}
 q_z &\sim& (\xi_z)^{-1}\, ,\nonumber\\
 q_{\perp} &\sim& \pm q_{\perp}^F
 \equiv\pm (\xi_z)^{-1}
 (\xi_{\perp}/\xi_{NL}^{\perp})^{1-{{\eta_K+\eta_3}\over 2}}\, ,
\end{eqnarray}
and $OE$ and $OE'$ thus connect with $EG$ and $E'G'$ respectively,
which are the crossover between the random tilt and random
compression fluctuations for big $q$'s (i.e.,
$q_{\perp}\gg\xi_{\perp}^{-1}$ or $q_z\gg\xi_z^{-1}$). The crossover
between the two terms in the common denominator has also been
calculated and given by Eq. (\ref{CrossOG}), which, in terms of
$\vec{q}$, is
\begin{eqnarray}
 q_z\xi_z=\left[(q_x+\Gamma q_z)\xi_{\perp}\right]^{\tilde{\zeta}_{z'}/
 \tilde{\zeta}_{x'}}\, .\label{}
\end{eqnarray}
It is easy to check that the loci defined by this function satisfies
$|q_x| \gg \Gamma q_z$. Thus this function can be further simplified
as
\begin{eqnarray}
 q_z\xi_z=\left(q_x\xi_{\perp}\right)^{\tilde{\zeta}_{z'}/
 \tilde{\zeta}_{x'}}\label{CrossOE}\, ,
\end{eqnarray}
which defines $OF$ and $OF'$ in Fig. \ref{fig: 10regions}. These two
Loci also serve as the scaling crossover separating the region for
small $q$'s into three distinct ones with different
wavevector-dependences of $\Delta_{x'}$. Now we calculate
$\vec{q}$-dependences of $C_{xx}(\vec{q})$ in different regions
separated by these crossovers. In region 4, $C_{xx}(\vec{q})$
is dominated by $\Delta_{z'}$-piece and given by
\begin{eqnarray}
 &~&\Delta_{z'}\left( T\right)q_x^2 \over \tilde{B}^2q^2_z\nonumber\\
 &=&{\Delta_n^0\over B_0^2}\left(\xi_{\perp}\over
 \xi_{NL}^{\perp}\right)^{\eta_c}
 \left(q_x\over q_z\right)^2\nonumber\\
 &\sim&{\Delta_n^0\over \Delta_t^0}{\lambda^5\over (\xi_{NL}^{\perp})^2}
 \left(\xi_{\perp}\over \xi_{NL}^{\perp}\right)^{\eta_c}
 \left(q_x\over q_z\right)^2\, ;
\end{eqnarray}
in regions 5 and 6, $C_{xx}(\vec{q})$ is dominated by
$\Delta_{x'}$-piece and given by
\begin{eqnarray}
 &~&\Delta_{x'}\left(\vec{q}\,', T \right)q_{x'}^2q_x^2 \over \tilde{B}^2q^4_z
 \nonumber\\
 &=& {\Delta_t^0\over B_0^2}
 \left(\xi_{\perp}\over\xi_{NL}^{\perp}\right)^{\eta_t}
 {q_{x'}^2q_x^2 \over q^4_z}(q_z\xi_z)^{-\tilde{\eta}_{x'}/\tilde{\zeta}_{z'}}
 \nonumber\\
 &\sim&{\lambda^5\over (\xi_{NL}^{\perp})^2}
 \left(\xi_{\perp}\over\xi_{NL}^{\perp}\right)^{\eta_t}
 {q_x^4 \over q^4_z}(q_z\xi_z)^{-\tilde{\eta}_{x'}/\tilde{\zeta}_{z'}}\, ;
 \label{Region56}
\end{eqnarray}
in regions 7 and 8, $C_{xx}(\vec{q})$ is also dominated by
$\Delta_{x'}$-piece but given by
\begin{eqnarray}
 &~&\Delta_{x'}\left(\vec{q}\,', T \right)q_x^2 \over
 \gamma^2 q_{x'}^2\nonumber\\
 &=& {\Delta_t^0\xi_{\perp}^4\over K_0^2}
 \left(\xi_{\perp}\over \xi_{NL}^{\perp}\right)^{\eta_t-2\eta_K}
 (q_x\xi_{\perp})^{-(\tilde{\eta}_{x'}+2\tilde{\eta}_{\gamma})/\tilde{\zeta}_{x'}}
 \nonumber\\
 &\sim&{\lambda\xi_{\perp}^4\over (\xi_{NL}^{\perp})^2}
 \left(\xi_{\perp}\over \xi_{NL}^{\perp}\right)^{\eta_t-2\eta_K}
 (q_x\xi_{\perp})^{-(\tilde{\eta}_{x'}+2\tilde{\eta}_{\gamma})/\tilde{\zeta}_{x'}}\,
 ,
 \nonumber\\
\end{eqnarray}
which is different from Eq. (\ref{Region56}) since regions 5 and 6
are separated from regions 7 and 8 by the scaling loci. In both of
the above two equations we approximated $q_{x'}$ as $q_x$, since we
have $|q_x|\gg \Gamma q_z$ in regions 5, 6, 7, 8.

The second special case we consider is when $\vec{q}$ is along
$\hat{q}_x$-axis. In this case both $\Delta_{z'}$-piece and
$\Delta_{x'}$-piece vanish, and $C_{ij}(\vec{q})$ is simply given by
\begin{eqnarray}
 C_{ss}(\vec{q})
 &=& \Delta_{s'}\over \tilde{K}^2q^4_s\nonumber\\
 &=& {\Delta_t^0\over K_0^2}
     \left(\xi_{\perp}\over\xi_{NL}^{\perp}\right)^{\eta_t-2\eta_K}
     (q_s\xi_{\perp})^{2\tilde{\eta}_K-\tilde{\eta}_t}
     \left(1\over q_s\right)^4
     \nonumber\\
 &\sim& {\lambda\over (\xi_{NL}^{\perp})^2}
     \left(\xi_{\perp}\over\xi_{NL}^{\perp}\right)^{\eta_t-2\eta_K}
     (q_s\xi_{\perp})^{2\tilde{\eta}_K-\tilde{\eta}_t}
     \left(1\over q_s\right)^4\, .\nonumber\\
\end{eqnarray}

\section{\label{sec: Stability}Stability of Transition Against Defects and Orientational
disorder}
The stability of the $A$ phase follows from the
stability of ``random field $XY$ model''; the stability of the $C$ phase follows from the
stability of ``$m=1$ smectic''. Both phases are
stable against orientational fluctuations and the unbinding of neutral pairs of topological
defects (i.e., smectic dislocation loops).
This implies the existence of the phase transition between
the two phases. However, in order for our theory for
the critical behavior to be valid, the
system also needs to be dislocation-bound and orientationally ordered right at
the critical point.

The orientational fluctuations are defined as the mean square
fluctuations of the layer normal $\hat{N}$. Assuming the system is orientationally
ordered, the dominating orientational fluctuations can be calculated as
\begin{eqnarray}
\left< \left|\vec{N}_{\perp} (x) \right|^2 \right> = \int {d^dq
\over  (2 \pi)^d} { \Delta(\vec{q}) q^4_{\perp} \over
\left(K(\vec{q}) q^4_{\perp} + B q^2_z\right)^2 }\, , \label{I15}
\end{eqnarray}
where $\vec{N}_{\perp}$ is the projection of $\hat{N}$ on the $\perp$ plane
(i.e., $x$-$y$ plane). We have only kept the fluctuations
contributed by the random tilt disorder, which are the most divergent
right at the critical point. The $\vec{q}$-dependences
of $\Delta(\vec{q})$ and $K(\vec{q})$ are given in Eqs. (\ref{K}) and
(\ref{Deltacrit}), respectively. $B$ is $\vec{q}$-independent.
Then requiring the fluctuations to be finite leads to
\begin{eqnarray}
\eta_t - {3\eta_K \over 2} < d - 3\, , \label{I16}
\end{eqnarray}
which is one of the conditions for our theory for the critical
behavior to be valid.

Now we check if the topological defects are bound right at the critical
point. The start point of the theory is the tilt only Hamiltonian:
\begin{eqnarray}
H &=& \int d^dr  \left[{ K \over 2}(\nabla^2_{\perp}u)^2 + {B
\over
2}(\partial_zu)^2+\vec{h}\left(\vec{r}\right)\cdot\vec{\nabla}u\right]
\, .
\nonumber\\
\label{Squadratic}
\end{eqnarray}
where the random field disorder is not included. It can be justified
that the random field disorder is also irrelevant when
dislocations are included. Since the model (\ref{Squadratic}) is virtually
the same as that for dislocations in smectic $A$ phase in
an isotropic random environment \cite{LC}, the theory is also
the same. Therefore, we will describe the theory very briefly.

When the smectic has a dislocation, the displacement field $u$ is
no longer single valued. Mathematically this can be represented as
\begin{eqnarray}
\vec{\nabla}\times\vec{\nabla} u=\vec{m} \label{defineM}
\end{eqnarray}
with
\begin{eqnarray}
\vec{m}(\vec{r})=\sum_i\int{aM_it(s_i)\delta^3\left(\vec{r}-
\vec{r}_i(s_i)\right)ds_i}\, ,
\end{eqnarray}
where $s_i$ stands for the whole configuration of the $i$'th
dislocation loop, $M_i$ is an integer giving the charge of that
loop, $t(s_i)$ is the local tangent of the loop, and $\vec{r}_i$
is a parametrization of the path of the loop. Furthermore equation
(\ref{defineM}) implies
\begin{eqnarray}
 \vec{\nabla}\cdot\vec{M}=0,
\end{eqnarray}
which means that dislocation lines can not end in the bulk of the
sample.

To obtain a dislocation Hamiltonian, we need to trace over field
$u$ which is constrained by equation (\ref{defineM}). This can be
done in the following standard way \cite{LC}. We separate the
field $\vec{v} = \vec{\nabla}u$ into
\begin{eqnarray}
 \vec{v} =  \vec{v}_d + \delta\vec{v}
 \label{decomposition},
\end{eqnarray}
where $\vec{v}_p$ minimizes Hamiltonian (\ref{Squadratic}) for a
given dislocation configuration $\vec{m}(\vec{r})$,
$\delta\vec{v}$ can be viewed as the fluctuation from the ground
state. Inserting (\ref{decomposition}) into Hamiltonian
(\ref{Squadratic}), we find that $\vec{v}_p$ and $\delta\vec{v}$
are decoupled due to the construction that $\vec{v}_d$
minimizes Hamiltonian (\ref{Squadratic}). Thus we obtain the
effective model for $\vec{m}(\vec{r})$.

Now let us go through the procedure. The Euler-Lagrange equation,
obtained by minimizing Hamiltonian (\ref{Squadratic}), is
\begin{eqnarray}
 (B\partial_z^2 + K\nabla^4_{\perp})u +
 \vec{\nabla}_{\perp}\cdot\vec{h}=0.
\end{eqnarray}
Rewriting this in terms of $\vec{v}_d=\vec{\nabla}u$ gives
\begin{eqnarray}
 \partial_z
 v_d^z-\lambda^2\nabla^2_{\perp}\vec{\nabla}_{\perp}\cdot
 \vec{v}_d^{\perp}+{1\over B}\vec{\nabla}_{\perp}\cdot\vec{h}=0\, ,
\end{eqnarray}
where $\lambda^2\equiv K/B$. In Fourier space, this equation becomes
\begin{eqnarray}
 q_zv_d^z+\lambda^2 q_{\perp}^2 \vec{q}_{\perp} \cdot
 \vec{v}_d^{\perp}+{1\over
 B}\vec{q}_{\perp}\cdot\vec{h}(\vec{q})=0\, ,
 \label{minimizeq}
\end{eqnarray}
and the solution for the constraint (\ref{defineM}) is
\begin{eqnarray}
 \vec{v}_d = {i\vec{q}\times\vec{m}\over q^2}+\vec{q}\phi\, ,
 \label{Solution}
\end{eqnarray}
where $\phi$ is the smooth elastic distortion around the
dislocation line. Inserting (\ref{Solution}) into equation
(\ref{minimizeq}) gives
\begin{eqnarray}
 \phi=-{iq_z(1-\lambda^2q_{\perp}^2)\epsilon_{zij}q_im_j\over\Gamma_qq^2}
 -{\vec{q}_{\perp}\cdot\vec{h}\over B\Gamma_q}\, ,
 \label{Phi}
\end{eqnarray}
where we have defined the inverse of the smectic propagator
\begin{eqnarray}
 \Gamma_q = q_z^2 + \lambda^2 q_{\perp}^4\, .
\end{eqnarray}
Inserting (\ref{Phi}) into the solution for $\vec{v}_d$ and plugging equation
(\ref{decomposition}) into Hamiltonian (\ref{Squadratic}), we obtain
the dislocation Hamiltonian
\begin{eqnarray}
H_d=\int{d^d q}\left[{Kq_{\perp}^2\over
2\Gamma_q}P_{ij}^{\perp}m_im_j+\vec{m}\cdot\vec{a}\right],
 \label{defectH}
\end{eqnarray}
where $P_{ij}^{\perp}=(1-\delta_{iz})(1-\delta_{iz})(\delta_{ij}-
q_i^{\perp}q_j^{\perp}/q_{\perp}^2)$,
$\Gamma_q=q_z^2+\lambda^2q_{\perp}^4$, and
\begin{eqnarray}
\vec{a}=i\left[{\vec{q}\times\vec{h}\over
q^2}-{\left(\hat{z}\times\vec{q}\right)\cdot\vec{h}\over\Gamma_qq^2}
q_z\left(1-\lambda^2q_{\perp}^2\right)\right].
 \label{fielda}
\end{eqnarray}

By putting the model on a a simple cubic lattice, the partition
function can be written as
\begin{eqnarray}
 Z=\sum_{\vec{m}{\vec{r}}}e^{-S[\vec{m}]}\, ,
 \label{Partition1}
\end{eqnarray}
where
\begin{eqnarray}
 \vec{m}(\vec{r}) &=& {a\over d^2}[n_x(\vec{r}), n_y(\vec{r}),
 n_z(\vec{r})],
 \label{discretem}
 \\
 S[\vec{m}] &=&{1\over T}\left[H_d[\vec{m}]+{E_cd^4\over
 a^2}\sum_{\vec{r}}|\vec{m}(\vec{r})|^2\right],
\end{eqnarray}
where the $n_i$'s are integers, $d$ is the cubic lattice
constants used in the discretization, and ${E_cd^4\over
 a^2}\sum_{\vec{r}}|\vec{m}(\vec{r})|^2$ is the core energy term
 coming from the core of the defect line that is not accurately
 treated by the continuum elastic theory.

To cope with the constraint $\vec{\nabla}\cdot\vec{m} = 0$, we
introduce an auxiliary field $\theta(\vec{r})$, rewriting the
partition function equation (\ref{Partition1}) as
\begin{eqnarray}
 Z = \prod_{\vec{r}}\int d\theta(\vec{r}) \sum_{\vec{m}(\vec{r})}
 e^{-S[\vec{m}]+i\sum_{\vec{r}}\theta(\vec{r})\vec{\nabla}\cdot
 \vec{m}(\vec{r})d^2/a},
\end{eqnarray}
where the sum over $\vec{m}(\vec{r})$ is now unconstrained.

Then we introduce a dummy vector field $\vec{A}$ to mediate the
long-range interaction between defect loops in the Hamiltonian
(\ref{defectH}). This is accomplished by rewriting the partition
function as
\begin{eqnarray}
 Z = \prod_{\vec{r}}\int d\theta(\vec{r})d\vec{A}(\vec{r})
 \sum_{\vec{m}(\vec{r})}
 e^{-S[\vec{m}, \theta, \vec{a}]}\delta(\vec{\nabla}\cdot\vec{A})
 \delta(A_z)\nonumber\\
 \label{Partition2}
\end{eqnarray}
with
\begin{eqnarray}
 S&=&{1\over T}\sum_{\vec{r}}\left[\vec{m}\cdot \left(-i{Td^2\over
 a}\vec{\nabla}\theta(\vec{r})+d^3[i\vec{A}(\vec{r})+\vec{a}(\vec{r})]\right)
 \right.\nonumber\\
 &~&\left.+E_c{d^4\over a^2}|\vec{m}|^2\right]+{1\over 2T}
 \sum_{\vec{q}}{\Gamma_q\over Kq_{\perp}^2}|\vec{A}|^2.
\end{eqnarray}

Now the sum over $\vec{m}(\vec{r})$ is recognized to be the
``periodic Gaussian'' made by Villain \cite{Villain}. Then the
partition function equation (\ref{Partition2}) can be rewritten as
\begin{eqnarray}
 Z &=& \prod_{\vec{r}}\int d\theta(\vec{r})d\vec{A}(\vec{r})
 \ \ \delta(\vec{\nabla}\cdot\vec{A})\delta(A_z)
 \nonumber\\
 &~&\times \exp\left[-\sum_{\vec{r}i}V_p
 \left[\theta(\vec{r}+\hat{x}_i)-\theta(\vec{r})
 -{ad\over T}[A_i(\vec{r})\right.\right.\nonumber\\
 &~&\left.\left.-ia_i(\vec{r})]\right]
 -{1\over 2T}\sum_{\vec{q}}{\Gamma_q\over
 Kq_{\perp}^2}|\vec{A}|^2\right],
 \label{dualmodel}
\end{eqnarray}
where the $2\pi$-period Villain potential $V_p(x)$ is defined as
\begin{eqnarray}
 e^{-V_p(x)} \equiv \sum_{-\infty}^{\infty}
 e^{-n^2E_c/T + ixn}.
\end{eqnarray}
Since $V_p(x)$ has sharper minima for {\it smaller} $E_c/T$,
which corresponds to higher temperature, raising the temperature
in the original model is equivalent to lowering temperature in the
dual model equation (\ref{dualmodel}).

Standard universality class arguments imply that the model
equation (\ref{dualmodel}) has the same universality class as the
``soft spin'', or Landau-Ginsburg-Wilson, with the {\it complex}
``action''
\begin{eqnarray}
S_r&=&\sum_{r,\alpha}\left[{c\over 2}[\vec{\nabla}+{ad\over
T}(i\vec{A}_{\alpha}+\vec{a})t]\phi_{\alpha}^*\cdot
[\vec{\nabla}-{ad\over T}(i\vec{A}_{\alpha}+\vec{a})]\right.\nonumber\\
&~&\left.\times\phi_{\alpha}+t_d|\phi_{\alpha}|^2+u_d|\phi_{\alpha}|^4\right]
+\sum_{q,\alpha}{\Gamma_q\over
2TKq_{\perp}^2}|\vec{A}_{\alpha}|^2\, ,\nonumber\\
 \label{softspin}
\end{eqnarray}
where $\phi$ is a complex order parameter whose phase is
$\theta(\vec{r})$. Because of the duality transformation's
inversion of the temperature axis, the reduced temperature $t_d$
is a monotonically decreasing function of the temperature $T$ (of
the original dislocation loop model), which vanishes at the
mean-field transition temperature of the dislocation model
({\ref{dualmodel}). Disorder is included in the model
(\ref{softspin}) through $\vec{a}(\vec{r})$, which is related to
the random tilt field $\vec{h}(\vec{r})$ by equation
(\ref{fielda}).

Because of the duality inversion of the temperature axis,
the ordered phase of the dual model (\ref{softspin})
corresponds to the disordered phase (dislocation loops unbound) of
the original dislocation model.

An complete analysis of the dislocation loop unbinding transition
described by the model (\ref{softspin}) is beyond this
paper. The goal here is to know if a dislocation bound state ever
exists near the critical region of the smectic $A$-to-$C$
transition. To answer that question, let us check the one-loop
graphical correction to the dual temperature:
\begin{eqnarray}
t_R=t_0-{(d-2)ca^2d^2\over T^2}\int{{d^dq\over
(2\pi)^d}{\Delta_tq_z^2q_{\perp}^2\over q^2\Gamma^2_q}}\, .
 \label{Correctiont}
\end{eqnarray}
It is easy to see that the integral diverges when $K$ and $\Delta_t$
are considered as constants. This implies that $t_R$ is always negative and the dual model
is always in its ordered state, which corresponds to the
dislocation unbound state. This implies that the system would have
become disorder {\it before} reaching the critical point, which
signals either a first order transition, or a reentrant
nematic phase intervening between the $A$ and $C$ phase. However,
this conclusion only holds within the {\it harmonic} approximation.
In section \ref{sec: ACTransition} we have shown that the anharmonic
effects are important. A crude way
to include the effects is to treat $K(\vec{q})$ and
$\Delta_t(\vec{q})$ anomalous. Implementing this in the integral
and requiring it to be finite, we obtain a
restriction on $\eta_K$ and $\eta_t$:
\begin{eqnarray}
2\eta_t-\eta_K<0\, ,
\end{eqnarray}
which is another condition for our theory to be valid.

We test the two conditions using the values of
$\eta_K$ and $\eta_t$ obtained by the $\epsilon$-expansion to $O(\epsilon^2)$.
Unfortunately, the two conditions are not satisfied in the physical
dimension $d=3$, which seems
to imply that the second-order phase transition is not stable. However, since
$\epsilon=2$ in $d=3$, $\epsilon$-expansion is not quantitatively reliable,
and therefore whether the second-order phase transition is stable in $d=3$ remains an
open question.


\section{CONCLUSION}
In summary, a theory of smectic $A-C$ phase transition in
anisotropic disordered media is developed. We show that the phase
transition can be second-order and calculated the critical
exponents to the first order in an $\epsilon = 5 - d$ expansion.
In addition, the elasticity and fluctuations of this system (at
the phase-transition temperature) are studied.  The implications
of these results are expected to be observable.

\appendix
\section{\label{sec: Average}}
In this appendix, we review the derivation in reference
\cite{Karl} of the numerical estimates of the anomalous elastic
exponents for the C phase. These are obtained from two different
$\epsilon-$expansions, based on two different analytic
continuations of the model to higher dimensions. In the ``hard''
continuation, the higher dimensional ($d>3$) model is chosen to
still have only one ``soft'' direction for all spatial dimensions
$d$, as it does in $d=3$. In this case, the exponents are given by
\begin{eqnarray}
\tilde{\zeta}_{x'}&=&2-{{\tilde{\eta}_{\gamma}+\tilde{\eta}_K}\over
2}\, ,\\
\tilde{\zeta}_{z'}&=&2-{\tilde{\eta}_K\over 2}\, ,\\
\tilde{\eta}_K&=&{16\over
15}\tilde{\epsilon}+O(\tilde{\epsilon}^2)\label{hard1}\, ,\\
\tilde{\eta}_{\gamma}&=&{4\over
5}\tilde{\epsilon}+O(\tilde{\epsilon}^2)\label{hard2}\, ,\\
\tilde{\eta}_{s'}&=&{4\over
15}\tilde{\epsilon}+O({\tilde{\epsilon}}^2)\label{hard3}
\end{eqnarray}
with $\tilde{\epsilon} = {7\over 2} - d$, and obey two exact
scaling relations:
\begin{eqnarray}
-{1\over2}\tilde{\eta}_{\gamma}-{{7-d}\over
2}\tilde{\eta}_K+\tilde{\eta}_{s'}=2d-7 \label{Ascaling}.
\end{eqnarray}
In addition, by analytically continuing the problem to higher
dimensions ($d>3$) in a different way (namely, by keeping the
number of {\it hard} directions fixed at 2, as opposed to the
$d={7\over 2}-\tilde{\epsilon}$ expansion, which is based on
keeping the number of the {\it soft} directions fixed at one),
this problem was been studied in \cite{Karl} using a
$d=4-\tilde{\tilde{\epsilon}}$ expansion, yielding the following
exponents and the exact scaling relation:
\begin{eqnarray}
\tilde{\eta}_K&=&{3\over
8}\tilde{\tilde{\epsilon}}+O({\tilde{\tilde{\epsilon}}}^2)\, ,\label{soft1
}\\
\tilde{\eta}_{\gamma}&=&{3\over
4}\tilde{\tilde{\epsilon}}+O({\tilde{\tilde{\epsilon}}}^2)\, ,\label{soft2
}\\
\tilde{\eta}_{s'}&=&{1\over
8}\tilde{\tilde{\epsilon}}+O({\tilde{\tilde{\epsilon}}}^2)\, ,\label{soft3
}\\
4-d+\tilde{\eta}_{s'}&=&{\eta_{\gamma}\over
2}+2\tilde{\eta}_K\label{soft4}\, .
\end{eqnarray}
As argued in \cite{Karl}, combining the results from the
$d=4-\tilde{\tilde{\epsilon}}$ and the earlier $d={7\over
2}-\tilde{\epsilon}$ expansion we see that the exponents
$\tilde{\eta}_K$ and $\tilde{\eta}_{s'}$ are more consistent
between the two. This suggests that the most accurate estimate of
all $3$ exponents will be obtained by taking $\tilde{\eta_K}$ and
$\eta_s$ from weighted average of the $4-\tilde{\tilde{\epsilon}}$
and ${7\over 2}-\tilde{\epsilon}$ exponents:
\begin{eqnarray}
\tilde{\eta}_K={{4\tilde{\eta}_K({7\over2}-\tilde{\epsilon})+\tilde{\eta}_K(4-\tilde{\tilde{\epsilon}})}\over
5}\label{average1}\, ,\\
\tilde{\eta}_{s'}={{4\tilde{\eta}_{s'}({7\over2}-\tilde{\epsilon})+\tilde{\eta}_{s'}(4-\tilde{\tilde{\epsilon}})}\over
5}\, ,\label{average2}
\end{eqnarray}
(the factor of 4 appearing because, a priori, we expect the
${7\over 2}-\tilde{\epsilon}$ expansion to be 4 times as accurate
as the $4-\tilde{\tilde{\epsilon}}$ expansion, since the errors in
both are $O(\epsilon^2)$, and $\tilde{\tilde{\epsilon}}(=1)$ is
twice as big as $\tilde{\epsilon}(={1\over 2})$), and then
obtaining $\eta_{\gamma}$ from the {\it exact} scaling relation
(\ref{Ascaling}) in $d=3$ (note that (\ref{soft4}) reduces to
(\ref{Ascaling}) in $d=3$). One can also estimate the errors in
the exponents obtained in this way  as the differences between the
weighted averages (\ref{average1}, \ref{average2}) and the
${7\over 2}-\tilde{\epsilon}$ expansion results. Doing all of
this \cite{Karl} obtains
\begin{eqnarray}
\tilde{\eta}_K&=&0.50\pm0.03\, ,\\
\tilde{\eta}_{\gamma}&=&0.26\pm0.12\, ,\\
\tilde{\eta}_{s'}&=&0.132\pm0.002\, .
\end{eqnarray}
In addition, inserting $\tilde{\eta}_K$, $\tilde{\eta}_{\gamma}$,
$\tilde{\eta}_{s'}$ into the {\it exact} scaling relation Eq.
(\ref{scaling3}) gives
\begin{eqnarray}
 \tilde{\eta}_{x'}=1.37\pm0.15\, .
\end{eqnarray}

\section{\label{Sec: Gamma}}
During the discussion of the universality class of the $C$ phase, we
have shown the advantage of writing the model in a special
coordinate. The relation between this coordinate and the lab one is
fully determined by the parameter $\Gamma$, which is defined by
(\ref{Gamma}). Now we show that near the critical point, $\Gamma$ is
also renormalized by the critical fluctuations and hence strongly
temperature-dependent. Using the critical RG, the renormalized
$\Gamma$ can be calculated as
\begin{eqnarray}
 \Gamma = e^{(\omega-1)\ell^*}\Gamma(\ell^*)\, ,
 \label{TrajecGamma}
\end{eqnarray}
where the RG is stopped at $\ell^*=\ln(\Lambda\xi_{\perp})$, and the
prefactor on the right-hand comes from the dimensional rescaling. To
calculate $\Gamma(\ell^*)$, it is convenient to reorganize
(\ref{Gamma}) as
\begin{eqnarray}
 \Gamma(\ell^*)=\sqrt{-{{g_3(\ell^*)B(\ell^*)}\over{g_4(\ell^*)D(\ell*)}}}
              \sqrt{w(\ell^*)\over w'(\ell^*)}\, ,
 \label{GammaFlow1}
\end{eqnarray}
where $g_3$ and $g_4$ are defined by (\ref{Defineg3}) and
(\ref{Defineg4}), respectively. Previous calculation shows that for
large $\ell^*$ (which is true near the critical region),
$g_4(\ell^*)=32\epsilon/15$ and $g_3(\ell^*)\sim e^{-\eta_3\ell^*}$,
where $\eta_3$ is defined via
\begin{eqnarray}
{{d g_3(\ell)}\over {d \ell}} = -\eta_3 g_3(\ell) \label{eta_3}.
\end{eqnarray}
According to the linearized flow equation of $g_3$ given by
(\ref{deltag3}), $\eta_3$ is given, to the leading order in
$\epsilon=5-d$, by
\begin{eqnarray}
 \eta_3={1\over 5}\epsilon+O(\epsilon^2)\, .
\end{eqnarray}
To calculate $w'(\ell^*)$, we rewrite (\ref{Definitionw'}) as
\begin{eqnarray}
w'(\ell^*)=w(\ell^*)\left[1-{g_3(\ell^*)\over g_4(\ell^*)}\right]\,
.
\end{eqnarray}
Since $g_3(\ell^*)/g_4(\ell^*)$ is much less than 1, the above
equation implies $w(\ell^*)/w'(\ell^*)$ = 1. Based on these
arguments, we obtain from (\ref{GammaFlow1})
\begin{eqnarray}
 \Gamma(\ell^*) \sim e^{-\eta_3\ell^*/2}\sqrt{B(\ell^*)\over D(\ell^*)}
        = {e^{-\eta_3\ell^*/2}\over \Lambda}\sqrt{B(\ell^*)\over K(\ell^*)}\,
        ,
 \label{GammaFlow2}
\end{eqnarray}
where $B(\ell^*)$ and $K(\ell^*)$ can be calculated by integrating
the RG flow equations (\ref{10}, \ref{11}). Plugging the above
equation into (\ref{TrajecGamma}), we get
\begin{eqnarray}
 \Gamma \sim e^{(1-(\eta_K+\eta_3)/2)\ell^*}
        \sim (\xi_{\perp}\Lambda)^{1-{\left(\eta_K+\eta_3\right)\over 2}}\,
        .
\end{eqnarray}

\section{\label{sec: Table}}
\begin{widetext}
\begin{table}[h]  \caption{\label{tab: TableB}Overview of the Elastic Constants and Disorder Variances}
\begin{tabular}
{|c|c|c|}\hline  Symbols & Description& Eq.
\\\hline\hline

 $B$&Compression modulus in Hamiltonian Eq. (\ref{H_AC})&$B=B_0$\\\hline

$K$ &Bend modulus in Hamiltonian Eq. (\ref{H_AC})& (\ref{K})\\\hline

$\Delta_t$ & Random tilt disorder variance in Hamiltonian Eq.
(\ref{H_AC})& (\ref{Deltacrit})\\\hline

$\Delta_c$ &Random compression disorder variance in Hamiltonian Eq.
(\ref{H_AC})& (\ref{Deltacrit})\\\hline

$B_0$ &Bare value of $B$&
\\\hline

$K_0$ &Bare value of $K$&
\\\hline

$\Delta_t^0$ &Bare value of $\Delta_t$&
\\\hline

$\Delta_c^0$ &Bare value of $\Delta_c$&
\\\hline

$\tilde{B}$ &Compression modulus in Hamiltonian Eq. (\ref{H_C})&
$\tilde{B}=B_0$\\\hline

$\tilde{K}$ &Bend modulus in Hamiltonian Eq. (\ref{H_C})&
(\ref{KanomC})\\\hline

$\gamma$ &Tilt modulus along $\hat{x'}$ in Hamiltonian Eq.
(\ref{H_C})& (\ref{gammaanomC})\\\hline

$\Delta_{s'}$ &Random tilt disorder variance along $\hat{s'}$ in
Hamiltonian Eq. (\ref{H_C})& (\ref{DeltaanomC})\\\hline

$\Delta_{x'}$ &Random tilt disorder variance along $\hat{x'}$ in
Hamiltonian Eq. (\ref{H_C})& (\ref{DeltaanomC})\\\hline

$\Delta_{z'}$ &Random compression disorder variance in Hamiltonian
Eq. (\ref{H_C})& $\Delta_{z'}=\Delta_{z'}^c$\\\hline

$\tilde{B}_c$ &Half-dressed value of $\tilde{B}$&$\tilde{B}_c=B_0$
\\\hline

$\tilde{K}_c$ &Half-dressed value of $\tilde{K}$&$(\ref{K_c})$
\\\hline

$\gamma_c$ &Half-dressed value of $\gamma$&(\ref{gamma_c})
 \\\hline

$\Delta_{s'}^c$ &Half-dressed value of
$\Delta_{s'}$&(\ref{Deltasx_c})
 \\\hline

$\Delta_{x'}^c$ &Half-dressed value of
$\Delta_{x'}$&(\ref{Deltasx_c})
 \\\hline

$\Delta_{z'}^c$ &Half-dressed value of
$\Delta_{z'}$&(\ref{Deltaz_c})
 \\\hline
\end{tabular}
\end{table}
\end{widetext}

\begin{widetext}
\begin{table}[h]  \caption{\label{tab: TableB}Overview of the exponents}
\begin{tabular}
{|c|c|c|}\hline  Symbols & Description& Eq.
\\\hline\hline

 $\eta_B$&Anomalous exponent of $B$&$\eta_B=0$\\\hline

$\eta_K$ &Anomalous exponent of $K$& (\ref{eta_K})\\\hline

$\eta_t$ & Anomalous exponent of $\Delta_t$&
(\ref{eta_Delta})\\\hline

$\eta_c$ &Anomalous exponent of $\Delta_c$& (\ref{eta_c})\\\hline

$\zeta$ &Anisotropy exponent for the model (\ref{H_AC})& (\ref{nuanis})\\\hline

$\tilde{\eta}_K$ &Anomalous exponent of $\tilde{K}$&
(\ref{Cnumexp1})\\\hline

$\tilde{\eta}_{\gamma}$ &Anomalous exponent of
$\gamma$&(\ref{Cnumexp2})
\\\hline

$\tilde{\eta}_{s'}$ &Anomalous exponent of
$\Delta_{s'}$&(\ref{Cnumexp3})
\\\hline

$\tilde{\eta}_{x'}$ &Anomalous exponent of
$\Delta_{x'}$&(\ref{etaxnum})
\\\hline

$\tilde{\zeta}_{x'}$ &Anisotropy exponent (between $q_{s'}$ and $q_{x'}$) for the model (\ref{H_C}) &(\ref{zetax'})
\\\hline

$\tilde{\zeta}_{z'}$ &Anisotropy exponent (between $q_{s'}$ and $q_{z'}$) for the model (\ref{H_C}) &(\ref{zetaz'})
\\\hline
\end{tabular}
\end{table}
\end{widetext}

\begin{widetext}
\begin{table}[h]  \caption{\label{tab: TableB}Overview of the characteristic lengths}
\begin{tabular}
{|c|c|c|}\hline  Symbols & Description& Eq.
\\\hline\hline

$\lambda$ & Smectic penetration length & $\lambda=\sqrt{K_0/B_0}$\\\hline

$\xi_{NL}^{\perp, z}$&Nonlinear crossover lengths&(\ref{NonlinearPerp}, \ref{NonlinearZ})\\\hline

$\xi_{\perp, z}$ & Correlation lengths for the phase transition&
(\ref{xis})\\\hline

$\xi_{\perp, z}^c$ & Crossover lengths between the power-law and broad x-ray scattering patters&
(\ref{xic})\\\hline

$\xi_{\perp, z}^x$ &Line widths of the broad x-ray scattering pattern& (\ref{fxwid}, \ref{fzwid})\\\hline

\end{tabular}
\end{table}
\end{widetext}

\  \


\begin{thebibliography}{99}
\bibitem{OP}
Strictly speaking, $\sin\theta_0$ is the magnitude of the order
parameter; this, of course, is very close to $\theta_0$ near the
transition, where $\theta_0$ is small since the transition is
continuous.
\end{thebibliography}
\end{document}